\pdfoutput=1
\documentclass[10pt,twocolumn,letterpaper]{article}
\usepackage[letterpaper,textwidth=6.75in,textheight=9.25in,columnsep=0.25in,centering,footskip=0.4in]{geometry}
\usepackage[round]{natbib}

\usepackage{amsmath,amssymb,amsthm,booktabs,graphicx,xcolor,microtype}
\usepackage[hidelinks]{hyperref}
\hypersetup{pdftitle={CoupleFlow: Null-Calibrated Auditory Attention Decoding With Time-Locked Brain-Speech Coupling and Flow-Matching Likelihood Ratios},pdfauthor={Ali Alavi, Donald S. Williamson}}
\usepackage{caption}
\usepackage{tikz}
\usepackage[section]{placeins}
\usetikzlibrary{positioning,arrows.meta,fit,patterns,calc,decorations.pathreplacing}

\newcommand{\abMax}{9}
\newcommand{\ablOtherMax}{0.8}
\newcommand{\ablRankDTU}{4.2}
\newcommand{\ablRankKUL}{3.0}
\newcommand{\ablRankMEG}{2.0}

\newcommand{\cfKUL}{72.7}

\newcommand{\deltaMax}{91}
\newcommand{\deltaMin}{71}
\newcommand{\fEff}{9.7}
\newcommand{\kEEGmax}{0.63}
\newcommand{\kEEGmin}{0.13}

\newcommand{\kMEG}{1.02}
\newcommand{\kNBrod}{0.181}
\newcommand{\kNUHD}{0.224}
\newcommand{\ksfBrod}{0.140}
\newcommand{\ksfUHD}{0.190}
\newcommand{\lagCtlAccDTU}{61.6}
\newcommand{\lagCtlAccKUL}{53.8}
\newcommand{\lagCtlAccMEG}{82.3}
\newcommand{\lagCtlDTU}{13.7}
\newcommand{\lagCtlKUL}{18.8}
\newcommand{\lagCtlMEG}{15.3}
\newcommand{\lagZeroDTU}{+1.2}
\newcommand{\lagZeroKUL}{+0.3}
\newcommand{\lagZeroMEG}{+1.5}
\newcommand{\lawLooMax}{2.0}
\newcommand{\lawLooMin}{0.4}
\newcommand{\lawMaeMax}{1.8}
\newcommand{\lawMaeMin}{0.2}
\newcommand{\leakMax}{+2.0}
\newcommand{\leakMin}{-0.3}
\newcommand{\maAblAux}{-1.3}
\newcommand{\maAblBanks}{-1.9}
\newcommand{\maAblGlimpse}{-1.4}
\newcommand{\maAblNce}{-3.6}
\newcommand{\maAblRank}{-8.7}

\newcommand{\maBeforeTrial}{46.3}

\newcommand{\maSafeMax}{0.8}

\newcommand{\maUnmatchedGap}{0.3}
\newcommand{\maeCompTrial}{70.2}

\newcommand{\maeLTrial}{56.9}

\newcommand{\maePLCon}{42.2}
\newcommand{\maePLTrial}{67.1}

\newcommand{\maePWCon}{53.6}
\newcommand{\maePWTrial}{78.3}

\newcommand{\maePkgTrial}{77.9}
\newcommand{\maePreTrial}{78.0}
\newcommand{\maeSnrP}{0.22}
\newcommand{\maeSnrRho}{0.03}

\newcommand{\maeWStThree}{53.2}
\newcommand{\maeWStTwo}{68.2}
\newcommand{\maeWTrial}{71.2}

\newcommand{\megDzPred}{2.14}
\newcommand{\megKcfObs}{1.62}
\newcommand{\megKlinObs}{2.03}
\newcommand{\megRratio}{2.10}
\newcommand{\mgBrod}{-0.1}
\newcommand{\mgDTU}{+3.8}
\newcommand{\mgESAA}{+3.8}
\newcommand{\mgFug}{+2.3}
\newcommand{\mgKUL}{+6.8}
\newcommand{\mgMEG}{+2.8}
\newcommand{\mgSparr}{+8.9}
\newcommand{\mgUHD}{+3.9}
\newcommand{\nCurves}{14}
\newcommand{\nWins}{seven}
\newcommand{\pmiGapMax}{25}
\newcommand{\pmiGapMin}{5}

\newcommand{\rSparr}{0.13}
\newcommand{\rTwoMax}{0.17}
\newcommand{\rTwoMin}{0.07}
\newcommand{\sdMax}{1.8}
\newcommand{\sdMin}{0.0}
\newcommand{\spAmax}{62}
\newcommand{\spAmin}{40}
\newcommand{\spFugMax}{71}
\newcommand{\spFugMin}{67}
\newcommand{\spKULBmax}{98.1}
\newcommand{\spKULBmin}{96.8}
\newcommand{\tEEGmax}{170}
\newcommand{\tEEGmin}{7}
\newcommand{\tMEG}{2.6}

\newcommand{\taskFacMax}{1.38}
\newcommand{\taskFacMean}{1.19}
\newcommand{\taskFacMin}{1.05}
\newcommand{\taskMaeCF}{0.07}
\newcommand{\taskMaeLin}{0.02}
\newcommand{\taskPLin}{0.004}
\newcommand{\taskRCF}{0.84}
\newcommand{\taskRLin}{0.95}
\newcommand{\thMape}{3.7}
\newcommand{\thR}{0.996}
\newcommand{\thetaMax}{22}
\newcommand{\thetaMin}{7}

\newcommand{\confEceCalMax}{0.028}
\newcommand{\confEceRawMax}{0.17}
\newcommand{\confEceRawMin}{0.08}
\newcommand{\confEnsMax}{2.1}
\newcommand{\confEnsMin}{1.6}
\newcommand{\confGainOneMax}{5.6}
\newcommand{\confGainOneMin}{5.1}
\newcommand{\confGainTwoMax}{7.9}
\newcommand{\confGainTwoMin}{5.3}
\newcommand{\confLagMax}{0.05}
\newcommand{\confLagMin}{0.05}
\newcommand{\confLinGainOneMax}{5.5}
\newcommand{\confLinGainOneMin}{3.8}
\newcommand{\confLinGainTwoMax}{8.0}
\newcommand{\confLinGainTwoMin}{3.3}
\newcommand{\confOvAdjMax}{2.7}
\newcommand{\confOvAdjMin}{0.5}
\newcommand{\confOvCorOne}{0.74}
\newcommand{\confOvCorThree}{0.25}
\newcommand{\confOvCorTwo}{0.48}
\newcommand{\confOvEceCorrMax}{0.02}
\newcommand{\confOvEceRawMax}{0.20}
\newcommand{\confOvFixedMax}{0.7}
\newcommand{\confOvSeqMax}{1.6}
\newcommand{\confOvSeqMin}{0.1}
\newcommand{\confSumMax}{4.2}
\newcommand{\confSumMin}{-0.1}
\newcommand{\confTheoryRatioMax}{1.08}
\newcommand{\confTheoryRatioMin}{0.92}
\newcommand{\confTlMixFiveSMax}{+2.6}
\newcommand{\confTlMixFiveSMin}{0.0}

\newcommand{\confTlMixKapMax}{1.22}
\newcommand{\confTlMixKapMin}{1.05}
\newcommand{\confTlMixOneSMax}{+3.8}
\newcommand{\confTlMixOneSMin}{+0.4}
\newcommand{\confTlMixOneSN}{8}
\newcommand{\confTlMixThirtySMax}{+3.8}
\newcommand{\confTlMixThirtySMin}{0.0}

\newcommand{\confTlMixTimeMax}{0.88}
\newcommand{\confTlMixTimeMin}{0.76}
\newcommand{\confTlOneFiveSMax}{+0.9}
\newcommand{\confTlOneFiveSMin}{-1.3}

\newcommand{\confTlOneOneSMax}{+5.9}
\newcommand{\confTlOneOneSMin}{+0.8}
\newcommand{\confTlOneOneSN}{8}
\newcommand{\confTlOneThirtySMax}{+0.8}
\newcommand{\confTlOneThirtySMin}{-1.8}

\newcommand{\confTlOneTimeMax}{0.84}
\newcommand{\confTlOneTimeMin}{0.74}
\newcommand{\confTlTenFiveSMax}{-0.1}
\newcommand{\confTlTenFiveSMin}{-1.4}

\newcommand{\confTlTenOneSMax}{-0.2}
\newcommand{\confTlTenOneSMin}{-4.3}

\newcommand{\confVarMax}{1.13}
\newcommand{\confVarMin}{0.99}
\newcommand{\confDetAboveMax}{+9.9}
\newcommand{\confDetAboveMin}{-2.2}
\newcommand{\confDetForcedAccMax}{79.6}
\newcommand{\confDetForcedAccMin}{58.4}
\newcommand{\confDetForcedWorst}{Broderick}
\newcommand{\confDetLongDevMax}{9.7}
\newcommand{\confDetOneAccMax}{80.2}
\newcommand{\confDetOneAccMin}{74.9}

\newcommand{\confDetOneMedMax}{7}
\newcommand{\confDetOneMedMin}{3}
\newcommand{\confDetOneNominal}{73.1}
\newcommand{\confDetOneOneSec}{3}
\newcommand{\confDetOneOneSecMax}{20}
\newcommand{\confDetOnePNineMax}{21}
\newcommand{\confDetOnePNineMin}{7}

\newcommand{\confDetShortDevMax}{1.9}
\newcommand{\confDetTwoAccMax}{90.6}
\newcommand{\confDetTwoAccMin}{86.1}
\newcommand{\confDetTwoForcedMax}{37.6}
\newcommand{\confDetTwoMedMax}{17}
\newcommand{\confDetTwoMedMin}{6}
\newcommand{\confDetTwoNominal}{88.1}

\newcommand{\confDetTwoPNineMax}{49}
\newcommand{\confDetTwoPNineMin}{14}

\newcommand{\netSOne}{+1.15}
\newcommand{\netSTwo}{-1.15}
\newcommand{\netTau}{0.063}

\newcommand{\oldSegKUL}{86.4}
\newcommand{\oldTrialKUL}{55.9}

\newcommand{\oldSegNJU}{86.0}

\newcommand{\oldDropAllMin}{16.7}
\newcommand{\oldDropAllMax}{31.5}

\newcommand{\grpKULTrials}{320}
\newcommand{\grpKULGroups}{4}

\newcommand{\grpDTUGroups}{60}
\newcommand{\grpFugTrials}{1401}
\newcommand{\grpFugGroups}{4}

\newcommand{\grpESAAGroups}{32}

\newcommand{\grpUHDGroups}{2}

\newcommand{\grpBrodGroups}{30}

\newcommand{\grpSparrGroups}{63}

\newcommand{\grpMEGGroups}{30}

\newcommand{\grpMaeGroups}{100}
\newcommand{\grpMinR}{0.964}

\newcommand{\grpNullMax}{0.93}
\newcommand{\grpNullMed}{0.53}

\newcommand{\losoMEGFive}{95.0}

\newcommand{\losoMEGpca}{54.1}
\newcommand{\losoDropMin}{3.4}
\newcommand{\losoDropMax}{9.2}

\newcommand{\nullShortMin}{49.7}
\newcommand{\nullShortMax}{50.5}
\newcommand{\nullLongMin}{42.5}
\newcommand{\nullLongMax}{61.1}

\newcommand{\cmpTwentyMax}{+8.5}
\newcommand{\cmpLosoMin}{2.0}
\newcommand{\cmpLosoMax}{8.3}
\newcommand{\rCFattMin}{0.082}
\newcommand{\rCFattMax}{0.148}
\newcommand{\rCFmeg}{0.396}

\newcommand{\cmpGrowN}{5}
\newcommand{\rCFhigherN}{6}
\newcommand{\cmpOneMM}{10.9}
\newcommand{\hiAcc}{80.6}
\newcommand{\nhAcc}{78.5}
\newcommand{\trainMinMin}{5}
\newcommand{\trainMinMax}{215}
\newcommand{\kulExFoldZero}{72.8}
\newcommand{\esaaLateFive}{71.2}

\newcommand{\maMainNullMin}{24.8}
\newcommand{\maMainNullMax}{25.3}
\newcommand{\pmiPriorMin}{47.0}
\newcommand{\pmiPriorMax}{50.9}
\newcommand{\dRuncMaxOther}{0.005}
\newcommand{\dRuncKUL}{-0.012}
\newcommand{\linInnerCV}{137}

\newcommand{\nceDropMin}{30}
\newcommand{\nceDropMax}{83}

\newcommand{\archPerListener}{256}

\newcommand{\archDecision}{25{,}689}
\newcommand{\archFlow}{85{,}505}
\newcommand{\archAudIn}{54}
\newcommand{\archAudInEnv}{4}
\newcommand{\archTotMin}{126{,}607}
\newcommand{\archTotMax}{148{,}879}

\newtheorem{proposition}{Proposition}

\newcommand{\PMI}{\mathrm{PMI}}

\begin{document}
\title{CoupleFlow: Null-Calibrated Auditory Attention Decoding\\ With Time-Locked
Brain--Speech Coupling\\ and Flow-Matching Likelihood Ratios}
\author{\begin{tabular}[t]{c}Ali Alavi\\ \normalsize\texttt{alavibajestan.1@osu.edu}\end{tabular}\hspace{3em}\begin{tabular}[t]{c}Donald S. Williamson\\ \normalsize\texttt{williamson.413@osu.edu}\end{tabular}\\[0.5em]
\normalsize Department of Computer Science and Engineering, The Ohio State University}
\date{}
\maketitle

\begin{abstract}

Auditory attention decoding (AAD) seeks to identify the attended talker from brief noninvasive brain recordings by comparing neural responses with competing speech streams. Recent electroencephalography (EEG) decoders report nearly perfect accuracy from one second of data, yet we show that this performance fails to generalize to unseen speech because models exploit trial leakage, eye gaze, and acoustic cues instead of speech evoked neural activity. We introduce \emph{CoupleFlow}, which decodes attention through temporal coupling between neural and acoustic embeddings. Its predictions remain at chance when the brain signal contains no informative coupling, preventing shortcuts based on speech content or talker identity. A conditional flow matching head implements a likelihood ratio decision rule based on pointwise mutual information between brain and speech. This rule is Bayes optimal when the recording depends only on the attended talker and removes preferences for speech that is likely a priori. We evaluate CoupleFlow with content disjoint splits recovered from the audio and null tests that distinguish genuine temporal coupling from trial recognition. Across nine EEG and magnetoencephalography (MEG) datasets processed through a common pipeline, spanning four languages, high density EEG recordings, listeners with hearing impairment, and a room with multiple talkers, CoupleFlow outperforms linear and deep baselines while all null tests remain at chance. Finally, we derive an evidence accumulation law that predicts accuracy across window lengths from a single parameter per dataset and predicts that parameter from the strength of neural speech tracking. This law identifies recording signal to noise ratio as a limit on decoding speed and reliability and estimates the window length required to reach a target accuracy.
\end{abstract}

\section{Introduction}

When several people talk at once, a listener naturally isolates one voice. The auditory cortex reflects this selection: low-frequency cortical activity tracks the attended talker's speech envelope far more strongly than ignored speech.
\citep{ding2012emergence,mesgarani2012selective,osullivan2015attentional}. Auditory
attention decoding (AAD) asks whether the attended talker can be identified from this
activity recorded non-invasively, with electroencephalography (EEG) or magnetoencephalography
(MEG). %
EEG is promising because it offers millisecond time scale resolution and works with a few electrodes near the ear \citep{mirkovic2015decoding,fiedler2017single}. This makes AAD a
candidate for neuro-steered hearing devices that amplify the attended talker
\citep{geirnaert2021eeg}, since hearing-impaired
listeners track the attended talker more strongly than the competing one, as normal-hearing
listeners do \citep{petersen2017neural,decruy2020hearing,fuglsang2020effects}. Building such a device is
beyond the scope of this paper, but we study the decoding problem in terms of its ability to generalize to unheard speech using limited data from unseen subjects, solely from brain activity. 

Using a few seconds of multichannel neural data and the speech of $K$ candidate talkers, the decoder identifies the attended voice.
The problem is difficult because the neural response to speech is weak (a linear decoder's reconstruction correlates with the attended envelope at only $r=\rTwoMin$--$\rTwoMax$ on our two-talker EEG datasets),
differs across listeners, and is mixed with activity unrelated to the audio stimulus, while the
candidates often differ in ways a decoder can exploit without reading the brain: loudness,
voice, or the identity of the recording session.

AAD decoders follow two strategies: Stimulus reconstruction and Direct classification. Stimulus
reconstruction uses the EEG to estimate of the attended speech envelope,  and it selects the candidate
talker that correlates best with this envelope. Direct classification maps EEG to a label, usually the
side of the attended talker. In both cases, however, performance usually fails short of expectations (details in App.~\ref{app:related}). With stimulus reconstruction, linear backward models and canonical correlation analysis \citep{osullivan2015attentional, crosse2016mtrf,decheveigne2018decoding} are robust,
but they require long signal durations (i.e., tens of seconds) for high accuracy, which is not practical. Deep nonlinear reconstructions have not changed this picture. The best of them still
reconstruct the envelope weakly ($r\approx0.19$ for VLAAI; \citealp{accou2023vlaai}), nonlinear
decoders did not consistently outperform linear ones in an independent comparison on two
public datasets \citep{geirnaert2021eeg}, and many reported gains come from evaluations in
which training and test data overlap \citep{puffay2023relating}.
Conversely, direct classifiers \citep{vandecappelle2021eeg, darnet2024, dbpnet2024, listennet2025} report 95--97\% at 1\,s, but their accuracy rests largely on shortcuts: training and test windows come from the same trial, so the classifier can recognise the trial rather than decode attention (leakage), and it can read eye gaze toward the attended talker. When training and test data come from different trials, their accuracy drops to near chance \citep{rotaru2024what}.
Furthermore, datasets replay the same stories to every listener, and test
windows are often cut from training trials \citep{puffay2023relating}. %
It is also not inherently clear that decoders are not overfitting to audio cues, such as loudness, spectral
differences, or trial identity, indicating that the decoder is ignoring the brain signal.%
Most studies report accuracy at only one or two window lengths, so the window needed for a given accuracy cannot be predicted. Each study also preprocesses each dataset its own way, so numbers cannot be compared across studies or datasets. Finally, in one widely used dataset (NJU) we find that the EEG is not time-aligned with the audio at any offset, so results on it cannot reflect speech tracking (App.~\ref{app:data}).

Our approach addresses the AAD problem by ensuring that decoding is driven only by the brain's time-locked response to speech, while quantifying the effects of window length and recording quality on accuracy. %
Our approach scores candidates based solely on the temporal covariance between learned acoustic and EEG embeddings. The EEG representation is centered so that signals without time-varying information assign equal scores to all candidates. A flow-matching model adds a likelihood-ratio decision rule, while content-disjoint splits and null analyses quantify the contribution of neural activity to decoding performance.

\textbf{Contributions.} %
(i) A decoder that can only use the brain's response to speech. CoupleFlow (Sec.~\ref{sec:method}) chooses the attended talker using only how the EEG follows each talker's speech over time. It is therefore guaranteed to be at chance when the EEG carries no information about the speech. Examples are EEG from a different trial, EEG shifted in time so that it no longer lines up with the speech, or random noise. Differences in loudness, voice or trial identity cannot raise its accuracy. A generative head adds a new decision rule for AAD, which is optimal under a standard assumption.

(ii) A leakage-free benchmark. We evaluate on nine public EEG and MEG datasets under one preprocessing pipeline. Test speech is never heard in training, and control tests check that accuracy comes from the brain. Under this protocol, CoupleFlow outperforms linear and deep baselines trained on the same data, and we show where previously reported accuracies came from.

(iii) A law for accuracy versus window length. A simple formula predicts decoding accuracy at any window length from one number per dataset, and predicts that number from how strongly the brain tracks speech. It shows that fast, reliable decoding is limited by the quality of the recording, and it gives the window needed for any target accuracy.

\section{CoupleFlow}\label{sec:method}
\begin{figure*}[!t]\centering
\resizebox{\textwidth}{!}{\input{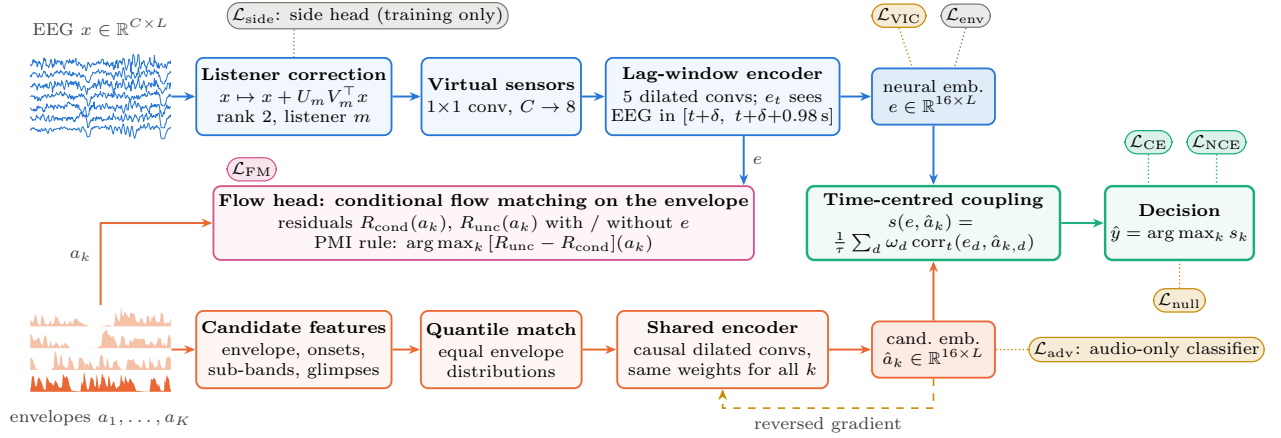}}
\caption{CoupleFlow. The two small signal plots show the input of one real MAESTRO test
window (listener S03): seven of its EEG channels and the envelopes of its $K=4$ candidate
talkers, with the attended talker in dark orange. The neural path (blue) maps the EEG $x$ to an
embedding $e$ whose value at time $t$ sees only EEG from $t+\delta$ to
$t+\delta+0.98$\,s after the sound ($\delta\in\{0,125\}$\,ms). The acoustic path (orange) embeds each
candidate $k$ with one shared encoder. The paths meet only in the coupling score $s_k$
(Eq.~\ref{eq:score}), which makes the decision. The flow head (pink) supplies the alternative
pointwise-mutual-information (PMI) rule (Prop.~\ref{prop:pmi}) and per-source residuals. Tags mark where each loss in
Eq.~\ref{eq:loss} acts, coloured by role: decoding (green), density (pink), anti-shortcut
(gold), auxiliary (grey). Dashed: reversed gradient of the audio-only adversary.}\label{fig:arch}
\end{figure*}
A decision window of $T$ seconds ($L=64T$ samples at 64\,Hz) gives the recording
$x\in\mathbb{R}^{C\times L}$ of $C$ channels, the speech envelopes
$a_1,\dots,a_K\in\mathbb R^{L}$ of $K$ candidate talkers, and the index $y$ of the attended
one. Fig.~\ref{fig:arch} shows how CoupleFlow turns them into a decision. A neural path
(blue) embeds the recording and an acoustic path (orange) embeds each candidate, both as
time series of the same size. The two paths meet only in a coupling score (green), which
makes the decision, and a flow head (pink) supplies a second decision rule.
App.~\ref{app:net} gives every layer.

The EEG first goes through a \emph{listener correction}, a small low-rank adjustment added
to a model shared by all listeners. It absorbs individual differences such as head shape and
cap placement without fitting a separate model to each listener's few minutes of data. A
layer of \emph{virtual sensors} then reduces the channels to a few learned spatial filters,
as backward models do, and a \emph{lag-window encoder} turns them into the neural embedding
$e$. Each time step of $e$ sees only the second of EEG that follows the corresponding sound
(from a lag $\delta\in\{0,125\}$\,ms on), which is where a cortical response to that sound
must lie. EEG before the sound cannot be a response to it, and serves as our lag control.

The acoustic path starts from \emph{candidate features}: each talker's envelope and onsets
and, where the audio is available, sub-band envelopes and \emph{glimpsed/masked} channels
that mark where the talker is louder or quieter than the competing one
\citep{brodbeck2020neural}. In training, a \emph{quantile match} gives the candidates'
envelopes identical value distributions, which removes loudness and level cues. A
\emph{shared encoder}, with the same weights for every candidate, then maps each candidate
to an embedding $\hat a_k$ of the same size as $e$. It sees only past audio, so the brain
response it is compared with must follow the sound.

The \emph{coupling score} measures how each candidate's embedding co-varies in time with
the neural embedding:
\begin{equation}
s(e,\hat a)=\frac{1}{\tau}\sum_{d=1}^{16} \omega_d\,
\frac{\langle e_{\cdot d}-\bar e_d,\ \hat a_{\cdot d}-\bar{\hat a}_{d}\rangle}
{\|e_{\cdot d}-\bar e_d\|\,\|\hat a_{\cdot d}-\bar{\hat a}_{d}\|},
\label{eq:score}
\end{equation}
where $e_{\cdot d}\in\mathbb R^{L}$ is dimension $d$ of $e$ over time and $\bar e_d$ its time
average (likewise for $\hat a$). Each fraction is the Pearson correlation over time between one
neural and one acoustic dimension; the weights $\omega_d$ and the temperature $\tau$ are
learned. Candidate $k$ scores $s_k=s(e,\hat a_k)-\frac1K\sum_{j}s(e,\hat a_j)$, and the
decision is $\hat y=\arg\max_k s_k$. Because the correlation removes each dimension's time
average, a neural embedding that does not vary in time gives every candidate the score 0.
The audio alone then cannot favour any candidate, and the decoder is at chance whenever the
EEG carries no information about the speech (Prop.~\ref{prop:null}).

Alongside the coupling score, the \emph{flow head}, a conditional flow-matching network
\citep{lipman2023flow}, learns to generate the attended envelope from $e$. Its condition is
dropped in part of the training windows \citep{ho2022classifierfree}, so the same network
also learns to generate envelopes without the brain. At test time each candidate passes
through it with and without $e$, and how much the recording lowers the network's error
approximates the pointwise mutual information $\PMI(a_k;x)=\log p(a_k\mid x)-\log p(a_k)$.
Picking the candidate with the largest PMI is a second decision rule. It is Bayes-optimal
when the recording depends only on the attended talker (Prop.~\ref{prop:pmi}), and since
$\log p(a_k)$ cancels, no candidate can win by being typical speech.

The model is trained end to end on
\begin{align}
\mathcal L={}&\underbrace{\mathcal L_{\rm CE}+\lambda_{\rm NCE}\mathcal L_{\rm NCE}}_{\text{decoding}}
+\underbrace{\lambda_{\rm FM}\mathcal L_{\rm FM}}_{\text{density}}\nonumber\\
&+\underbrace{\lambda_{\rm null}\mathcal L_{\rm null}+\lambda_{\rm adv}\mathcal L_{\rm adv}
+\lambda_{\rm VIC}\mathcal L_{\rm VIC}}_{\text{anti-shortcut}}\nonumber\\
&+\underbrace{\lambda_{\rm side}\mathcal L_{\rm side}+\lambda_{\rm env}\mathcal L_{\rm env}}_{\text{auxiliary}},
\label{eq:loss}
\end{align}
and the tags in Fig.~\ref{fig:arch} mark where each term acts. The decoding terms ask for
the right decision, through cross-entropy over the candidates and a contrastive term that
matches each window's EEG to its own attended speech among all windows of the batch. The
anti-shortcut terms close routes around the brain: a hinge requires the attended speech to
score higher with its own recording than with another window's, an audio-only adversary,
through a reversed gradient, strips acoustic cues to the attended talker from the shared
encoder, and a VICReg term keeps the neural embedding from collapsing. The density term trains the
flow head. The auxiliary terms train a side head, which predicts the attended side from the
EEG in training only, and an envelope read-out that reports tracking
(Table~\ref{tab:track}); neither enters a reported decision. Checkpoints are selected on
validation \emph{contribution}, accuracy minus the across-trial null, which rewards what the
brain adds rather than what the audio or trial identity allows. App.~\ref{app:loss} defines
every term, gives its weight (set once for all datasets) and reports the effect of removing
it.

\section{Theory}\label{sec:theory}
\textbf{Setting.} A decision window gives a recording $x$, candidate envelopes
$a_{1:K}=(a_1,\dots,a_K)$ and the attended index $y\in\{1,\dots,K\}$, drawn jointly. A
decoder gives candidate $k$ a score $s_k$ and predicts $\hat y=\arg\max_k s_k$; a tie for the
maximum counts as an error.
\begin{proposition}[null calibration]\label{prop:null}
Assume (a) $s_k=g(x,a_k)$ for one function $g$ shared by all candidates; (b) $H_0$: $x$ is
independent of $(y,a_{1:K})$; (c) given $y$, the joint law of $a_{1:K}$ is invariant to
permuting the candidate indices. Then under $H_0$, $P(\hat y=y)\le 1/K$.
\end{proposition}
Eq.~\ref{eq:score} satisfies (a); the across-trial and circular nulls realise (b).
\begin{proposition}[PMI is Bayes-optimal]\label{prop:pmi}
Let $p(a)$ be the density of the attended envelope $a_y$ and $p(a\mid x)$ its density given
the recording, and $\PMI(a;x)=\log p(a\mid x)-\log p(a)$. Assume (a) the recording depends on
the candidates only through the attended one, $p(x\mid a_{1:K},y)=p(x\mid a_y)$; (b) $y$ is
uniform on $\{1,\dots,K\}$ and independent of $a_{1:K}$. Then the maximum a posteriori
(MAP) decision is $\hat y=\arg\max_k\PMI(a_k;x)$. Scoring by $\log p(a_k\mid x)$ instead adds
$\log p(a_k)$, a term that depends on the audio alone.
\end{proposition}
\begin{proposition}[evidence accumulation]\label{prop:law}
Let a decoder output a time series $\tilde a$ over a window of $T$\,s (e.g.\ a reconstructed
envelope), let $r_{\rm att},r_{\rm neg}$ be its sample Pearson correlations with the attended and
the other (competing, or match--mismatch) envelope and $\rho_{\rm att},\rho_{\rm neg}$ their
population values, and let the decision be correct when $r_{\rm att}>r_{\rm neg}$. Assume (i)
the window holds $n_{\rm eff}=fT$ effectively independent samples ($f$ per second); (ii) the
Fisher transforms $z_j=\operatorname{atanh}r_j\sim\mathcal N(\zeta_j,\frac{1}{n_{\rm eff}-3})$ with
$\zeta_j=\operatorname{atanh}\rho_j$, $j\in\{{\rm att},{\rm neg}\}$; (iii) $z_{\rm att},z_{\rm neg}$
are independent. With $\Delta\zeta=\zeta_{\rm att}-\zeta_{\rm neg}$,
\[P(\mathrm{correct}\mid T)=\Phi\big(\Delta\zeta\sqrt{(fT-3)/2}\big)\approx\Phi(\kappa\sqrt T),\]
with $\kappa=\Delta\zeta\sqrt{f/2}$, and accuracy $p$ needs $T_p=(\Phi^{-1}(p)/\kappa)^2$\,s. If further $\tilde a$ is a signal plus
independent noise with per-sample signal-to-noise ratio (SNR) $\sigma$ (variance ratio), and $\rho^\ast_j$
are the noise-free correlations, then $\rho_j=\rho^\ast_j\sqrt{\sigma/(1+\sigma)}$ and, for
small $\rho_j$, $\kappa^2\propto\sigma/(1+\sigma)$.
\end{proposition}
Proofs are in App.~\ref{app:proofs}. Prop.~\ref{prop:law} makes $\kappa$ a single,
window-independent figure of merit; for CoupleFlow, whose score is not a single
correlation, we fit $\kappa$ per curve. Model quality and training data act through the
noiseless correlation, and recording quality through $\sigma$.

\section{Experiments}\label{sec:exp}
\textbf{Data.} We test on nine public datasets (Table~\ref{tab:data}), chosen so that a gain
cannot come from one recording setup. Six are two-talker EEG studies in Danish, Dutch,
Mandarin and English: KU Leuven \citep{das2016kul}, DTU \citep{fuglsang2017dtu}, Fuglsang'20
\citep{fuglsang2020effects}, whose 44 older listeners include 22 with hearing loss, ESAA
\citep{li2022esaa}, and two high-density sets with 255 \citep{narayanan2021uhd} and 128
electrodes \citep{broderick2018semantic}. SparrKULee \citep{accou2024sparrkulee} and MEG-MASC
\citep{gwilliams2023megmasc} have a single talker. They test whether a decoder can tell the
heard speech from other speech, and MEG-MASC shows how far decoding goes when the recording
is much cleaner than EEG. MAESTRO \citep{maestro2026} adds a room in which the listener
attends one of four talkers.

\textbf{Preprocessing.} Published results preprocess each dataset differently, so
differences between datasets partly reflect the pipeline. We therefore run every raw
recording through one pipeline (App.~\ref{app:data}). It keeps 0.5--32\,Hz, which contains the
slow cortical tracking of the speech envelope \citep{ding2012emergence}, and removes drift
and most muscle noise. It maps every cap onto the same 64 channels, so one network serves all
datasets, and resamples to 64\,Hz, enough for this band. We compute the speech envelope with
a gammatone filterbank, a standard model of the cochlea; two datasets ship no audio, and for
them we use the authors' envelopes. Finally, we check that the EEG of every dataset tracks
its audio at a plausible latency. One widely used dataset fails this check and is excluded
(App.~\ref{app:data}).

\textbf{Protocol.} Test speech is never heard in training. Most datasets play the same
stories to every listener, and some replay them, so holding out trials does not hold out
speech. We therefore group the trials that contain the same speech, found by correlating
their envelopes since the releases lack a reliable stimulus key, and build each test fold
from whole groups (\emph{content-disjoint} folds; App.~\ref{app:data}). We report three
candidate sets: \emph{competing} (comp.), the attended vs.\ the other talker; \emph{matched},
the same pair after quantile matching, which removes loudness cues; and \emph{same-talker}
(match--mismatch, mm), the attended speech vs.\ the same talker's speech from another window
of the trial, the only task for single-talker data. Every accuracy comes with nulls that
make the EEG uninformative: another trial's recording of the same listener
(\emph{across-trial}) and, at 5\,s, the trial's own recording shifted by at least 2\,s
(\emph{circular}), which keeps trial identity but breaks time-locking.
\textbf{Baselines.} We compare with one decoder from each main family, all trained on the
same folds, windows and inputs and stopped early on validation data: the per-listener linear
backward model, still the standard for AAD; VLAAI \citep{accou2023vlaai}, a deep
reconstruction model trained across listeners; and a dilated-convolution match--mismatch
model (Dil.\ MM; \citealp{accou2021dilated}), which, like CoupleFlow, compares the EEG with
each candidate directly. A recursive decoder (DECAF) stays near chance on the two-talker sets
and appears only in App.~Table~\ref{tab:app_baselines}; training details are in
App.~\ref{app:data}.

\begin{table*}[!t]\centering\small
\caption{Accuracy (\%) of CoupleFlow (Ours) and the baselines at 1, 5 and 20\,s on the
content-disjoint folds (every model trained and tested on the same folds), and on unseen
listeners (LOSO: listener and content held out, 5\,s) against the pooled linear backward
model. comp.: attended vs.\ competing talker; mm: match--mismatch. Ours: mean of 3 seeds
(s.d.\ \sdMin--\sdMax{} at 5\,s); LOSO: one seed. Bold: best per group. CoupleFlow's
EEG-shuffle nulls lie in [\nullShortMin, \nullShortMax] up to 10\,s. All windows and per-trial accuracy: App.~Table~\ref{tab:app_baselines};
full LOSO results: App.~Table~\ref{tab:loso}.}\label{tab:main}
\setlength{\tabcolsep}{3.2pt}
\resizebox{\textwidth}{!}{\begin{tabular}{ll|cccc|cccc|cccc|cc}\toprule
 & & \multicolumn{4}{c|}{1\,s} & \multicolumn{4}{c|}{5\,s} & \multicolumn{4}{c|}{20\,s} & \multicolumn{2}{c}{LOSO, 5\,s} \\
Dataset & task & Linear & VLAAI & Dil.\ MM & Ours & Linear & VLAAI & Dil.\ MM & Ours & Linear & VLAAI & Dil.\ MM & Ours & Linear & Ours \\\midrule
KU Leuven & comp. & 57.0 & 56.9 & 55.3 & \textbf{60.5} & 64.9 & 63.8 & 65.9 & \textbf{72.7} & 75.1 & 73.0 & 78.2 & \textbf{86.7} & 59.7 & \textbf{68.0} \\
DTU & comp. & 59.1 & 57.5 & 56.0 & \textbf{61.4} & 72.4 & 64.8 & 69.1 & \textbf{76.2} & 86.1 & 74.1 & 83.7 & \textbf{90.6} & 64.0 & \textbf{70.4} \\
Fuglsang'20 & comp. & 63.4 & 60.6 & 57.9 & \textbf{63.8} & 77.3 & 68.9 & 73.0 & \textbf{79.6} & 91.9 & 81.7 & 89.0 & \textbf{93.3} & 67.5 & \textbf{73.7} \\
ESAA & comp. & 61.3 & 58.3 & 57.5 & \textbf{63.2} & 74.1 & 66.4 & 70.6 & \textbf{77.9} & 87.6 & 78.8 & 84.7 & \textbf{90.6} & 66.1 & \textbf{68.6} \\
KUL-255ch & comp. & 57.0 & 55.9 & 54.7 & \textbf{58.9} & 65.5 & 62.1 & 65.0 & \textbf{69.4} & 76.2 & 72.1 & 77.6 & \textbf{83.0} & 64.0 & \textbf{66.0} \\
Broderick & comp. & \textbf{56.5} & 51.0 & 50.5 & 55.8 & \textbf{63.7} & 52.2 & 52.3 & 63.6 & \textbf{75.9} & 53.1 & 54.4 & 75.2 & 49.2 & \textbf{50.1} \\
\midrule
SparrKULee & mm & 61.8 & 59.9 & 62.6 & \textbf{73.5} & 76.3 & 70.5 & 83.3 & \textbf{92.2} & 90.5 & 83.9 & 96.4 & \textbf{99.3} & -- & \textbf{85.9} \\
MEG-MASC & mm & 72.7 & 71.0 & 72.0 & \textbf{83.7} & 92.0 & 89.1 & 96.2 & \textbf{99.0} & 99.4 & 98.6 & \textbf{100.0} & \textbf{100.0} & -- & \textbf{95.0} \\
\bottomrule\end{tabular}
}
\end{table*}

\textbf{Main results} (Table~\ref{tab:main}; leave-one-listener-out, LOSO: listener and
content held out together). At 5\,s CoupleFlow is best on \nWins{} of eight
datasets. Its margin over the strongest baseline is \mgKUL{} points on KU Leuven, \mgDTU{} on
DTU, \mgFug{} on Fuglsang'20, \mgESAA{} on ESAA, \mgUHD{} on the 255-channel set, \mgSparr{} on
SparrKULee and \mgMEG{} on MEG. On Broderick it ties the per-listener linear decoder
(\mgBrod, within one seed s.d.) and stays within one point of it at every window.  On the
\cmpGrowN{} other two-talker EEG sets the margin grows from 1 to 20\,s, to at most
\cmpTwentyMax{} points. On the single-talker
match--mismatch sets the margin is largest at 1\,s (at least \cmpOneMM{} points) and shrinks
as both models approach 100\%. On unseen listeners CoupleFlow keeps
\cmpLosoMin--\cmpLosoMax{} points over the pooled linear model wherever either transfers; on
Broderick neither does (App.~\ref{app:loso}). The input features were chosen during
development on KU Leuven fold~0 (App.~\ref{app:data}); without that fold KU Leuven scores
\kulExFoldZero\% at 5\,s.
\textbf{Lag window.} The lag $\delta$ is the one hyperparameter set per dataset, on validation
accuracy only (App.~\ref{app:data}): 0\,ms for DTU, ESAA, the 255-channel set, SparrKULee,
MEG and MAESTRO, 125\,ms otherwise. It matters most on ESAA, whose audio onset had to be
estimated (\esaaLateFive\% at 5\,s with 125\,ms; App.~Table~\ref{tab:app_variants}).

\begin{figure*}[!t]\centering
\includegraphics[width=0.85\textwidth]{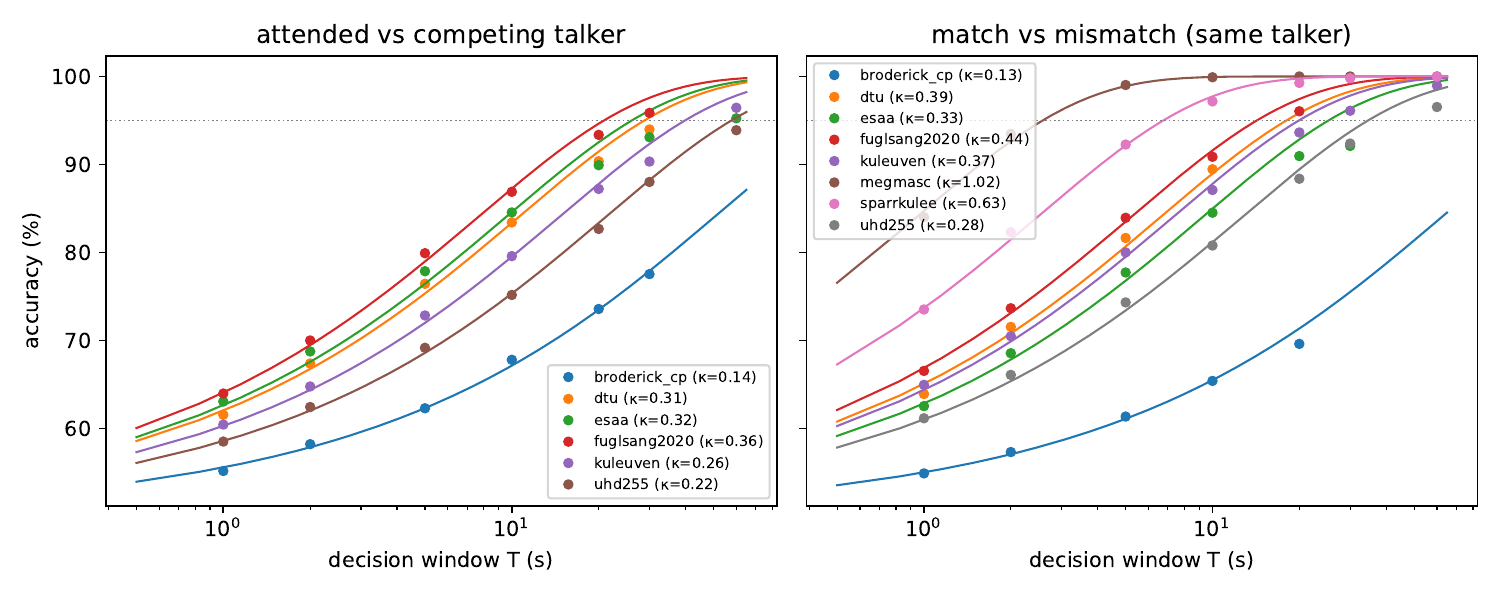}
\caption{Accuracy vs.\ window length (log scale). Markers: CoupleFlow (seed 0), pooled over folds.
Lines: $\Phi(\kappa\sqrt T)$ with one $\kappa$ per curve (Prop.~\ref{prop:law}).}
\label{fig:law}
\end{figure*}

\textbf{The accumulation law holds.} Fig.~\ref{fig:law}: one $\kappa$ per curve fits all
\nCurves{} dataset--task curves from 1 to 60\,s with mean absolute error \lawMaeMin--\lawMaeMax{}
points, and \lawLooMin--\lawLooMax{} points when predicting a held-out window. EEG gives
$\kappa$ = \kEEGmin--\kEEGmax{} ($T_{95}$, the window for 95\% accuracy, = \tEEGmin--\tEEGmax\,s). MEG gives \kMEG{}
($T_{95}\!\approx\!\tMEG$\,s). All native
channels raise $\kappa$ from \ksfBrod{} to \kNBrod{} (Broderick) and from \ksfUHD{} to
\kNUHD{} (253 channels; App.~Table~\ref{tab:channels}).

\begin{figure*}[!t]\centering
\begin{minipage}[b]{0.55\textwidth}\centering
\includegraphics[width=\linewidth]{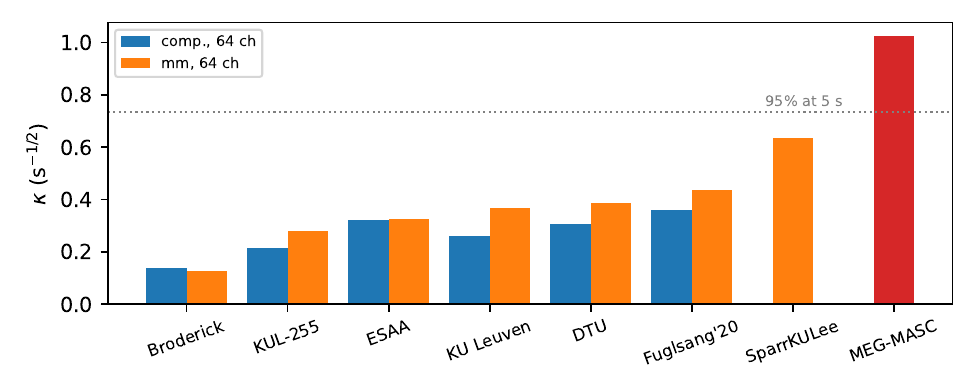}
\caption{Accumulation-law $\kappa$ per dataset. Bars: 64 canonical
channels, MEG-MASC 208 sensors (comp.: attended vs.\ competing; mm: match--mismatch). Dotted:
the $\kappa$ needed for 95\% at 5\,s.}\label{fig:kappa}
\end{minipage}\hfill
\begin{minipage}[b]{0.42\textwidth}\centering
\captionof{table}{Envelope tracking $r$ with the attended and the competing envelope.
Linear: per-listener backward model, held-out trials (App.~Table~\ref{tab:kth}). Ours:
linear read-out of the neural embedding ($\mathcal L_{\rm env}$), 5-s test windows,
content-disjoint folds, mean of 3 seeds.}\label{tab:track}
{\small\setlength{\tabcolsep}{4pt}\begin{tabular}{lcccc}\toprule
 & \multicolumn{2}{c}{Linear (trials)} & \multicolumn{2}{c}{Ours (5\,s)} \\
\cmidrule(lr){2-3}\cmidrule(lr){4-5}
Dataset & $r_{\rm att}$ & $r_{\rm comp}$ & $r_{\rm att}$ & $r_{\rm comp}$ \\\midrule
KU Leuven & 0.099 & 0.027 & 0.133 & 0.024 \\
DTU & 0.130 & 0.023 & 0.142 & 0.025 \\
Fuglsang'20 & 0.171 & 0.018 & 0.148 & 0.006 \\
ESAA & 0.139 & 0.019 & 0.136 & 0.021 \\
KUL-255ch & 0.101 & 0.018 & 0.115 & 0.024 \\
Broderick & 0.073 & 0.003 & 0.082 & 0.048 \\
\midrule
SparrKULee & 0.132 & -- & 0.172 & -- \\
MEG-MASC & 0.278 & -- & 0.396 & -- \\
\bottomrule\end{tabular}
}
\end{minipage}
\end{figure*}

\begin{figure}[!t]\centering
\includegraphics[width=0.85\columnwidth]{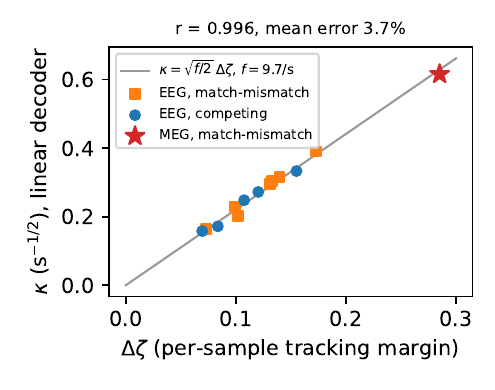}
\caption{Prop.~\ref{prop:law} predicts $\kappa$ from per-sample tracking. Each point is one
dataset and task (linear decoder); $\Delta\zeta$ is the Fisher-$z$ margin between the
attended and the negative candidate's correlation with the reconstruction. One shared
constant $f$ fits EEG and MEG, competing and match--mismatch alike.}\label{fig:kth}
\end{figure}

\textbf{Why does MEG reach $\kappa\approx1$: the sensor, or the easier task?}
Prop.~\ref{prop:law} separates the two. A match--mismatch negative is uncorrelated with the
response ($\rho_{\rm neg}=0$), while a competing talker is weakly tracked,
$\rho_{\rm neg}=\gamma\rho_{\rm att}$ with $\gamma$ the ignored-talker ratio, so for small $\rho$
the task alone predicts $\kappa_{\rm comp}/\kappa_{\rm mm}=1-\gamma$. With $\gamma$ measured by
the per-listener linear decoder (Table~\ref{tab:track}), this holds on all six two-talker
datasets (mean error \taskMaeLin, $r=\taskRLin$): the task raises $\kappa$ by a factor
\taskFacMin--\taskFacMax, not more. With one effective rate $f=\fEff$\,samples/s,
$\kappa=\sqrt{f/2}\,\Delta\zeta$ fits all 14 dataset--task values of the linear decoder
($r=\thR$; Fig.~\ref{fig:kth}), and MEG lies on the same line as EEG. Its advantage is
therefore a per-sample signal-to-noise effect (details in App.~\ref{app:lawdetails}).
CoupleFlow's envelope read-out, an auxiliary head evaluated on the stricter content-disjoint
folds, tracks the attended envelope better than the backward model on \rCFhigherN{} of eight
datasets (Table~\ref{tab:track}). Whether the sensor or MEG-MASC's clean synthetic speech
raises tracking we cannot say; the lag control (\lagCtlAccMEG\% with pre-stimulus data)
suggests the stimulus contributes.

\textbf{Flow rule and per-source residuals.} Table~\ref{tab:flow}. The prior-free PMI rule
beats conditional likelihood by \pmiGapMin--\pmiGapMax{} points on every dataset, while the audio prior alone
is near chance (\pmiPriorMin--\pmiPriorMax\%), as Prop.~\ref{prop:pmi} predicts. Conditioned on the brain, the attended
talker's flow residual is lower than the competing talker's
($\Delta R_{\rm cond}>0$ on all datasets), while the unconditional residuals are nearly equal
($|\Delta R_{\rm unc}|\le\dRuncMaxOther$ on all datasets but KU Leuven, \dRuncKUL): the attention effect
lies mainly in the brain-conditioned term, carried by the delta and theta bands of the
envelope (App.~\ref{app:further}).

\begin{table}[!t]\centering\small
\caption{Flow-head decision rules at 5\,s (\%): prior-free PMI, conditional likelihood
only, audio prior only, and the PMI rule's EEG-shuffle null.
$\Delta R=R(\text{comp.})-R(\text{att.})$ for the conditional ($R_{\rm cond}$) and
unconditional ($R_{\rm unc}$) residuals; for SparrKULee and MEG-MASC the negative is the
mismatched window. Seed 0. NH/HI: normal-hearing/hearing-impaired listeners;
zh: Mandarin.}\label{tab:flow}
\resizebox{\columnwidth}{!}{\begin{tabular}{lcccccc}\toprule
Dataset & PMI & $\log p(a|x)$ & $\log p(a)$ & null & $\Delta R_{\rm cond}$ & $\Delta R_{\rm unc}$ \\\midrule
KU Leuven & 69.8 & 51.8 & 47.0 & 50.3 & +0.003 & -0.012 \\
DTU & 72.5 & 52.6 & 49.9 & 50.0 & +0.015 & -0.002 \\
Fuglsang'20 (NH+HI) & 78.1 & 57.8 & 50.9 & 50.0 & +0.024 & -0.005 \\
ESAA (zh) & 73.9 & 59.1 & 50.2 & 49.8 & +0.021 & +0.001 \\
KUL-255ch & 66.8 & 55.4 & 50.1 & 49.0 & +0.016 & +0.001 \\
Broderick & 57.4 & 52.0 & 49.9 & 50.7 & +0.006 & -0.001 \\
SparrKULee & 85.2 & 60.1 & 50.2 & 50.2 & +0.039 & +0.000 \\
MEG-MASC (MEG) & 97.1 & 77.2 & 49.7 & 49.3 & +0.158 & -0.000 \\
\bottomrule\end{tabular}
}
\end{table}

\textbf{Four talkers in a room (MAESTRO).} MAESTRO \citep{maestro2026} is harder in
every respect. There are six loudspeakers in a room, four of them attendable; the EEG has
32 channels; listeners move freely. We follow the protocol of the concurrent MERIT
study \citep{merit2026}: 4-way
matching over the attendable talkers with quantile-matched candidates (chance 25\%), one
decision per trial or per window, within listeners (content-disjoint folds) and on unseen
listeners (one listener and 20\% of the content held out). Table~\ref{tab:maestro}. Under
the release's timing, which MERIT and the benchmark use, CoupleFlow reaches \maePWTrial\% per
trial within listeners (+\maePWCon) and \maePLTrial\% on unseen listeners (+\maePLCon). It is
above MERIT's EEG-alone model and above the benchmark in both protocols. With the playback
latency corrected, so that the window truly follows the sound, it reaches \maeWTrial\% and
\maeLTrial\%. This causal result is our headline number. The same-talker controls stay above chance (causal, 5-s K=2:
\maeWStTwo\%; K=3: \maeWStThree\%), and every 5-s null lies in [\maMainNullMin, \maMainNullMax]\%.
\emph{Synchronisation} (App.~\ref{app:data}). The release's timestamps precede the sound
by 0.78\,s; we add this latency, confirmed without labels. Without it, windows cover EEG
from before the sound and per-trial accuracy rises (seed 0: \maePkgTrial\%, vs.\ \maePreTrial\%
for a correctly aligned model given pre-sound EEG): our lag rule excludes that EEG by design.

\begin{table}[!t]\centering\small
\caption{MAESTRO, 4-way matched, EEG only (chance 25\%). CoupleFlow: mean$\pm$s.d.\ over 3
seeds. \emph{Release timing}: audio anchored at the earliest playback timestamp, as in MERIT
and the benchmark, whose reported numbers are listed. \emph{Latency-corrected}: our causal
alignment. Contribution: per-trial accuracy minus the EEG-shuffle null.}\label{tab:maestro}
\resizebox{\columnwidth}{!}{\begin{tabular}{lccc}\toprule
Model & 5-s window & per trial & contribution \\\midrule
\multicolumn{4}{l}{\emph{within listeners}} \\
CoupleFlow, release timing & 50.1{\scriptsize$\pm$0.8} & 78.3{\scriptsize$\pm$1.0} & +53.6{\scriptsize$\pm$0.0} \\
MERIT (EEG alone), release timing & -- & 69.2 & +39.7 \\
Benchmark (EEG, 30\,s), release timing & -- & 59.1 & +32.0 \\
CoupleFlow, latency-corrected (causal) & 46.4{\scriptsize$\pm$0.4} & 71.2{\scriptsize$\pm$1.1} & +45.7{\scriptsize$\pm$1.2} \\
\multicolumn{4}{l}{\emph{unseen listeners (LOSO)}} \\
CoupleFlow, release timing & 42.5{\scriptsize$\pm$0.3} & 67.1{\scriptsize$\pm$0.8} & +42.2{\scriptsize$\pm$4.7} \\
MERIT (EEG alone), release timing & -- & 60.0 & +31.3 \\
Benchmark (EEG, 30\,s), release timing & -- & 61.9 & +36.3 \\
CoupleFlow, latency-corrected (causal) & 39.7{\scriptsize$\pm$0.8} & 56.9{\scriptsize$\pm$2.0} & +30.9{\scriptsize$\pm$3.0} \\
\bottomrule\end{tabular}
}
\end{table}

\textbf{Recent side classifiers.} DARNet \citep{darnet2024}, DBPNet \citep{dbpnet2024}
and ListenNet \citep{listennet2025} classify the attended \emph{side} from EEG alone and report
95--97\% at 1\,s on KU Leuven. We ran the authors' models in our training pipeline
(Table~\ref{tab:spatial}). Under their within-trial split (the last 10\% of each trial held
out for testing) they reproduce the published numbers on KU Leuven (\spKULBmin--\spKULBmax\% at 1\,s). On content-disjoint
folds (subject-dependent, as in the papers) they fall to \spAmin--\spAmax\% on DTU, ESAA and
the 255-channel set and stays flat from 1 to 5\,s: the signature of a classifier keyed to
trial-specific signal, not to attention. Fuglsang'20 is the exception (\spFugMin--\spFugMax\%),
possibly alpha lateralisation or uncontrolled gaze \citep{rotaru2024what}. On KU Leuven the
side task is ill-posed under content-disjoint folds (one class per listener in each test
fold).

\begin{table}[!t]\centering\small
\caption{Recent side classifiers, accuracy (\%) under their within-trial split / our
content-disjoint folds (subject-dependent, as in the papers), next to CoupleFlow on the
content-disjoint folds (seed 0). $^\dagger$Ill-posed under content-disjoint folds (one class per
listener in each test fold).}\label{tab:spatial}
\resizebox{\columnwidth}{!}{\begin{tabular}{lcccccc|cc}\toprule
 & \multicolumn{2}{c}{DARNet} & \multicolumn{2}{c}{DBPNet} & \multicolumn{2}{c}{ListenNet} & \multicolumn{2}{c}{CoupleFlow} \\
Dataset & 1\,s & 5\,s & 1\,s & 5\,s & 1\,s & 5\,s & 1\,s & 5\,s \\\midrule
DTU & 68/54 & 67/53 & 66/51 & 69/51 & 75/54 & 71/55 & 62 & 76 \\
ESAA & 85/56 & 86/59 & 88/60 & 89/62 & 92/56 & 88/59 & 63 & 78 \\
Fuglsang'20 & 83/67 & 83/69 & 84/68 & 82/69 & 89/71 & 85/71 & 64 & 80 \\
KUL-255 & 99/42 & 99/40 & 99/41 & 100/42 & 99/47 & 98/46 & 59 & 69 \\
KU Leuven$^\dagger$ & 97/52 & 98/54 & 97/47 & 98/45 & 98/42 & 97/44 & 60 & 73 \\
\bottomrule\end{tabular}
}
\end{table}

\textbf{Leakage control} (App.~\ref{app:leak}). A trial split that shares \emph{stories}
across folds inflates no model by more than \leakMax{} points on KU Leuven and DTU;
splitting \emph{windows within a trial} inflated our earlier decoders by
\oldDropAllMin--\oldDropAllMax{} points (Table~\ref{tab:earlier}), because test windows then
neighbour training EEG of the same trial and label.
\textbf{Ablations} (App.~\ref{app:ablation}). The per-listener correction is the one
component that carries accuracy (\ablRankKUL, \ablRankDTU{} and \ablRankMEG{} points on KU
Leuven, DTU and MEG); removing any other changes 5-s accuracy there by at most \ablOtherMax{}
points (one seed per variant; on MAESTRO, InfoNCE also matters, App.~\ref{app:mabl}). Reading EEG from \emph{before} the sound costs \lagCtlDTU--\lagCtlKUL{} points.

\section{Discussion and Limitations}
Leakage-free evaluation lowers reported AAD accuracies, but the remaining signal is real and
time-locked across nine datasets. The accumulation law turns the 5-s target many applications ask for into a statement about the
signal-to-noise ratio (SNR) of the recording: EEG would need $\kappa\approx0.74$, which only our MEG data reach ($\kappa=\kMEG$).  The decoder transfers to unseen listeners at a cost of
\losoDropMin--\losoDropMax{} points (App.~\ref{app:loso}), and sequential decisions on
calibrated 1-s scores gain \confGainTwoMin--\confGainTwoMax{} points at the same mean
decision time (App.~\ref{app:conf}). On MAESTRO, where the attended talker is always the
loudest, removing any one safeguard leaves every null at chance (App.~Table~\ref{tab:mabl}): the
time-centred score closes that shortcut by construction.
\emph{Limitations.}
Prop.~\ref{prop:pmi} assumes the recording depends on the attended stream only;
unattended speech is weakly tracked, which the matched and same-talker sets partly
address. Two datasets use published envelopes; AV-GC (no content-disjoint split) and NJU (no
time-locking) are excluded; the 255-channel set has two content folds; MEG is evaluated by
match--mismatch only.

\subsection*{AI use statement}
We used a generative AI coding assistant (Claude, Anthropic) to write and run code for data
processing, models, baselines and experiments, to produce tables and figures, and to draft
parts of the text and proofs. It was not used to collect data, and no reported number was
entered by hand: all results are generated by scripts from logged runs. The authors set the
research questions and evaluation protocol, verified code with tests and data checks,
checked all proofs and references, and take full responsibility for the content.

\bibliographystyle{apalike}
\bibliography{refs}

\clearpage
\appendix
\onecolumn
\section{Extended Related Work}\label{app:related}
\paragraph{Neural basis.} Cortical activity phase-locks to the slow (1--8\,Hz) envelope of
speech. With two competing talkers the attended talker's envelope is represented more
strongly, in MEG \citep{ding2012emergence}, intracranial recordings
\citep{mesgarani2012selective} and single-trial EEG \citep{osullivan2015attentional}. The
ignored talker is still weakly represented, most clearly where it is masked
\citep{brodbeck2020neural}. This is the premise of every envelope-based decoder, and the
weak representation of the ignored talker is why the attended-only assumption of
Prop.~\ref{prop:pmi} is an approximation. We measure its strength $\gamma$ on every
dataset (Table~\ref{tab:kth}).

\paragraph{Stimulus reconstruction.} Backward (decoding) models map time-lagged EEG to the
envelope by ridge regression \citep{osullivan2015attentional,crosse2016mtrf}, or project
EEG and audio into a shared space by canonical correlation analysis (CCA)
\citep{decheveigne2018decoding}. The talker whose envelope correlates better with the
reconstruction is chosen. These methods are interpretable, need little data and transfer
across listeners, and they cannot use spatial or identity shortcuts. Their weakness is
speed: one reconstructed correlation is a noisy statistic, and high accuracy needs tens of
seconds \citep{geirnaert2021eeg}. Prop.~\ref{prop:law} explains why: accuracy grows with
$\sqrt T$ at a rate set by the per-sample tracking margin. Deep reconstructions such as
VLAAI \citep{accou2023vlaai} raise the reconstructed correlation modestly. Under our
content-disjoint folds VLAAI stays below the per-listener linear decoder on every
two-talker dataset (Table~\ref{tab:main}), plausibly because its subject-independent
training cannot use the listener-specific channel weights that the linear model fits.
CoupleFlow keeps the time-locked principle of reconstruction but learns the neural and
acoustic representations jointly, with a per-listener correction; this is the component
that carries its accuracy (Table~\ref{tab:ablation}).

\paragraph{Match--mismatch.} The match--mismatch task asks whether a speech segment is
the one that evoked a given EEG segment \citep{accou2021dilated,puffay2023relating}. It is
free of spatial confounds and applies to single-talker data. It is usually evaluated on
unseen time windows of seen stimuli, and its negatives are often taken from nearby in the
same recording. We use it on held-out stimuli as our same-talker candidate set and as the
task for single-talker datasets.

\paragraph{Direct classifiers of the attended side.} Convolutional networks
\citep{vandecappelle2021eeg} and their successors with attention, graph and dual-branch
architectures \citep{darnet2024,dbpnet2024,listennet2025} predict whether the attended
talker is left or right from EEG alone, with 95--97\% at 1\,s on KU Leuven. No audio is
used, so any EEG signal that differs between left and right trials can be exploited. This
includes eye movements toward the attended loudspeaker, lateralised muscle activity and
slow trial-specific drifts. On a dataset where gaze and attention were dissociated,
accuracy followed gaze \citep{rotaru2024what}. Under their within-trial split we reproduce
the published numbers. On held-out stimuli, the same code falls to \spAmin--\spAmax\% on
DTU, ESAA and the 255-channel set (Table~\ref{tab:app_spatial}). On KU Leuven, attended
side and story are confounded, so the side task is ill-posed there without shared
stories.

\paragraph{Brain decoding with learned representations and generative models.}
Contrastive alignment between M/EEG and pretrained speech representations identifies
which of many speech segments was heard \citep{defossez2023decoding}. Flow matching
\citep{lipman2023flow} and diffusion models give tractable conditional generative models,
and conditioning dropout lets one network represent both $p(a\mid x)$ and $p(a)$
\citep{ho2022classifierfree}. Likelihood-ratio classification with generative models is
classical. We use the ratio for a different purpose: it cancels the audio prior $p(a)$,
which would otherwise favour acoustically typical candidates regardless of the brain
(Prop.~\ref{prop:pmi}; Table~\ref{tab:flow}). Our InfoNCE term follows contrastive
decoding, restricted to windows of one listener so that identity cannot solve it.

\paragraph{Evaluation validity.} Data leakage through non-independent splits inflates
results across machine-learning (ML) based science \citep{kapoor2023leakage}. In AAD, reviews have noted
within-trial splits and repeated stimuli \citep{puffay2023relating}, and
\citet{rotaru2024what} demonstrated gaze and trial-identity confounds. Table~\ref{tab:failure}
lists the failure modes we found or controlled, and the control we use for each.

\begin{table}[h]\centering\small
\caption{Failure modes in AAD evaluation, what they inflate, and the control in this
work.}\label{tab:failure}
\begin{tabular}{p{0.2\textwidth}p{0.33\textwidth}p{0.38\textwidth}}\toprule
Failure mode & Effect & Control in this work\\\midrule
Within-trial window split & test windows neighbour training windows of the same trial;
any trial-specific signal is rewarded & content-disjoint folds (Prop.~\ref{prop:split});
\oldSegKUL\%$\to$\oldTrialKUL\% for our earlier decoder
(Table~\ref{tab:earlier})\\
Repeated stimuli & the same story in training and test, across or within listeners &
content groups recovered from the audio; trial-split control (Table~\ref{tab:leak})\\
Gaze and lateralised artefacts & side classifiers read eye position rather than attention &
no side information in the decision path; side head reported separately\\
Audio-only cues & loudness or spectral differences mark the attended talker (MAESTRO: always
loudest) & time-centred score (Prop.~\ref{prop:null}); quantile-matched candidates;
adversary; Table~\ref{tab:mabl}\\
Trial identity and slow state & EEG identifies the trial, not the attended talker &
across-trial and circular-shift nulls\\
Wrong timing & the ``post-stimulus'' window contains pre-stimulus EEG, or there is no
time-locking at all & offset scans on every bundle; MAESTRO latency correction; NJU
excluded (App.~\ref{app:data})\\
Selection on test data & hyperparameters tuned on the test fold & lag window and
checkpoints chosen on validation only\\
\bottomrule\end{tabular}
\end{table}

\paragraph{Multi-talker rooms.} MAESTRO \citep{maestro2026} records EEG while listeners
attend one of several loudspeakers in a room, and the concurrent MERIT
\citep{merit2026} evaluates multimodal decoders on it. We follow MERIT's 4-way protocol and
show that the release timing places the decision window partly before the sound
(Sec.~\ref{sec:exp}, App.~\ref{app:data}).

\section{Proofs}\label{app:proofs}
\paragraph{Proof of Proposition~\ref{prop:split}.} Suppose a test stream of trial $i$ were
linked to a training or validation stream of trial $j$. Then $\mathcal G$ has an edge
$(i,j)$, so $i$ and $j$ lie in the same connected component. Test folds are unions of whole
components and training and validation data come only from the other components, a
contradiction. Content shared only in chunks shorter than 10\,s or
below the threshold is not covered. \qed

\paragraph{Proof of Proposition~\ref{prop:null}.} Condition on $y$. By (b), $x$ is independent of $a_{1:K}$ given $y$, so by (c)
the law of $(x,a_{\pi(1)},\dots,a_{\pi(K)})$ given $y$ is invariant to every permutation
$\pi$, and by (a) so is the law of $(s_1,\dots,s_K)=(g(x,a_1),\dots,g(x,a_K))$. Each index is
then the unique maximiser with the same probability, $(1-P(\text{tie}\mid y))/K$; counting
ties as errors, $P(\hat y=y\mid y)\le 1/K$, and averaging over $y$ gives the claim. For the time-centred score (Eq.~\ref{eq:score}) a time-constant
embedding yields $s_k=0$ for all $k$, a tie, hence an error: the acoustic pathway alone
cannot exceed chance. Assumption (c) holds exactly for the same-talker set (the negative is
the attended talker's own speech) and approximately for the competing set in
counterbalanced designs; the across-trial and circular nulls instantiate (b). \qed

\paragraph{Proof of Proposition~\ref{prop:pmi}.} By Bayes' rule, (a) and (b),
$P(y=k\mid x,a_{1:K})\propto p(x\mid a_{1:K},y=k)\,P(y=k\mid a_{1:K})=p(x\mid a_k)/K$.
Bayes' rule for the pair $(a_y,x)$ gives $p(x\mid a)=p(a\mid x)p(x)/p(a)$, evaluated at
$a=a_k$. The factor $p(x)$ is common to all $k$, so
$\log P(y=k\mid\cdot)=\PMI(a_k;x)+\text{const}$ and the MAP decision maximises PMI.
Replacing PMI by $\log p(a_k\mid x)$ adds $\log p(a_k)$ to candidate $k$'s score, which
depends on the audio only. \qed

\paragraph{Proof of Proposition~\ref{prop:law}.} By (i) and (ii),
$z_j\sim\mathcal N(\zeta_j,1/(n_{\rm eff}-3))$ with $n_{\rm eff}=fT$, and by (iii)
$z_{\rm att}-z_{\rm neg}\sim\mathcal N(\Delta\zeta,2/(n_{\rm eff}-3))$. Since $\operatorname{atanh}$ is increasing, the
decision is correct iff $z_{\rm att}>z_{\rm neg}$, so
$P(\mathrm{correct}\mid T)=\Phi(\Delta\zeta\sqrt{(n_{\rm eff}-3)/2})$. For $n_{\rm eff}\gg3$ this is
$\Phi(\kappa\sqrt T)$ with $\kappa=\Delta\zeta\sqrt{f/2}$; solving $\Phi(\kappa\sqrt{T_p})=p$
gives $T_p$. For the noise model, write $\tilde a=v+\eta$ with $\eta$ independent of $v$ and
of the envelopes, and $\sigma=\mathrm{Var}(v)/\mathrm{Var}(\eta)$. Then
$\mathrm{Cov}(\tilde a,a_j)=\mathrm{Cov}(v,a_j)$ and
$\mathrm{Var}(\tilde a)=\mathrm{Var}(v)(1+\sigma)/\sigma$, so
$\rho_j=\rho_j^\ast\sqrt{\sigma/(1+\sigma)}$ with $\rho^\ast_j=\mathrm{corr}(v,a_j)$. For small
$\rho_j$, $\operatorname{atanh}\rho_j\approx\rho_j$, so
$\Delta\zeta\approx(\rho^\ast_{\rm att}-\rho^\ast_{\rm neg})\sqrt{\sigma/(1+\sigma)}$ and
$\kappa^2\propto\sigma/(1+\sigma)$. Multiplying $\sigma$ by a factor $b>0$ multiplies $\kappa^2$ by
$b(1+\sigma)/(1+b\sigma)$: linear at low SNR, saturating at high SNR. Assumption (iii) is
exact when the two envelopes are uncorrelated and approximate when they are weakly
correlated. \qed

\section{Datasets and Preprocessing}\label{app:data}
\begin{table}[h]\centering\small
\caption{Datasets after unified preprocessing. Groups: content groups recovered from the
audio; folds $=\min(5,\text{groups})$. Env.: gammatone envelope computed from the audio
(ours) or published by the dataset authors (prov.). HI: hearing-impaired. 1-/2-talker:
single talker or two competing talkers.}\label{tab:data}
\resizebox{\textwidth}{!}{%
\begin{tabular}{lcccccccc}\toprule
Dataset & listeners & sensors & language & task & hours & groups & folds & env.\\\midrule
KU Leuven \citep{das2016kul,biesmans2017auditory} & 16 & 64 EEG & Dutch & 2-talker & 20.6 & \grpKULGroups & 4 & ours\\
DTU \citep{fuglsang2017dtu} & 18 & 64 EEG & Danish & 2-talker & 15.0 & \grpDTUGroups & 5 & ours\\
Fuglsang'20 \citep{fuglsang2020effects} & 44 (22 HI) & 64 EEG & Danish & 2-talker & 20.1 & \grpFugGroups & 4 & prov.\\
ESAA \citep{li2022esaa} & 17 & 64 EEG & Mandarin & 2-talker & 10.9 & \grpESAAGroups & 5 & ours\\
KUL-255ch \citep{narayanan2021uhd} & 30 & 253 EEG & Dutch & 2-talker & 13.2 & \grpUHDGroups & 2 & ours\\
Broderick \citep{broderick2018semantic} & 33 & 128 EEG & English & 2-talker & 16.4 & \grpBrodGroups & 5 & prov.\\
SparrKULee \citep{accou2024sparrkulee} & 85 & 64 EEG & Dutch & 1-talker & 132.4 & \grpSparrGroups & 5 & ours\\
MEG-MASC \citep{gwilliams2023megmasc} & 27 & 208 MEG & English & 1-talker & 51.4 & \grpMEGGroups & 5 & ours\\
MAESTRO \citep{maestro2026} & 16 & 32 EEG & English & 4 of 6 talkers & 13.1 & \grpMaeGroups & 5 / LOSO & ours\\
\bottomrule\end{tabular}}
\end{table}

\paragraph{Datasets.} All data are public releases, recorded by their authors with ethics
approval and informed consent; we collected no new data. Table~\ref{tab:licenses} lists every
release, the version we used and its license. ``Our use'' states what we changed from the
release.

\emph{KU Leuven} \citep{das2016kul,biesmans2017auditory}. 16 normal-hearing listeners,
64-channel BioSemi ActiveTwo, released at 128\,Hz. Four Dutch short stories read by male
narrators, played at 60\,dB(A) (A-weighted sound level) through insert phones, either dichotically
(dry) or filtered with head-related transfer functions (HRTFs) to $\pm90^\circ$. 20 trials per listener (8 of about 6\,min and 12 two-minute
repetitions, about 72\,min); the attended ear is fixed within a trial. Our use: all 20
trials, envelopes from the dry story recordings, content groups = the four stories.

\emph{DTU} \citep{fuglsang2017dtu}. 18 normal-hearing listeners, 64-channel BioSemi at
512\,Hz. Danish fictional stories read by a male and a female professional speaker, about
50\,s per trial, 60 trials per listener, talkers at $\pm60^\circ$ through simulated room
impulse responses (anechoic, mildly and highly reverberant rooms) and insert earphones; the
target talker is cued before each trial. Our use: the two-talker trials only (single-talker
trials and one trial that stops after 19.6\,s are dropped).

\emph{Fuglsang'20} \citep{fuglsang2020effects}. 44 listeners aged 51--76, 22 with
sensorineural hearing loss and 22 with normal hearing; 64-channel BioSemi at 512\,Hz. Two
audiobooks read by a male and a female talker at $\pm90^\circ$ (non-individualised HRTFs,
insert phones), 48 trials of about 50\,s, of which 32 are two-talker; the competing talker
starts 3--5\,s after the target. The release contains no audio, only the authors'
envelopes. Our use: the two-talker trials (32 per listener; 25 for one listener) and the provided envelopes of the
unamplified stimuli (hearing-impaired listeners heard an amplified version), so every listener
shares one envelope per stimulus.

\emph{ESAA} \citep{li2022esaa}. The release holds 17 normal-hearing listeners (the paper
reports 20), 64-channel EEG recorded at 1\,kHz (BrainProducts files). Mandarin short
stories told by a female and a male professional storyteller, HRTF-filtered to $\pm90^\circ$,
32 trials of 55--90\,s per listener (one recording ends after 16\,s). Our use: all trials, including those the authors flag
for poor comprehension answers; the stimulus reaches the ear about 0.3\,s after the start
marker, which we correct (App.~\ref{app:data}, alignment checks).

\emph{KUL-255ch} \citep{narayanan2021uhd}. 30 normal-hearing male listeners aged 22--35,
255-electrode 10-5 cap (253 EEG channels in the files), Compumedics SynAmps RT at 1\,kHz.
Two parts of two of the KU Leuven Dutch stories, HRTF-filtered to the left and right, four
trials of about 6.5\,min, one per attended ear and story part. Our use: all four trials.
Because the stories are those of KU Leuven, the two datasets are never mixed across training
and test. The authors corrected their stimulus table in August 2026; our attended labels are
checked on the data, as the labelled attended envelope is the one the EEG tracks
(Table~\ref{tab:main}).

\emph{Broderick} \citep{broderick2018semantic}. 33 native English listeners (34 recorded),
128-channel BioSemi ActiveTwo plus mastoids at 512\,Hz, released at 128\,Hz. Two Jules Verne
novels read by two male speakers, one per ear over headphones (\emph{Twenty Thousand Leagues
Under the Sea} left, \emph{Journey to the Centre of the Earth} right), 30 trials of about
60\,s; each listener attends the same side in every trial (listeners 1--17 left, 18--33
right). The release contains envelopes, not audio. Our use: the ``Cocktail Party'' part,
the provided envelopes, and the 64 channels nearest the canonical layout (all 128 in
Table~\ref{tab:channels}).

\emph{SparrKULee} \citep{accou2024sparrkulee}. 85 normal-hearing native Flemish listeners,
64-channel BioSemi ActiveTwo (recorded at 8192\,Hz, distributed at 1024\,Hz). Single-talker
Flemish audiobooks and podcasts, 6--10 trials of about 15\,min per listener. Our use: the
recordings of the training set of the auditory-EEG challenge at the 2024 IEEE International
Conference on Acoustics, Speech and Signal Processing (ICASSP); we drop the stories that
also occur in KU Leuven and the two speech-in-noise runs, and re-time the audio to the EEG
clock with the recorded trigger pulses.

\emph{MEG-MASC} \citep{gwilliams2023megmasc}. 27 English speakers, 208-channel KIT axial
gradiometer MEG at 1\,kHz. Four synthesised English fictional stories (about 2\,h, one or two
sessions), interleaved with short word lists, with comprehension questions. Our use: every
played audio file as one trial, sound onsets corrected by the delay measured on the recorded
audio channel, and all 208 sensors, read by a 208-channel input layer. A per-listener
principal component analysis (PCA) to 64 components, which we used first, fits different components for each listener; it
matched the 208-sensor model on the content folds but cannot transfer to an unseen listener
(\losoMEGpca\% instead of \losoMEGFive\% under leave-one-listener-out evaluation), so no model in the paper uses it.

\emph{MAESTRO} \citep{maestro2026}. 16 listeners aged 18--30, 32-channel ANT Neuro eego at
500\,Hz, plus eye gaze, pupil, egocentric video and head motion (not used here). English
speech from six loudspeakers in a room: four attendable talkers and two distractors, with
background noise; 100 evaluation trials of about 30\,s and 5 familiarisation trials. Our use:
the EEG of the 100 evaluation trials, with the synchronisation described below.

\emph{Excluded.} NJU \citep{zhang2023learnable} and AV-GC \citep{rotaru2024what}, for the
reasons given below; their licenses are in Table~\ref{tab:licenses}.

\paragraph{Licenses.} Every license permits our use, which is non-commercial research. We
redistribute no data, stimuli or derived features; the released code contains no data and
reads the original files, so the share-alike terms (KU Leuven, MAESTRO) do not apply to it. SparrKULee's stimuli
are licensed for non-commercial use only, and 107 of its recordings are available only on
request; we used the recordings distributed for the ICASSP 2024 challenge.
\begin{table*}[h]\centering\small
\caption{Releases, versions and licenses (checked on the release pages, September 2026).
CC: Creative Commons; BY: attribution; NC: non-commercial; SA: share-alike; CC0: public-domain
dedication. BIDS: Brain Imaging Data Structure; RDR: Research Data Repository; OSF: Open
Science Framework. Excluded datasets are listed for completeness.}
\label{tab:licenses}
\resizebox{\textwidth}{!}{%
\begin{tabular}{llllll}\toprule
Dataset & repository & identifier & version used & license & access\\\midrule
KU Leuven & Zenodo & 10.5281/zenodo.4004271 & v2.0 (2019)$^a$ & CC BY-NC-SA 4.0 & open\\
DTU & Zenodo & 10.5281/zenodo.1199011 & v1 (2018) & CC BY-NC 4.0 & open\\
Fuglsang'20 & Zenodo & 10.5281/zenodo.3618205 & 2020 (BIDS 1.3.0) & CC BY 4.0 & open (audio on request)\\
ESAA & Zenodo & 10.5281/zenodo.7795585 & v1.0 (2022) & CC BY 4.0 & open\\
KUL-255ch & Zenodo & 10.5281/zenodo.4518754 & v1.1.0 (2021) & CC BY-NC 4.0 & open\\
Broderick & Dryad & 10.5061/dryad.070jc & v3 (2019) & CC0 1.0 & open\\
SparrKULee & KU Leuven RDR & 10.48804/K3VSND & v3.1 (2023) & CC BY-NC 4.0 & open; 107 recordings on request\\
MEG-MASC & OSF & 10.17605/OSF.IO/AG3KJ & 2022 & CC0 1.0 & open\\
MAESTRO & Hugging Face & aspire-osu/maestro-eeg-dataset & 2026 & CC BY-NC-SA 4.0 & open\\\midrule
NJU (excl.) & Zenodo & 10.5281/zenodo.7253438 & v1 (2022) & CC BY 4.0 & open\\
AV-GC (excl.) & Zenodo & 10.5281/zenodo.11058711 & v1 (2024) & CC BY-SA 4.0 & open (13 of 16 listeners)\\
\bottomrule\end{tabular}}
\par\smallskip\raggedright\footnotesize $^a$Version 2.1 (July 2026, 10.5281/zenodo.21415059)
corrects a 70-ms misalignment between EEG and stimuli, after our experiments. Our lag window
spans about 1\,s and its start was chosen on validation data (125\,ms, consistent with the
offset scan of our bundle, which peaks one 125-ms step after 0\,s); we have not re-run the
experiments on version 2.1.
\end{table*}

\paragraph{Content groups.} Table~\ref{tab:groups} summarises the groups behind the
content-disjoint folds (Sec.~\ref{sec:exp}). Holding out trials does not hold out audio: most
datasets play the same stories to all listeners, and KU Leuven replays each story pair five
times per listener, so a trial split puts the same speech in training and test, where a
decoder with an audio pathway can memorise it. The audio, not the metadata, defines the groups
because stimulus keys are missing, inconsistent or per-listener in the releases (e.g.\
Fuglsang'20 numbers its files per listener, yet its \grpFugTrials{} trials contain only
\grpFugGroups{} distinct contents). The search has two stages. First, every candidate
envelope is averaged to 8\,Hz, cut into non-overlapping 10-s chunks, and each chunk is
correlated with every 10-s window, at every $1/8$-s offset, of every other envelope; pairs
with $r>0.9$ are candidates. Second, each candidate is re-correlated at 64\,Hz over the full
overlap at the implied offset ($\pm1/8$\,s) and kept if $r>0.95$. A copy of one recording
gives $r\approx1$ (every kept link has $r\ge\grpMinR$). Different contents stay well below the
threshold: the 64-Hz correlation at the best 8-Hz match of a 10-s chunk in another group has
a median of at most \grpNullMed{} per dataset (one listener's chunks against all envelopes of a second
listener). The closest cases (Table~\ref{tab:groups}, last column), up to \grpNullMax, have
identifiable causes that are not shared speech: in Fuglsang'20 every competing-talker envelope
starts with 3--5\,s of zero padding, and in SparrKULee one 10-s stretch of a podcast
resembles stretches of three other podcasts. Two envelopes are \emph{linked} if both stages pass. Let $\mathcal G$ be the graph whose
nodes are trials, with an edge between two trials whenever any of their envelopes are
linked; its connected components, found with union--find, are the content \emph{groups}
(e.g.\ the \grpKULTrials{} KU Leuven trials form \grpKULGroups{} groups).
\begin{proposition}[content disjointness]\label{prop:split}
No test stream is linked to any training or validation stream.
\end{proposition}
The proof is immediate from the component construction (App.~\ref{app:proofs}). Folds are unions of whole groups
($\min(5,\text{groups})$ folds); validation uses whole training groups when at least 10
remain, otherwise the last 15\% of each training trial of at least 200\,s (Fuglsang'20,
whose trials last about 50\,s: the last 20\% of each trial of at least 30\,s), separated
from the training part by a gap of one context margin on each side. Such validation data share
content with training and serve only for early stopping and the lag choice; no reported
number uses them.
\begin{table*}[h]\centering\small
\caption{Content groups recovered from the audio. Groups per listener: how many groups one
listener heard. Single-trial groups: content heard in one trial only. Links: 64-Hz-confirmed links that
merged two components (trials $-$ groups). Last two columns: best 8-Hz match of a 10-s
chunk in another group, and the 64-Hz correlation at that location: the chance level the
0.95 threshold must exceed.}\label{tab:groups}
\resizebox{\textwidth}{!}{\begin{tabular}{lcccccccc}\toprule
& & & & & \multicolumn{2}{c}{copies (links)} & \multicolumn{2}{c}{best match, other groups} \\
\cmidrule(lr){6-7}\cmidrule(lr){8-9}
Dataset & trials & groups & groups per listener & single-trial groups & number & min.\ $r$ (64\,Hz) & max.\ $r$ (8\,Hz) & max.\ $r$ (64\,Hz) \\\midrule
KU Leuven & 320 & 4 & 4 & 0 & 316 & 1.000 & 0.71 & 0.60 \\
DTU & 1079 & 60 & 58--59 & 0 & 1019 & 0.988 & 0.88 & 0.86 \\
Fuglsang'20 & 1401 & 4 & 4 & 0 & 1397 & 0.977 & 0.94 & 0.88 \\
ESAA & 544 & 32 & 32 & 0 & 512 & 1.000 & 0.74 & 0.57 \\
KUL-255ch & 120 & 2 & 2 & 0 & 118 & 1.000 & 0.69 & 0.57 \\
Broderick & 984 & 30 & 29--30 & 0 & 954 & 1.000 & 0.77 & 0.66 \\
SparrKULee & 550 & 63 & 2--9 & 0 & 487 & 0.964 & 0.94 & 0.93 \\
MEG-MASC & 1470 & 30 & 30 & 0 & 1440 & 1.000 & 0.82 & 0.67 \\
MAESTRO & 1600 & 100 & 100 & 0 & 1500 & 1.000 & 0.72 & 0.59 \\
\bottomrule\end{tabular}
}
\end{table*}

\paragraph{Alignment checks.} For every bundle, a per-listener ridge backward model
(lags 0--250\,ms, leave-trials-out) tracks the attended envelope
(held-out $r$ = 0.07--0.27, against $r\approx0$ for the envelope shifted by 1--4\,s), and a
global offset scan (125-ms steps) peaks at 0\,s, or at $+125$\,ms on KU Leuven and Broderick.
For SparrKULee the audio was re-timed to the EEG clock using the recorded 1-Hz trigger
pulses (drift $3.4$--$5.5\times10^{-5}$). For MEG-MASC, sound onsets were corrected by the
delay measured on the audio channel recorded in the MEG file (40--56\,ms).

\paragraph{Preprocessing details.} The steps, in order: bad-channel removal, average
reference, resampling to 128\,Hz, notch, 0.5--32\,Hz band-pass, mapping to 64 canonical
channels by name or, for high-density caps, nearest position (missing channels are zero; MEG
keeps its 208 sensors), resampling to 64\,Hz, and a z-score per trial and channel. Bad channels: standard deviation outside 0.2--5 times the
median channel's (MAESTRO: the release's list, flat channels, and channels at the clipping
level in at least 1\% of samples); they are zeroed. Reference: mean of the good channels
after removing each channel's mean (MEG is not re-referenced; its bad sensors are the union
over a listener's runs). Notch: second-order IIR at the mains frequency ($Q=30$). Band-pass:
fourth-order Butterworth, 0.5--32\,Hz, applied forwards and backwards (zero phase). Envelope:
mean over 28 gammatone bands (ERB-spaced, 50\,Hz--5\,kHz) of the compressed magnitude
$|\cdot|^{0.6}$. The published envelopes differ: Fuglsang'20 provides 24 gammatone bands,
100--4000\,Hz, $|\cdot|^{0.3}$, of the unamplified stimuli; Broderick provides its own
broadband envelope. SparrKULee stories that overlap the KU Leuven stimuli were removed.

\paragraph{MAESTRO synchronisation.} Each stream is trimmed to the release's common window,
from the latest start among EEG and audio playback to the earliest end. The playback start is
known only from software timestamps. The glasses' scene microphone recorded the room audio,
and a label-free search by generalised cross-correlation with phase transform (GCC-PHAT,
300--4000\,Hz) located every talker's audio in it. On 1{,}240
trials with clear peaks, the three playback devices start within a median 14\,ms of each
other (90th percentile 118\,ms), while their timestamps differ by a median 162\,ms. Slicing
each device from its own timestamp therefore adds misalignment; it lowers tracking. We place
all talkers at the earliest timestamp plus one global latency of 0.781\,s (50 samples at
64\,Hz). Odd and even listeners give the same estimate (0.75 and 0.81\,s). EEG tracking of
the six-talker mixture, which uses no attention label, peaks at 0.000\,s after the
correction. There is no per-trial or per-listener adjustment. Without the correction, a window
that nominally follows the sound in fact covers EEG from 0.78\,s \emph{before} it, and
per-trial accuracy rises to \maePkgTrial\%. A correctly aligned model given the same pre-sound
EEG reaches the same \maePreTrial\% (vs.\ \maeCompTrial\% causal), so the extra accuracy comes
from EEG that precedes the sound. It is consistent with the lag control: pre-sound EEG alone
gives \maBeforeTrial\% per trial on MAESTRO. The attended talker is always the loudest, by
3--18\,dB, but per-trial accuracy does not depend on that margin (Spearman
$\rho=\maeSnrRho$, $p=\maeSnrP$). Bad channels are those listed
in the release, plus flat channels and channels at the amplifier's clipping level in at
least 1\% of samples. M1/M2 are dropped by the 64-channel mapping, leaving 28.0 channels on
average.

\paragraph{Excluded datasets.}
\emph{NJU} \citep{zhang2023learnable} (Mandarin, 32-channel): no attended-envelope tracking at any global offset
($\pm60$\,s), time stretch (0.8--1.25), trial pairing, or alignment to the recorded
playback audio. A decoder trained on KU Leuven, which transfers to DTU ($r=0.030$), gives
$r=0.005$ on NJU. The released audio is spliced from the playback recording with removed
stretches, so its timing cannot be restored.
\emph{AV-GC} \citep{rotaru2024what}: its podcasts are re-paired across listeners, so all
trials form one content component and no content-disjoint split exists.

\paragraph{DECAF.} Trained with the same folds and selection rule, the recursive decoder
reached test $r\approx0$ on the two-talker sets (also its EEG branch alone) and only weak
tracking on the single-talker sets (Table~\ref{tab:app_baselines}); on Fuglsang'20 it falls
below chance, and in 3 of 33 folds validation selected the untrained model. We therefore do
not report it as a baseline.

\paragraph{Baseline training details.} Linear: per-listener ridge regression from EEG at lags
0--250\,ms to the attended envelope; penalty from $\{10^{-2},\dots,10^{2}\}$ times
$\mathrm{tr}(X^\top X)/D$ ($D$: number of lagged features) chosen on the listener's validation
data, then refit on training and validation data. Where whole-group validation holds none of a
listener's trials (\linInnerCV{} SparrKULee listener-folds), the penalty is chosen by
leave-one-segment-out cross-validation on that listener's training segments. VLAAI and Dil.\ MM are trained across all training listeners of a fold:
5-s windows with 80\% overlap, Adam (learning rate $10^{-3}$), batch 64, early stopping on
validation (patience 5), at most 1{,}500 steps per epoch and 95 minutes; VLAAI minimises
$-r$, Dil.\ MM the cross-entropy against a mismatched window of the same trial. All baselines use CoupleFlow's splits, validation data and test windows. On MEG-MASC they
read all 208 sensors; VLAAI's output-context layer then has 208 instead of 64 filters, so that
its skip connection matches the input width.

\paragraph{Training details.} AdamW (learning rate $10^{-3}$, weight decay $10^{-4}$),
gradient-norm clipping at 1, batches of up to 32 windows from one listener (a listener's
last partial batch is kept if it has at least 4), 5-s training crops, up to 100 epochs with
early stopping (patience 20) on validation contribution, learning rate halved when it has not
improved for more than 5 epochs. \emph{Nulls.} The across-trial null uses 20 permutations at 5\,s and 3 at other windows;
when a listener has a single test trial, windows of that trial are permuted instead. The
circular null (5\,s only) uses 5 random shifts of at least 2\,s. Shapley values split the
accuracy above the null between the two EEG pathways (coupling and side head). Validation contribution is measured on the matched
candidates (same-talker on SparrKULee and MEG-MASC) with 3 across-trial permutations. Loss
weights: side 0.3, InfoNCE 1, null hinge 0.5 (margin 0.5), VICReg 0.1, adversary 0.3, flow
0.3, envelope 0.2; they were set once and are identical in every reported run. What differs
between datasets: the lag window (chosen on validation), the validation tails of Fuglsang'20,
and MEG-MASC's 208-channel input. \emph{Development.} The input features (8- and 28-band
sub-bands and glimpses) and the training length were chosen during development on KU Leuven
fold~0, whose test accuracy we inspected; no other dataset or fold was used for design
choices. Without fold~0, KU Leuven scores \kulExFoldZero\% at 5\,s (all folds:
\cfKUL\%). Each run used one NVIDIA A100 graphics processing unit (GPU); a fold takes
\trainMinMin--\trainMinMax{} minutes including test evaluation.

\begin{table}[h]\centering\small
\caption{Canonical 64 vs.\ all native channels (5-s and trial accuracy, \%; 125-ms lag
window, seed 0).}
\label{tab:channels}\begin{tabular}{lccc}\toprule
Dataset & channels & 5\,s & trial \\\midrule
Broderick & 64 / 128 & 62.3 / 66.4 & 81.5 / 86.5 \\
KUL-255ch & 64 / 253 & 67.1 / 69.9 & 98.3 / 100.0 \\
\bottomrule\end{tabular}

\end{table}

\section{Network Details}\label{app:net}
\definecolor{cfN}{HTML}{2A78D6}\definecolor{cfA}{HTML}{EB6834}
\definecolor{cfC}{HTML}{1BAF7A}\definecolor{cfF}{HTML}{D6558A}
\definecolor{cfG}{HTML}{C98A00}\definecolor{cfX}{HTML}{7A7974}
\definecolor{cfInk}{HTML}{2B2B2B}
\tikzset{
  nb/.style={draw=#1,fill=#1!7,rounded corners=2pt,line width=0.6pt,align=center,
    inner sep=2pt,font=\scriptsize,minimum height=0.8cm},
  nbw/.style 2 args={nb=#1,minimum width=#2cm},
  op/.style={circle,draw=#1,fill=white,line width=0.6pt,inner sep=0.5pt,font=\scriptsize,
    minimum size=0.42cm},
  tens/.style={draw=#1,fill=#1!18,line width=0.5pt,inner sep=1.5pt,font=\scriptsize,
    align=center},
  ar/.style={->,line width=0.6pt,draw=#1},
  arl/.style={font=\tiny,text=cfInk,fill=white,inner sep=0.8pt},
  note/.style={font=\tiny,text=cfX,align=left},
  ttl/.style={font=\footnotesize\bfseries,text=cfInk,anchor=west},
  lossn/.style={draw=#1,fill=#1!14,rounded corners=4pt,line width=0.5pt,inner sep=1.6pt,
    font=\scriptsize},
  >={Stealth[length=3.6pt,width=3pt]}}

Table~\ref{tab:arch} lists every module with its parameter count, read from the trained
checkpoints. All datasets share one architecture and one set of hyperparameters, except the lag window and
Fuglsang'20's validation tails (App.~\ref{app:data}). Only the
number of listeners, and hence of listener corrections, and the width of the audio input
change. The input has \archAudIn{} channels per candidate with sub-band features and
\archAudInEnv{} for the two datasets with published envelopes only. The full model has
\archTotMin--\archTotMax{} parameters per dataset. The part that makes the decision,
listener correction, virtual sensors, both encoders and the score, has only \archDecision{}
shared parameters plus \archPerListener{} per listener. $L$ is the number of samples in the decision window
(320 at 5\,s) and $L'=L+200$ includes 100 context samples on each side.

\begin{table}[h]\centering\small
\caption{CoupleFlow modules and parameter counts (KU Leuven checkpoint; $^\ast$envelope-only
input). GN: GroupNorm; GELU: Gaussian error linear unit; MLP: multilayer perceptron. Only the rows in the decision path are needed at test time for the
coupling decision; the flow head is needed for the PMI rule.}\label{tab:arch}
\begin{tabular}{lp{0.43\textwidth}lr}\toprule
Module & Specification & Output & Parameters\\\midrule
Listener correction & $x+U_m V_m^{\top}x$, rank 2, $U_m$ init.\ 0 & $64\times L'$ & 256 / listener\\
Virtual sensors & $1{\times}1$ conv, $64\to8$ & $8\times L'$ & 520\\
EEG encoder & 5 conv, kernel 3, dilation 1--16, width 32, GN+GELU, no padding & $16\times L$ & 11{,}952\\
Audio encoder (per $k$) & 4 conv, kernel 3, dilation 1--8, width 32, causal, shared & $16\times L$ & 13{,}200 (8{,}400$^\ast$)\\
Coupling score & weights $w_d$, temperature $\tau$ & $K$ & 17\\
Flow head & $1{\times}1$ in ($34\to64$), 5 residual dilated blocks, zero-init out & $1\times L$ & 85{,}505\\
Side head & $1{\times}1$ ($64\to16$), 4 dilated conv, mean/log-s.d., linear & 2 & 10{,}659\\
Audio adversary & MLP $32\to32\to1$ per candidate, gradient reversal & $K$ & 1{,}089\\
Envelope read-out & $1{\times}1$ conv, $16\to1$ & $1\times L$ & 17\\
\midrule
Shared total & & & 122{,}959\\
Decision path (coupling) & listener correction, virtual sensors, both encoders, score & & 25{,}689 + 256/listener\\
\bottomrule\end{tabular}

\end{table}

\begin{figure*}[t]\centering
\resizebox{\textwidth}{!}{\input{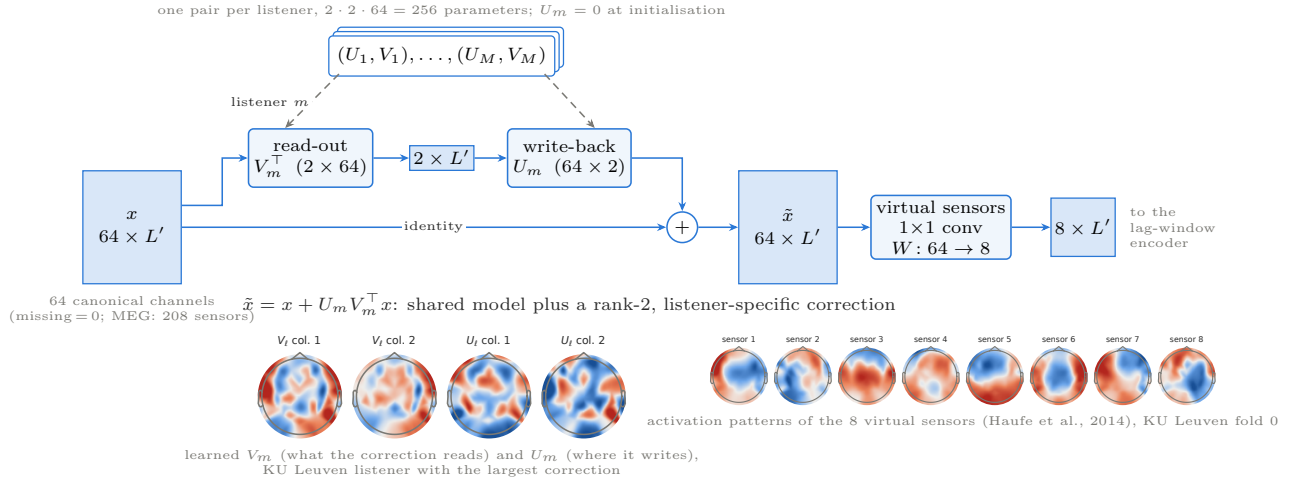}}
\caption{Input, listener correction and virtual sensors. The correction reads two spatial
patterns of the EEG with $V_m$ and writes them back with $U_m$; the identity path
keeps the shared model, and $U_m=0$ at initialisation, so a listener departs from it only
as far as the data require. Bottom: maps from the KU Leuven fold-0 checkpoint, namely
$V_m$ and $U_m$ of the listener with the largest correction, and the activation
patterns $\Sigma W^{\top}(W\Sigma W^{\top})^{-1}$ of the eight virtual sensors, with $W$ their
weights and $\Sigma$ the covariance
of the corrected training EEG \citep{haufe2014interpretation}. Diverging colour scale per
map, grey at zero.}\label{fig:net_input}
\end{figure*}

\paragraph{Input and listener correction.} Every EEG dataset is mapped to the same 64
canonical channels (missing channels are zero), so one network serves all caps. MEG-MASC
keeps its 208 sensors, in the same order for every listener (one helmet), and is the only
dataset with a different input layer: the listener correction, the virtual sensors and the
side head's first layer take 208 instead of 64 channels; nothing else changes. Listeners
differ in head shape, cap placement and source orientation. A full per-listener spatial
filter ($64\times64$) would overfit the few minutes of data per listener, so we add a
rank-2 correction (Fig.~\ref{fig:net_input}) $x+U_m V_m^{\top}x$ for listener $m=1,\dots,M$
($M$: number of listeners in the dataset), with \archPerListener{} parameters per listener. $U_m$ starts at zero, so training starts from one shared
model and each listener departs from it only as far as its data support. This is the
component that carries accuracy: removing it lowers 5-s accuracy by \ablRankKUL, \ablRankDTU{} and \ablRankMEG{} points on
KU Leuven, DTU and MEG-MASC
(Table~\ref{tab:ablation}) and changes MAESTRO per-trial accuracy by \maAblRank{} points
(Table~\ref{tab:mabl}).

\paragraph{Virtual sensors.} A $1{\times}1$ convolution maps the 64 channels to 8
virtual sensors, the learned analogue of the spatial filters of backward models and CCA
\citep{crosse2016mtrf,decheveigne2018decoding}. The narrow bottleneck forces the encoder
to use a few spatial patterns, which suits the low spatial resolution of EEG and limits
overfitting. Their activation patterns (Fig.~\ref{fig:net_input}, bottom) are smooth and
spatially broad.

\begin{figure*}[t]\centering
\resizebox{0.95\textwidth}{!}{\input{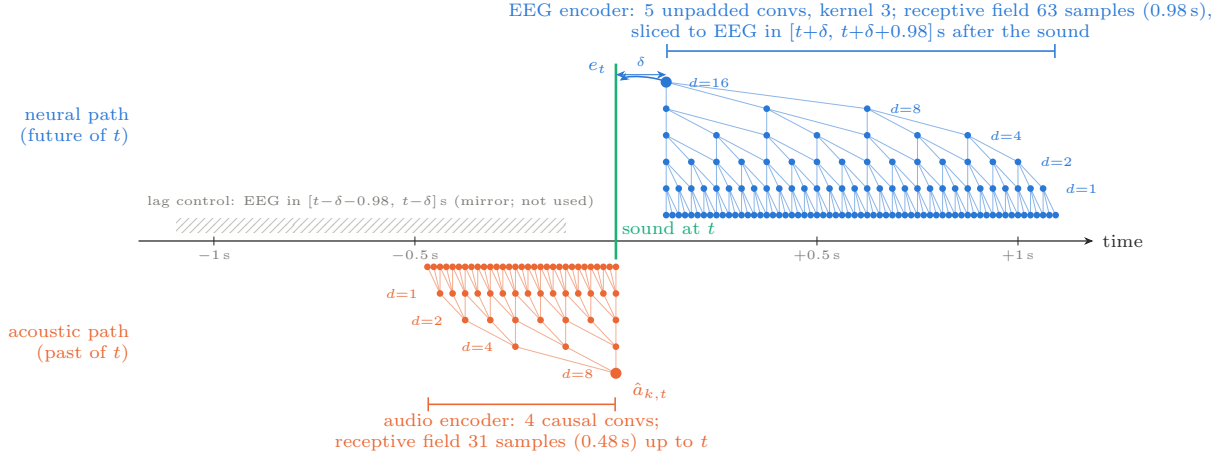}}
\caption{Receptive fields of both encoders on one time axis, drawn to scale (one dot per
sample at 64\,Hz; each node depends on three nodes of the layer below, spaced by the
dilation $d$). The neural embedding $e_t$ reads EEG from $\delta$ to
$\delta+0.98$\,s after the sound at $t$ (blue; $\delta=125$\,ms shown, 0\,ms on six
datasets, chosen on validation), and the candidate embedding $\hat a_{k,t}$ reads the
0.48\,s of audio up to $t$ (orange). Matching the two therefore fixes the lag direction:
the brain response must follow the sound. The lag control uses the mirrored window before
the sound (hatched). Each decision window carries 100 samples (1.56\,s) of context on both
sides, more than either receptive field needs, so no output depends on zero
padding.}\label{fig:net_enc}
\end{figure*}

\paragraph{EEG encoder and lag window.} Five convolutions (Fig.~\ref{fig:net_enc}) with kernel 3 and dilations
1, 2, 4, 8, 16 give a receptive field of 63 samples (0.98\,s at 64\,Hz). This covers the
latencies at which temporal response functions find cortical responses to speech, with
room for slower components. The
convolutions are unpadded, so no output depends on zero padding, and the output is sliced
so that $e_t$ depends only on EEG from $t+\delta$ to $t+\delta+0.98$\,s after the
sound at $t$ ($\delta\in\{0,125\}$\,ms, chosen on validation). EEG before the sound cannot
be a response to it, so this window excludes anticipation and slow drifts shared with the
past. The mirror window, EEG before the sound, is our lag control. GroupNorm (4 groups)
is used because batches are small and come from one listener. The last layer has no
activation, so the embedding is not restricted to positive values, and dropout 0.1
follows every hidden layer.

\begin{figure*}[t]\centering
\resizebox{\textwidth}{!}{\input{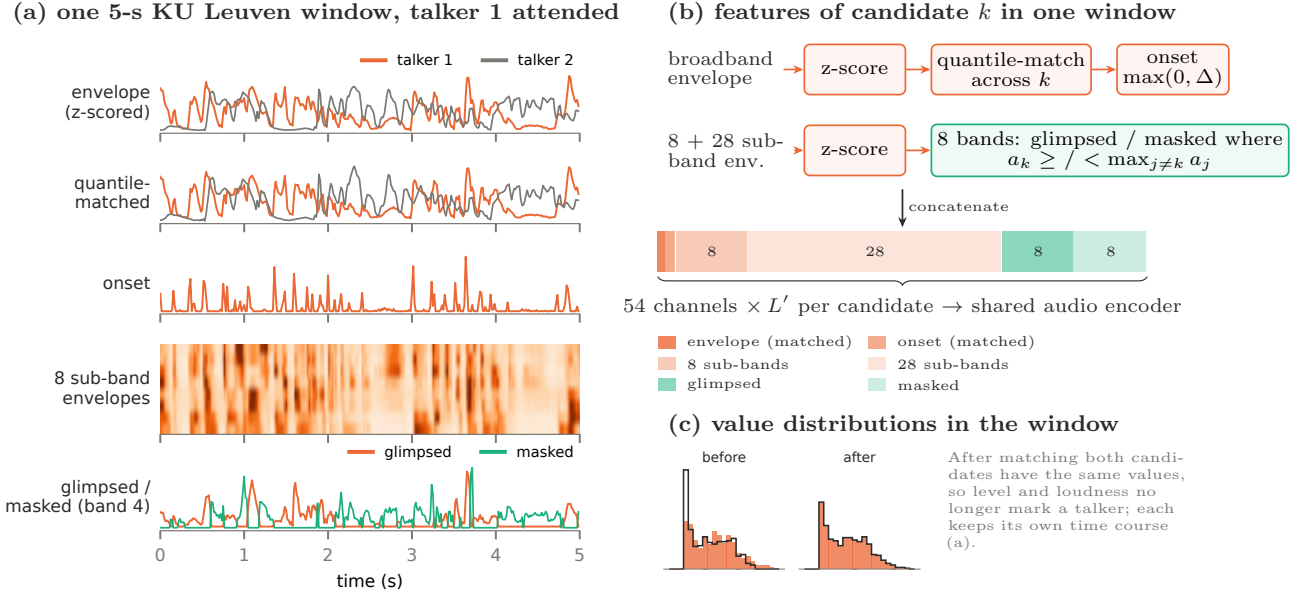}}
\caption{Candidate features on one 5-s KU Leuven test window. (a) The broadband envelope
before and after quantile matching, its onset channel, the 8-band sub-band envelopes and the
glimpsed and masked parts of one band. (b) How the 54 input channels of each candidate are
formed; the 8- and 28-band envelopes enter as they are, and glimpses are computed on the
8-band envelopes. (c) Matching gives both candidates the same value distribution over the
window while each keeps its own time course.}\label{fig:net_feat}
\end{figure*}

\paragraph{Candidate features and quantile matching.} Each candidate is described by its
broadband gammatone envelope ($|\cdot|^{0.6}$), its onsets (half-wave-rectified
derivative, since cortex responds strongly to acoustic edges), 8- and 28-band sub-band
envelopes (spectral detail), and glimpsed and masked channels: the candidate's window-standardised
sub-band envelope where it exceeds, or falls below, the largest standardised envelope of the
other candidates (on the broadband envelope for the two envelope-only datasets). Ignored speech is tracked mainly
where it is glimpsed \citep{brodbeck2020neural}, so these channels let the model treat
the two cases differently. In training, and for the matched and same-talker candidate sets, the broadband envelope
and onset of each window (with its context) are quantile-matched across candidates; every
other channel is z-scored within the window. The competing set is evaluated unmatched. The candidates
then have identical envelope value distributions, which removes loudness and level cues that
would identify the attended talker without the brain (Fig.~\ref{fig:net_feat}).

\paragraph{Audio encoder.} Four causal convolutions (Fig.~\ref{fig:net_enc}, orange) (dilations 1--8, receptive field 31
samples, 0.48\,s) map each candidate to $\hat{a}_k\in\mathbb R^{16\times L}$. Causal
means $\hat a_{k,t}$ depends only on audio up to $t$, so together with the EEG window the
lag direction is fixed by the architecture: EEG must follow the sound it is matched to. One
set of weights is shared by all candidates. This is required by Prop.~\ref{prop:null}
(assumption (a)) and means no candidate can be favoured by its position.

\begin{figure*}[t]\centering
\resizebox{\textwidth}{!}{\input{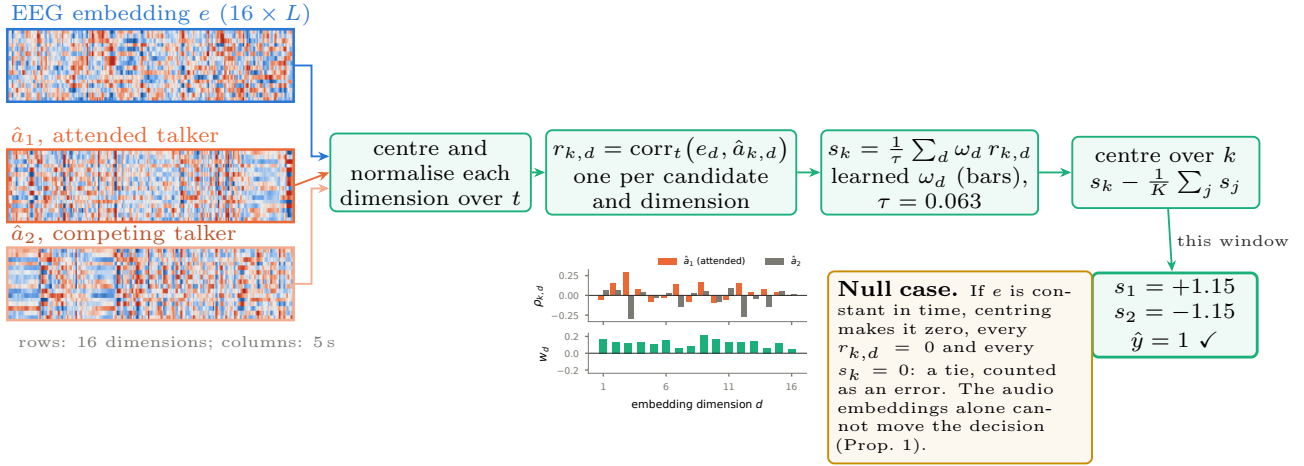}}
\caption{The time-centred coupling score on one KU Leuven test window (fold-0 model). The
EEG embedding and both candidate embeddings (heat maps, each dimension standardised) are
centred and normalised over time; each dimension gives one correlation per candidate
(bars: attended in orange, competing in grey), the learned weights $\omega_d$ combine them, and
centring over candidates yields the scores, which pick the attended talker here. Inset:
without time-varying neural input every score is zero.}\label{fig:net_coupling}
\end{figure*}

\paragraph{Coupling score.} The score (Eq.~\ref{eq:score}) is a weighted sum of
per-dimension Pearson correlations over time, scaled by a learned temperature (initialised
at 0.07) and centred over candidates. Correlation is invariant to the scale and offset of
either embedding, so the score depends only on how the two signals co-vary in time.
Centring the embeddings over time makes the score exactly zero for a time-constant neural
embedding. This is what makes the decoder null-calibrated: without time-varying neural
input there is nothing for the audio pathway to use. The weights $\omega_d$ start equal
($1/D$) and learn which embedding dimensions carry coupling. Fig.~\ref{fig:net_coupling}
follows one test window through the score.

\paragraph{Flow head.} The flow head models the attended envelope $a\in\mathbb R^{L}$ (the
z-scored broadband envelope channel of one window) given the neural embedding.
Flow matching joins Gaussian noise $\varepsilon$ to the data
along the straight path $a^{(u)}=(1-u)\,\varepsilon+u\,a$, indexed by the flow time
$u\in[0,1]$ ($\varepsilon\sim\mathcal N(0,I_L)$), whose velocity is $a-\varepsilon$, and trains a
network $v_\theta(a^{(u)},u,c)$ with parameters $\theta$ and condition $c$ to predict that
velocity:
\begin{equation}
\mathcal L_{\rm FM}=\mathbb E_{u\sim\mathcal U[0,1],\,\varepsilon}\big\|v_\theta(a^{(u)},u,c)-(a-\varepsilon)\big\|^2 .
\label{eq:fm}
\end{equation}
The condition is $c=e$; with probability 0.2 it is set to zero and a flag marks it as
absent (written $c=\varnothing$), as in classifier-free guidance \citep{ho2022classifierfree}. Integrating $v_\theta$ from $u=0$ to 1 carries noise to
envelopes, so one network defines both $p(a\mid x)$ (with $c=e$) and $p(a)$ (with
$c=\varnothing$).
A candidate is scored by how well each model explains it: its residuals
$R_{\rm cond}(a_k)=\mathbb E_{u,\varepsilon}\|v_\theta(a_k^{(u)},u,e)-(a_k-\varepsilon)\|^2$ and
$R_{\rm unc}(a_k)$, the same with $c=\varnothing$. A lower residual means a more typical sample,
since the expected flow-matching residual acts as a (weighted) surrogate for the negative
log-likelihood. The flow score $R_{\rm unc}(a_k)-R_{\rm cond}(a_k)$, how much the recording lowers the error
on $a_k$, is therefore a proxy for the pointwise mutual information
$\PMI(a_k;x)=\log p(a_k\mid x)-\log p(a_k)$, which is the Bayes-optimal score (Prop.~\ref{prop:pmi}). The audio-only term $\log p(a_k)$ cancels, and
a network that ignores $e$ scores every candidate 0, as Eq.~\ref{eq:score} does.
We also report the per-source residuals of the attended and the competing talker.

The network's input at each time step concatenates the noisy envelope $a^{(u)}$ (1 channel),
$e$ (16), a flag indicating whether the condition is present (1) and a sinusoidal
embedding of flow time (16). A $1{\times}1$ convolution maps these 34 channels to 64,
followed by five residual blocks (dilated convolution 1--16 with same padding, GroupNorm,
GELU, $1{\times}1$ convolution) and a zero-initialised output layer, so training starts
from zero velocity. At test time the residuals $R_{\rm cond},R_{\rm unc}$ are averaged over 8
Monte-Carlo draws of flow time (stratified) and noise. The draws are shared across
candidates and across the conditional and unconditional passes, so their noise cancels in
$R_{\rm unc}-R_{\rm cond}$. It is the largest module (\archFlow{} parameters) but lies outside the coupling
path. Removing it changes coupling accuracy by at most \ablOtherMax{} points
(Table~\ref{tab:ablation}). We keep it because it supplies the PMI rule and the
per-source residuals (Table~\ref{tab:flow}). Fig.~\ref{fig:net_flow} shows both uses.

\begin{figure*}[t]\centering
\resizebox{\textwidth}{!}{\begin{tikzpicture}[x=1cm,y=1cm]
\node[ttl] at (0,2.55) {(a) velocity network, trained on the attended envelope};
\node[tens=cfF,minimum width=1.45cm] (x1) at (0.75,1.55) {$a$ (attended)\\$1\times L$};
\node[tens=cfX,minimum width=1.45cm] (x0) at (0.75,0.35) {$\varepsilon\sim\mathcal N(0,I)$\\$1\times L$};
\node[nb=cfF,text width=2.3cm] (mix) at (3.35,0.95) {$a^{(u)}=(1-u)\,\varepsilon+u\,a$\\$u\sim\mathcal U(0,1)$};
\draw[ar=cfF] (x1.east) -- ++(0.2,0) |- ([yshift=0.15cm]mix.west);
\draw[ar=cfX] (x0.east) -- ++(0.2,0) |- ([yshift=-0.15cm]mix.west);

\def\sx{5.55}
\foreach \h/\c/\lab/\n [count=\i] in {0.14/cfF!65/{$a^{(u)}$}/1, 0.85/cfN!30/{$e$, or $0$}/16,
                                       0.14/cfX!55/{condition flag}/1, 0.85/cfF!20/{sinusoidal emb.\ of $u$}/16} {
  \ifnum\i=1 \coordinate (T\i) at (\sx,1.95); \else
    \pgfmathtruncatemacro{\j}{\i-1}\coordinate (T\i) at (B\j); \fi
  \coordinate (B\i) at ($(T\i)+(0,-\h)$);
  \fill[\c] (T\i) rectangle ($(B\i)+(0.45,0)$);
  \draw[white,line width=0.7pt] (B\i) -- ++(0.45,0);
  \node[font=\tiny,anchor=west,text=cfInk] at ($(T\i)!0.5!(B\i)+(0.5,0)$) {\lab\ (\n)};
}
\draw[decorate,decoration={brace,amplitude=2.5pt},cfInk,line width=0.4pt]
  ($(B4)+(-0.07,0)$) -- ($(T1)+(-0.07,0)$) node[midway,left=3pt,font=\tiny] {34 ch};
\draw[ar=cfF] (mix.east) to[out=0,in=180] ($(T1)!0.5!(B1)$);
\node[note,anchor=north west,text width=3.1cm] at ($(B4)+(0,-0.07)$)
  {with probability 0.2: $e\to0$ and flag $=0$, so one network learns
   $p(a\mid x)$ and $p(a)$};

\node[nb=cfF,text width=1.15cm] (in) at (9.0,0.95) {$1{\times}1$ conv\\$34\to64$};
\draw[ar=cfF] ($(T1)!0.5!(B4)+(2.65,0)$) -- (in.west);
\node[draw=cfF,fill=cfF!4,rounded corners=2pt,line width=0.6pt,minimum width=4.1cm,
      minimum height=1.45cm] (rb) at (12.0,0.95) {};
\node[font=\scriptsize,anchor=north,text=cfInk,inner sep=2pt] at (rb.north) {5 residual blocks, $d=1,2,4,8,16$};
\node[font=\scriptsize,text=cfInk,align=center] at ($(rb.center)+(0,-0.22)$)
  {residual block $d$:\\[1pt]{\tiny$\mathrm{conv}_{3,d}\to\mathrm{GN}\to\mathrm{GELU}\to\mathrm{conv}_{1\times1}$, $+$\,input}};
\draw[ar=cfF] (in) -- (rb);
\node[nb=cfF,text width=1.25cm] (outl) at (15.35,0.95) {$1{\times}1\to v_\theta$\\zero-init};
\draw[ar=cfF] (rb) -- (outl);
\node[lossn=cfF,anchor=north east] (lfm) at ($(outl.south east)+(0,-0.35)$)
  {$\mathcal L_{\rm FM}=\big\|v_\theta-(a-\varepsilon)\big\|^2$};
\draw[ar=cfF] (outl.south) -- (outl.south|-lfm.north);

\node[ttl] at (0,-1.3) {(b) test time: estimated pointwise mutual information of every candidate};
\node[tens=cfA,minimum width=1.45cm] (ak) at (0.75,-2.55) {$a_k$\\candidate $k$};
\node[nb=cfX,text width=2.55cm] (draw) at (3.45,-2.55)
  {8 draws $(u_j,\varepsilon_j)$, stratified $u$, identical for every $k$};
\draw[ar=cfA] (ak) -- (draw);
\node[nb=cfF,text width=5.3cm] (rc) at (8.55,-1.95)
  {$R_{\rm cond}(a_k)=\tfrac18\sum_j\big\|v_\theta(a_k^{(u_j)},u_j,e)-(a_k-\varepsilon_j)\big\|^2$};
\node[nb=cfF,text width=5.3cm] (ru) at (8.55,-3.15)
  {$R_{\rm unc}(a_k)=\tfrac18\sum_j\big\|v_\theta(a_k^{(u_j)},u_j,\varnothing)-(a_k-\varepsilon_j)\big\|^2$};
\coordinate (sp) at ($(draw.east)+(0.75,0)$);
\draw[ar=cfF] (draw.east) -- (sp) |- (rc.west);
\draw[ar=cfF] (sp) |- (ru.west);
\node[arl,anchor=east] at ($(sp|-rc.west)+(-0.05,0)$) {with $e$};
\node[arl,anchor=east] at ($(sp|-ru.west)+(-0.05,0)$) {without};
\node[nb=cfC,text width=2.75cm] (pmi) at (13.05,-2.55)
  {$\widehat{\mathrm{PMI}}_k=$\\$R_{\rm unc}(a_k)-R_{\rm cond}(a_k)$:\\how much $e$ explains $a_k$};
\draw[ar=cfF] (rc.east) -- ++(0.3,0) |- ([yshift=0.15cm]pmi.west);
\draw[ar=cfF] (ru.east) -- ++(0.3,0) |- ([yshift=-0.15cm]pmi.west);
\node[nb=cfC,line width=0.9pt,text width=1.55cm] (dec) at (15.6,-2.55)
  {$\hat y=\arg\max_k$\\$\widehat{\mathrm{PMI}}_k$};
\draw[ar=cfC] (pmi) -- (dec);
\node[note,anchor=north west,text width=15.6cm] at (0,-3.75)
  {Sharing the draws across candidates and across the two passes cancels their noise in
   $R_{\rm unc}-R_{\rm cond}$. Scoring by $R_{\rm cond}$ alone would add the audio-only term $\log p(a_k)$
   (Prop.~\ref{prop:pmi}); if the network ignores $e$, then $R_{\rm cond}=R_{\rm unc}$ and every
   $\widehat{\mathrm{PMI}}_k=0$.};
\end{tikzpicture}}
\caption{Flow head. (a) Training: the velocity network sees the noisy envelope $a^{(u)}$ on the
straight path from noise $\varepsilon$ to the attended envelope $a$, the neural embedding
(dropped with probability 0.2), a flag saying whether it is present, and the flow time $u$,
and regresses the path velocity $a-\varepsilon$. (b) Test: every candidate is scored with
and without the neural embedding on the same Monte-Carlo draws; the difference of the
residuals estimates its pointwise mutual information with the recording
(Prop.~\ref{prop:pmi}).}\label{fig:net_flow}
\end{figure*}

\paragraph{Side head.} A separate EEG-only branch (Fig.~\ref{fig:net_aux}a) ($1{\times}1$ convolution to 16
channels, four dilated convolutions, mean and log standard deviation over time, linear
layer) predicts the attended side. When the candidates' sides are known, its
log-probability of each candidate's side is centred over candidates and scaled by a learned gain ($\mathrm{softplus}(g_{\rm side})$,
$g_{\rm side}$ initialised at $-2$). During training this side score is added to $s_k$ inside
$\mathcal L_{\rm CE}$. All reported accuracies use the coupling score alone, and the side
head's contribution is reported separately. The branch gives lateralised EEG a separate,
measurable route instead of letting the coupling path absorb it. Removing it changes
coupling accuracy by at most \ablOtherMax{} points.

\paragraph{Audio adversary and envelope read-out.} The adversary (Fig.~\ref{fig:net_aux}b) summarises each candidate
embedding by its temporal mean and standard deviation (32 values) and scores it with a
two-layer MLP. A softmax over candidates gives $\mathcal L_{\rm adv}$, and a
gradient-reversal layer (coefficient 1) makes the audio encoder remove any feature that
identifies the attended talker from audio alone \citep{ganin2016domain}. The envelope
read-out, a $1{\times}1$ convolution from $e$, reports how well the neural embedding
reconstructs the attended envelope; it does not enter the decision (Fig.~\ref{fig:net_aux}c).

\begin{figure*}[t]\centering
\resizebox{\textwidth}{!}{\begin{tikzpicture}[x=1cm,y=1cm]
\node[ttl] at (0,1.25) {(a) side head: a separate, measured route for lateralised EEG};
\node[tens=cfN,minimum width=1.2cm] (xs) at (0.6,0.35) {$\tilde{x}$\\$64\times L'$};
\node[nb=cfX,text width=1.25cm] (s1) at (2.4,0.35) {$1{\times}1$ conv\\$64\to16$};
\node[nb=cfX,text width=1.45cm] (s2) at (4.35,0.35) {4 dilated\\convs};
\node[nb=cfX,text width=1.7cm] (s3) at (6.45,0.35) {mean, log s.d.\\over $t$ (32)};
\node[nb=cfX,text width=2.2cm] (s4) at (8.85,0.35) {linear $\to$\\$\log p(\mathrm L),\,\log p(\mathrm R)$};
\draw[ar=cfX] (xs) -- (s1);
\draw[ar=cfX] (s1) -- (s2);
\draw[ar=cfX] (s2) -- (s3);
\draw[ar=cfX] (s3) -- (s4);
\node[nb=cfX,text width=2.75cm] (s5) at (11.95,0.35)
  {when sides are known, candidate $k$ gets the log-prob.\ of its side, centred over $k$,
   $\times\,\mathrm{softplus}(g_{\rm side})$};
\draw[ar=cfX] (s4) -- (s5);
\node[lossn=cfC,anchor=west] (lce) at (13.75,0.72) {$+\,s_k$ inside $\mathcal L_{\rm CE}$ (training)};
\node[lossn=cfX,anchor=west] (rep) at (13.75,0.0) {not in reported accuracy};
\draw[ar=cfX] (s5.east) -- ++(0.15,0) |- (lce.west);
\draw[cfX,line width=0.6pt,densely dashed] (s5.east) ++(0.15,0) |- (rep.west);
\node[lossn=cfX,anchor=north] (lsd) at ($(s4.south)+(0,-0.3)$) {$\mathcal L_{\rm side}$: CE vs.\ attended side};
\draw[ar=cfX] (s4.south) -- (lsd.north);

\node[ttl] at (0,-1.35) {(b) audio-only adversary behind gradient reversal};
\node[nb=cfA,text width=1.6cm] (aenc) at (1.0,-2.3) {shared audio\\encoder};
\node[tens=cfA,minimum width=1.2cm] (ak) at (3.05,-2.3) {$\hat{a}_k$\\$16\times L$};
\draw[ar=cfA] (aenc) -- (ak);
\node[nb=cfG,text width=1.2cm] (grl) at (4.95,-2.3) {GRL\\$\times(-1)$};
\node[nb=cfG,text width=1.55cm] (f) at (6.95,-2.3) {mean, s.d.\\over $t$ (32)};
\node[nb=cfG,text width=1.8cm] (mlp) at (9.15,-2.3) {MLP\\$32\to32\to1$};
\node[nb=cfG,text width=1.55cm] (sm) at (11.3,-2.3) {softmax\\over $k$};
\node[lossn=cfG,anchor=west] (ladv) at (12.6,-2.3) {$\mathcal L_{\rm adv}$: pick the attended $k$ from audio};
\draw[ar=cfA] (ak) -- (grl);
\draw[ar=cfG] (grl) -- (f);
\draw[ar=cfG] (f) -- (mlp);
\draw[ar=cfG] (mlp) -- (sm);
\draw[ar=cfG] (sm) -- (ladv);
\draw[->,line width=0.7pt,draw=cfG,dashed] (ladv.south) -- ++(0,-0.55) -| (aenc.south)
  node[pos=0.25,below,font=\tiny,text=cfInk]
  {gradient reversed at the GRL: the encoder is trained to \emph{remove} whatever lets
   audio alone identify the attended talker};

\node[ttl] at (0,-3.95) {(c) envelope read-out (reporting only)};
\node[tens=cfN,minimum width=1.2cm] (e) at (0.6,-4.8) {$e$\\$16\times L$};
\node[nb=cfX,text width=1.25cm] (r1) at (2.45,-4.8) {$1{\times}1$ conv\\$16\to1$};
\node[tens=cfX,minimum width=0.9cm] (eh) at (4.1,-4.8) {$\hat a$};
\node[lossn=cfX,anchor=west] (lenv) at (5.0,-4.8) {$\mathcal L_{\rm env}=1-r(\hat a,a_{\rm att})$;
  gives the reconstruction $r$; not used by any decision rule};
\draw[ar=cfN] (e) -- (r1);
\draw[ar=cfX] (r1) -- (eh);
\draw[ar=cfX] (eh) -- (lenv);
\end{tikzpicture}}
\caption{Auxiliary heads. (a) The side head reads only the listener-corrected EEG; its side
log-probabilities enter the training loss through the candidate scores but never the
reported accuracies, which use the coupling score alone. (b) The audio-only adversary tries
to find the attended candidate from the audio embeddings; the gradient-reversal layer (GRL)
passes its gradient to the audio encoder with the sign flipped, so the encoder removes any
such cue. (c) The envelope read-out reports reconstruction accuracy and is not used by any
decision rule.}\label{fig:net_aux}
\end{figure*}

\paragraph{Size and compute.} The decision path has \archDecision{} shared parameters, a
size that suits wearable hardware and the limited data per listener. All modules are convolutional and run in time linear in the
window length. Training runs on one GPU in minutes to a few hours per fold
(App.~\ref{app:data}).

\section{Training Objective}\label{app:loss}
The objective (Eq.~\ref{eq:loss}) has one term that asks for the decision and several that
shape \emph{how} the decision is reached. This section explains each term's role in plain
terms, how the weights were set, and how large each term is in practice
(Table~\ref{tab:loss_terms}).

\paragraph{Definitions.} A training batch holds $B\le32$ windows $i=1,\dots,B$, all from \emph{one}
listener. Window $i$ has the recording $x^{(i)}$, the neural embedding $e^{(i)}$, the
acoustic embeddings $\hat a^{(i)}_k$ of its $K$ candidates, the attended index $y_i$ and the
attended envelope $a^{(i)}=a^{(i)}_{y_i}$. We write $s_{ik}=s(e^{(i)},\hat a^{(i)}_k)$ for the
score of Eq.~\ref{eq:score} before centring and $\bar s_{ik}$ for the centred score $s_k$ of
window $i$. The terms of Eq.~\ref{eq:loss} are as follows, with $\mathrm{sp}(z)=\log(1+e^z)$ the softplus function.
\begin{itemize}\setlength\itemsep{1pt}
\item \emph{Task.} $\mathcal L_{\rm CE}=-\frac1B\sum_i\sum_k q_{ik}\log\pi_{ik}$, the
cross-entropy (CE) between the predicted distribution $\pi_{ik}=\mathrm{softmax}_k(\bar s_{ik}+
h_{ik})$ and the label $q_{ik}=0.9\,[k{=}y_i]+0.1/K$ ($[k{=}y_i]$ is 1 if $k=y_i$, else 0), smoothed because the label is the
instruction, not what the listener did at every moment (attention lapses). $h_{ik}$ is the
side head's score for candidate $k$ (see $\mathcal L_{\rm side}$; 0 when the candidates'
sides are unknown); it is used only in training, and every reported decision uses
$\bar s_{ik}$ alone.
\item \emph{Contrastive matching.} With $G_{ij}=s(e^{(i)},\hat a^{(j)}_{y_j})$, the score
between window $i$'s brain response and window $j$'s attended speech,
$\mathcal L_{\rm NCE}=-\frac{1}{2B}\sum_i\big[\log\frac{e^{G_{ii}}}{\sum_j e^{G_{ij}}}+
\log\frac{e^{G_{ii}}}{\sum_j e^{G_{ji}}}\big]$ (symmetric InfoNCE). It asks which of the $B$
speech segments evoked each response, and vice versa: a 32-way problem, which gives the
encoders far more signal per window than the 2-way task. Since all windows come from one
listener, only the response's time course, not who is listening, can solve it.
\item \emph{Null hinge.} $\mathcal L_{\rm null}=\frac1B\sum_i\big[\mathrm{sp}(s(e^{(i-1)},
\hat a^{(i)}_{y_i})-s_{i,y_i}+\xi)+\mathrm{sp}(\xi-s_{i,y_i})\big]$, margin $\xi=0.5$,
$e^{(0)}=e^{(B)}$. The attended speech must score higher with its own recording than with
another window's recording (first term) and than with a constant recording, which scores 0
(second term). $\mathcal L_{\rm CE}$ only compares candidates \emph{within} a window; this
term ties the score to the recording itself, a training-time version of the null test.
\item \emph{Audio-only adversary.} A small network reads only the acoustic embeddings: the
time mean and standard deviation (s.d.) of each $\hat a^{(i)}_k$ (32 numbers) pass through a two-layer
perceptron to one logit per candidate, and $\mathcal L_{\rm adv}$ is the cross-entropy of its
softmax against $y_i$. The adversary minimises $\mathcal L_{\rm adv}$; between it and the
acoustic encoder sits a gradient-reversal layer \citep{ganin2016domain}, which flips the
sign of the gradient, so the encoder is trained to make the adversary fail. This removes
any acoustic cue (loudness, voice) that reveals the attended talker without the brain.
\item \emph{Collapse guard.} $\mathcal L_{\rm VIC}=\frac{1}{16B}\sum_{i,d}\big[\mathrm{sd}_{\min}-
\mathrm{sd}_t(e^{(i)}_{\cdot d})\big]_+ +\frac{0.04}{16}\sum_{d\neq d'}C_{dd'}^2$, with
$\mathrm{sd}_t$ the s.d.\ over time, $\mathrm{sd}_{\min}=0.5$,
$[z]_+=\max(z,0)$, and $C$ the $16\times16$ covariance of the time-centred embedding over
all samples of the batch: the variance and covariance terms of VICReg
\citep{bardes2022vicreg}. A constant embedding scores every candidate 0 and stops learning;
copies of one dimension waste the other 15. The term stays negligible unless either starts.
\item \emph{Flow matching.} $\mathcal L_{\rm FM}$ is Eq.~\ref{eq:fm} with $a=a^{(i)}$ and
$c=e^{(i)}$ (or $c=\varnothing$), averaged over the batch with one draw of $(u,\varepsilon)$
per window. It trains the flow head; through $c$ it also asks $e^{(i)}$ to carry what predicts
the attended envelope.
\item \emph{Side head.} A separate EEG-only network (App.~\ref{app:net}) gives probabilities
$p_{\rm side}(\text{left}\mid x^{(i)})$ and $p_{\rm side}(\text{right}\mid x^{(i)})$ that the
attended talker is on that side, and $\mathcal L_{\rm side}=-\frac{1}{|\mathcal B|}
\sum_{i\in\mathcal B}\log p_{\rm side}(z_i\mid x^{(i)})$, with $z_i$ the attended side and
$\mathcal B$ the windows where it is known. $h_{ik}$ is this log-probability for candidate
$k$'s side, centred over candidates and scaled by a learned gain. Lateralised EEG can carry
spatial attention, but also gaze or listener identity; a separate head lets us measure its
contribution and keep it out of reported decisions.
\item \emph{Envelope read-out.} $\mathcal L_{\rm env}=\frac1B\sum_i\big[1-r(\psi^\top
e^{(i)},a^{(i)})\big]$, with $\psi\in\mathbb R^{16}$ a learned linear map and $r$ the Pearson
correlation over time. It reports how well the neural embedding reconstructs the attended
envelope (Table~\ref{tab:track}), comparable to backward models, and is not used in decisions.
\end{itemize}
The weights, $\lambda_{\rm NCE}=1$, $\lambda_{\rm null}=0.5$,
$\lambda_{\rm adv}=\lambda_{\rm FM}=\lambda_{\rm side}=0.3$, $\lambda_{\rm VIC}=0.1$,
$\lambda_{\rm env}=0.2$, were set once and are the same for every dataset: the two decoding
terms at full weight, the others below them so that the decision shapes the embedding.

\paragraph{Why more than one term.} Cross-entropy over the candidates is all the task
requires, but on its own it accepts any route to the label. AAD data offer several routes
that do not go through the brain response: the attended talker may be louder or have a
recognisable voice, every window of a trial has the same label, and a listener may attend the
same side throughout. A decoder that takes such a route scores well on a leaky split and
fails on new content. The architecture already closes the main route: the time-centred
score is zero for every candidate without time-varying neural input (Prop.~\ref{prop:null}).
The remaining terms make the useful route easier to find (InfoNCE), guard against
degenerate solutions (null hinge, VICReg, adversary), and add outputs that the decision does
not use but that we report (flow head, side head, envelope read-out).

\paragraph{Decoding terms.} $\mathcal L_{\rm CE}$ is the task: given one window and its
candidates, put the highest score on the attended one. Label smoothing (0.1) reflects that
the label is an instruction, not an observation: listeners lapse, so a fraction of windows
is mislabelled and should not be fitted with full confidence. With two candidates
$\mathcal L_{\rm CE}$ gives one bit of supervision per window, too little to teach two
encoders from scratch. $\mathcal L_{\rm NCE}$ supplies more: it asks which of the $B=32$
windows' attended speech goes with which window's recording, a $B$-way matching problem
($\ln 32=3.47$ nats at chance against $\ln 2=0.69$ for the task). Because all windows in a
batch come from one listener, recognising the listener cannot solve it; only the time
course of the response can. It is the largest term at the first epoch
(Table~\ref{tab:loss_terms}) and falls by \nceDropMin--\nceDropMax\% as the encoders learn to align
brain and speech. Both terms act on the same score (Eq.~\ref{eq:score}), so what InfoNCE
learns transfers directly to the decision.

\paragraph{Anti-shortcut terms.} $\mathcal L_{\rm null}$ states the null hypothesis as a
constraint: the true recording must give the attended speech a higher score than the
recording of another window, and than a constant recording (score 0), each by a margin
$\xi=0.5$. A decoder that scores the attended talker high whatever the brain did violates
it. The softplus keeps a small gradient after the margin is met. $\mathcal L_{\rm adv}$
trains an audio-only classifier to find the attended candidate from summary statistics of
the candidate embeddings; the reversed gradient \citep{ganin2016domain} then removes from
the acoustic encoder whatever let it succeed. Its value is a direct check: it is
$0.3\ln 2=0.21$, chance, at the first and at the selected epoch on every dataset, so after
quantile matching the audio carries no usable cue to the attended role. $\mathcal L_{\rm VIC}$ is a
guard against a collapsed neural embedding (constant over time, or with all dimensions
alike), which would score every candidate 0 and stall training. Its variance part is zero while every
dimension's temporal s.d.\ exceeds $\mathrm{sd}_{\min}=0.5$; the whole term is below 0.005 at the first
and at the selected epoch on every dataset. These
three terms are therefore insurance: they cost nothing when the model behaves and act only
when it starts to take a shortcut. On MAESTRO, where a loudness shortcut exists, removing
any one of them leaves every null at chance (Table~\ref{tab:mabl}), because the score already closes
the shortcut.

\paragraph{Density and auxiliary terms.} $\mathcal L_{\rm FM}$ (Eq.~\ref{eq:fm}) trains the
flow head, which supplies the PMI rule and the per-source residuals (Table~\ref{tab:flow}).
Its gradient also reaches the neural encoder through the condition, and asks the embedding to
carry what predicts the attended envelope, which agrees with the coupling objective.
$\mathcal L_{\rm side}$ trains the EEG-only side head. Lateralised EEG can carry attention,
but also gaze and, when a listener attends one side throughout, listener identity: on
Broderick the term drops to 0 because side and listener coincide. We therefore measure its
credit separately and never use it in a reported decision. $\mathcal L_{\rm env}=1-r$
trains a linear read-out of the embedding to the attended envelope; it gives the tracking
values of Table~\ref{tab:track} and slightly regularises the embedding towards the envelope.

\paragraph{How the weights were set.} The weights were set once and not tuned on any dataset;
every dataset uses the same values. Where the concurrent MERIT
objective \citep{merit2026} has a corresponding term we took its weight; the others follow
from scale. The task has weight 1 and is the reference. InfoNCE also has weight 1: it is
the same log-softmax on the same score, and should dominate early. The null hinge (0.5) is
a constraint that should bind only when violated. The adversary (0.3) is kept moderate
because a gradient-reversal game becomes unstable when the reversed gradient dominates; the
flow, side and envelope terms (0.3, 0.3, 0.2) are kept below the decoding terms so that the
neural embedding is shaped mainly by the decision; VICReg (0.1) is a guard. Two facts make
the choice uncritical. First, the selected model is chosen on validation
\emph{contribution} (accuracy minus the across-trial null), not on the loss, so the
weights trade off training signals, not the selection criterion. Second, removing any one
term changes 5-s accuracy on KU Leuven, DTU and MEG-MASC by at most \ablOtherMax{} points
(Table~\ref{tab:ablation}; one seed per variant); only the per-listener correction, an
architectural component, matters there. On MAESTRO, InfoNCE does matter
(\maAblNce{} points per trial, Table~\ref{tab:mabl}).

\begin{table*}[h]\centering\small
\caption{Weighted loss terms during training (each already multiplied by its weight; mean
over folds and seeds of the logged epoch average), at the first epoch and at the selected
epoch. Chance: value of a term for a decoder at chance ($K=2$ candidates, batch $B=32$).
``--'': the side head needs two talkers in known positions. Five Broderick runs evaluated
from saved checkpoints have no logged history and are omitted.}\label{tab:loss_terms}
\begin{tabular}{llcccccccc}\toprule
Dataset & epoch & $\mathcal L_{\rm CE}$ & $\mathcal L_{\rm NCE}$ & $\mathcal L_{\rm null}$ & $\mathcal L_{\rm VIC}$ & $\mathcal L_{\rm adv}$ & $\mathcal L_{\rm FM}$ & $\mathcal L_{\rm side}$ & $\mathcal L_{\rm env}$ \\\midrule
KU Leuven & first & 0.67 & 3.36 & 0.90 & 0.00 & 0.21 & 0.29 & 0.20 & 0.19 \\
 & selected & 0.40 & 2.05 & 0.27 & 0.00 & 0.21 & 0.26 & 0.12 & 0.17 \\
DTU & first & 0.66 & 3.20 & 0.82 & 0.00 & 0.21 & 0.29 & 0.21 & 0.19 \\
 & selected & 0.43 & 1.78 & 0.21 & 0.00 & 0.21 & 0.23 & 0.19 & 0.17 \\
Fuglsang'20 & first & 0.65 & 3.15 & 0.87 & 0.00 & 0.21 & 0.30 & 0.20 & 0.19 \\
 & selected & 0.40 & 1.87 & 0.26 & 0.00 & 0.21 & 0.26 & 0.15 & 0.16 \\
ESAA & first & 0.68 & 3.32 & 0.93 & 0.00 & 0.21 & 0.31 & 0.21 & 0.19 \\
 & selected & 0.39 & 1.97 & 0.25 & 0.00 & 0.21 & 0.25 & 0.15 & 0.17 \\
KUL-255ch & first & 0.69 & 3.10 & 0.93 & 0.00 & 0.21 & 0.31 & 0.21 & 0.20 \\
 & selected & 0.38 & 1.90 & 0.29 & 0.00 & 0.21 & 0.25 & 0.09 & 0.17 \\
Broderick & first & 0.71 & 3.35 & 0.92 & 0.00 & 0.21 & 0.30 & 0.22 & 0.20 \\
 & selected & 0.27 & 2.35 & 0.35 & 0.00 & 0.21 & 0.26 & 0.00 & 0.18 \\
SparrKULee & first & 0.53 & 2.79 & 0.57 & 0.00 & 0.21 & 0.27 & -- & 0.18 \\
 & selected & 0.38 & 1.37 & 0.14 & 0.00 & 0.21 & 0.26 & -- & 0.17 \\
MEG-MASC & first & 0.42 & 1.75 & 0.27 & 0.00 & 0.21 & 0.27 & -- & 0.16 \\
 & selected & 0.32 & 0.29 & 0.01 & 0.00 & 0.21 & 0.23 & -- & 0.12 \\
\midrule
chance & & 0.69 & 3.47 & -- & -- & 0.21 & -- & 0.21 & -- \\
\bottomrule\end{tabular}

\end{table*}

\section{Additional Results}\label{app:results}
Table~\ref{tab:app_baselines} gives CoupleFlow and every baseline at every decision window.
Sec.~\ref{app:ablation} removes one component at a time, Sec.~\ref{app:leak} compares two
kinds of leakage, Sec.~\ref{app:mabl} repeats the ablation on MAESTRO, and
Sec.~\ref{app:loso} holds out listeners. Table~\ref{tab:app_variants} lists CoupleFlow
variants, and Table~\ref{tab:kth} tests the accumulation law (Prop.~\ref{prop:law}).
\begin{table}[h]\centering\small
\caption{CoupleFlow and every baseline at every window (\%, content-disjoint folds, pooled over
all test windows; competing talker for two-talker sets, match--mismatch for SparrKULee and
MEG-MASC). CoupleFlow: mean of 3 seeds with the validation-chosen lag window, as in
Table~\ref{tab:main}; ``CoupleFlow, mm'': match--mismatch on the two-talker recordings. trial:
one decision per trial. ``--'': window longer than the trials. CoupleFlow nulls:
\nullShortMin--\nullShortMax\% up to 10\,s; \nullLongMin--\nullLongMax\% at 20--60\,s, where
few windows exist. DECAF, trained with the same folds and selection rule, stays near chance on the
two-talker sets.}
\label{tab:app_baselines}
\begin{tabular}{llcccccccc}\toprule
Dataset & model & 1\,s & 2\,s & 5\,s & 10\,s & 20\,s & 30\,s & 60\,s & trial \\\midrule
KU Leuven & Linear & 57.0 & 60.0 & 64.9 & 69.9 & 75.1 & 79.2 & 84.2 & 87.8 \\
 & VLAAI & 56.9 & 59.3 & 63.8 & 68.5 & 73.0 & 77.4 & 82.6 & 88.1 \\
 & Dilated MM & 55.3 & 59.8 & 65.9 & 71.6 & 78.2 & 82.6 & 89.8 & 96.9 \\
 & DECAF & 50.9 & 51.6 & 51.3 & 52.0 & 55.4 & 54.1 & 58.3 & 65.0 \\
 & DECAF (EEG branch) & 51.3 & 51.4 & 49.1 & 50.2 & 51.0 & 51.2 & 50.3 & 55.3 \\
 & CoupleFlow & 60.5 & 65.0 & 72.7 & 79.3 & 86.7 & 90.5 & 96.4 & 99.3 \\
 & CoupleFlow, mm & 64.8 & 70.6 & 79.7 & 87.3 & 93.8 & 96.6 & 99.0 & -- \\
\midrule
DTU & Linear & 59.1 & 63.9 & 72.4 & 79.0 & 86.1 & 90.0 & -- & 93.9 \\
 & VLAAI & 57.5 & 60.4 & 64.8 & 69.7 & 74.1 & 78.2 & -- & 82.2 \\
 & Dilated MM & 56.0 & 61.4 & 69.1 & 75.3 & 83.7 & 88.5 & -- & 92.9 \\
 & DECAF & 50.0 & 50.0 & 50.3 & 50.1 & 50.1 & 48.8 & -- & 49.6 \\
 & DECAF (EEG branch) & 51.0 & 51.1 & 49.7 & 50.0 & 51.0 & 51.2 & -- & 52.5 \\
 & CoupleFlow & 61.4 & 67.4 & 76.2 & 83.4 & 90.6 & 94.5 & -- & 97.0 \\
 & CoupleFlow, mm & 64.1 & 71.0 & 81.9 & 89.1 & -- & -- & -- & -- \\
\midrule
Fuglsang'20 & Linear & 63.4 & 68.5 & 77.3 & 85.1 & 91.9 & 94.6 & -- & 98.3 \\
 & VLAAI & 60.6 & 63.7 & 68.9 & 74.5 & 81.7 & 85.4 & -- & 89.8 \\
 & Dilated MM & 57.9 & 64.3 & 73.0 & 80.2 & 89.0 & 92.7 & -- & 95.4 \\
 & DECAF & 48.7 & 47.9 & 41.4 & 37.0 & 32.5 & 19.8 & -- & 39.7 \\
 & DECAF (EEG branch) & 51.2 & 51.3 & 46.6 & 45.3 & 43.8 & 41.7 & -- & 39.5 \\
 & CoupleFlow & 63.8 & 69.8 & 79.6 & 86.7 & 93.3 & 95.9 & -- & 98.5 \\
 & CoupleFlow, mm & 66.3 & 73.1 & 83.0 & 90.2 & 95.7 & -- & -- & -- \\
\midrule
ESAA & Linear & 61.3 & 65.6 & 74.1 & 80.3 & 87.6 & 90.2 & 93.7 & 94.7 \\
 & VLAAI & 58.3 & 61.5 & 66.4 & 71.8 & 78.8 & 81.7 & 88.9 & 91.0 \\
 & Dilated MM & 57.5 & 62.7 & 70.6 & 77.3 & 84.7 & 87.3 & 92.7 & 94.3 \\
 & DECAF & 50.0 & 49.9 & 49.9 & 50.0 & 50.0 & 50.5 & 50.3 & 49.8 \\
 & DECAF (EEG branch) & 51.1 & 51.5 & 49.7 & 49.4 & 49.7 & 48.9 & 52.2 & 50.7 \\
 & CoupleFlow & 63.2 & 68.9 & 77.9 & 84.9 & 90.6 & 93.4 & 96.1 & 97.1 \\
 & CoupleFlow, mm & 62.6 & 68.7 & 77.1 & 84.6 & 91.2 & 92.8 & -- & -- \\
\midrule
KUL-255 & Linear & 57.0 & 60.5 & 65.5 & 70.9 & 76.2 & 81.5 & 87.4 & 98.3 \\
 & VLAAI & 55.9 & 58.0 & 62.1 & 67.0 & 72.1 & 75.4 & 82.2 & 95.0 \\
 & Dilated MM & 54.7 & 58.8 & 65.0 & 70.4 & 77.6 & 82.5 & 88.5 & 98.3 \\
 & DECAF & 50.1 & 50.1 & 50.4 & 50.9 & 49.8 & 50.1 & 51.2 & 52.5 \\
 & DECAF (EEG branch) & 51.0 & 50.9 & 49.9 & 50.9 & 51.1 & 53.1 & 52.2 & 52.5 \\
 & CoupleFlow & 58.9 & 62.8 & 69.4 & 75.9 & 83.0 & 87.7 & 93.9 & 99.4 \\
 & CoupleFlow, mm & 61.1 & 65.8 & 73.5 & 80.5 & 87.6 & 91.9 & 96.0 & -- \\
\midrule
Broderick & Linear & 56.5 & 59.1 & 63.7 & 68.8 & 75.9 & 78.8 & -- & 84.3 \\
 & VLAAI & 51.0 & 51.3 & 52.2 & 53.1 & 53.1 & 53.9 & -- & 55.4 \\
 & Dilated MM & 50.5 & 51.3 & 52.3 & 52.7 & 54.4 & 53.6 & -- & 56.2 \\
 & DECAF & 50.1 & 50.0 & 50.3 & 50.2 & 50.8 & 50.5 & -- & 50.6 \\
 & DECAF (EEG branch) & 50.6 & 50.6 & 50.3 & 50.0 & 52.2 & 50.8 & -- & 49.2 \\
 & CoupleFlow & 55.8 & 59.0 & 63.6 & 69.0 & 75.2 & 79.2 & -- & 83.1 \\
 & CoupleFlow, mm & 55.1 & 57.8 & 62.3 & 65.6 & 70.2 & -- & -- & -- \\
\midrule
SparrKULee & Linear & 61.8 & 67.2 & 76.3 & 83.7 & 90.5 & 93.5 & 97.4 & -- \\
 & VLAAI & 59.9 & 64.0 & 70.5 & 77.1 & 83.9 & 87.5 & 93.3 & -- \\
 & Dilated MM & 62.6 & 71.5 & 83.3 & 91.0 & 96.4 & 98.3 & 99.7 & -- \\
 & DECAF & 53.1 & 54.6 & 56.0 & 57.8 & 60.9 & 63.0 & 66.7 & -- \\
 & DECAF (EEG branch) & 54.5 & 56.5 & 59.2 & 61.8 & 66.6 & 69.0 & 73.2 & -- \\
 & CoupleFlow & 73.5 & 82.3 & 92.2 & 97.1 & 99.3 & 99.7 & 100.0 & -- \\
\midrule
MEG-MASC & Linear & 72.7 & 81.3 & 92.0 & 97.3 & 99.4 & 99.8 & 100.0 & -- \\
 & VLAAI & 71.0 & 78.7 & 89.1 & 95.6 & 98.6 & 99.6 & 100.0 & -- \\
 & Dilated MM & 72.0 & 85.0 & 96.2 & 99.3 & 100.0 & 100.0 & 100.0 & -- \\
 & DECAF & 62.9 & 69.9 & 80.6 & 89.0 & 95.8 & 97.4 & 99.7 & -- \\
 & DECAF (EEG branch) & 63.2 & 69.9 & 80.6 & 89.6 & 95.9 & 97.8 & 99.7 & -- \\
 & CoupleFlow & 83.7 & 93.3 & 99.0 & 99.9 & 100.0 & 100.0 & 100.0 & -- \\
\bottomrule\end{tabular}

\end{table}

\FloatBarrier
\subsection{Ablations}\label{app:ablation}
Table~\ref{tab:ablation}, 5-s accuracy, all folds, relative to the
125-ms model with 3 seeds; each variant is one run (seed 0). The per-listener low-rank correction is the component that
carries accuracy: without it the model loses \ablRankKUL{} (KU Leuven), \ablRankDTU{} (DTU) and
\ablRankMEG{} points (MEG). Starting the lag window at 0\,ms adds \lagZeroKUL, \lagZeroDTU{}
and \lagZeroMEG. Every other component changes 5-s accuracy by at most \ablOtherMax{} points
(one seed per variant). This includes the flow head, matched and glimpsed candidates,
the adversary, the null hinge, InfoNCE, contribution-based selection and the side head. These
components are therefore not accuracy devices. The flow head supplies the PMI decision rule
and the per-source residuals (Table~\ref{tab:flow}); the adversary, null hinge and selection
rule are safeguards, whose effect is on the null and credit diagnostics rather than on
accuracy. The lag control, which reads neural data from 1.1\,s to 125\,ms \emph{before} the
sound, loses \lagCtlKUL, \lagCtlDTU{} and \lagCtlMEG{} points (\lagCtlAccKUL, \lagCtlAccDTU{} and
\lagCtlAccMEG\%). MEG remains well above chance in this control. We attribute this, as a
hypothesis, to the regular rhythm of its synthetic speech, which makes the envelope
predictable from the preceding response.
\begin{table}[h]\centering\small
\caption{Change in 5-s accuracy (points) when one component is removed; full model:
mean$\pm$s.d.\ over 3 seeds, 125-ms lag window; variants: seed 0.}\label{tab:ablation}
\begin{tabular}{lccc}\toprule
Variant & KU Leuven & DTU & MEG-MASC (MEG) \\\midrule
full model & 72.7{\scriptsize$\pm$0.2} & 75.3{\scriptsize$\pm$0.3} & 97.6{\scriptsize$\pm$0.1} \\\midrule
lag window from 0\,ms & +0.3 & +1.2 & +1.5 \\
no flow head & +0.8 & +0.2 & +0.0 \\
no listener correction & -3.0 & -4.2 & -2.0 \\
no matched candidates & +0.7 & -0.4 & +0.1 \\
no glimpse features & -0.1 & +0.7 & +0.2 \\
no audio adversary & -0.4 & +0.5 & +0.1 \\
no null hinge & +0.7 & -0.3 & +0.0 \\
no InfoNCE & -0.6 & -0.4 & -0.2 \\
select on accuracy & +0.4 & -0.0 & +0.1 \\
no side head & +0.6 & -0.7 & -0.1 \\
no env/VICReg aux. & +0.1 & +0.1 & -0.1 \\
lag control (EEG before sound) & -18.8 & -13.7 & -15.3 \\
\bottomrule\end{tabular}

\end{table}

\FloatBarrier
\subsection{Two kinds of leakage}\label{app:leak}
Table~\ref{tab:leak} and Table~\ref{tab:earlier} can look contradictory: one finds that a
leaky split inflates accuracy by at most \leakMax{} points, the other by
\oldDropAllMin--\oldDropAllMax. They measure different leaks.

\emph{Content leakage} (Table~\ref{tab:leak}). In a trial split, whole trials are held
out, so no test sample, and no EEG from a test trial, is seen in training. Only the
\emph{stories} repeat: a test trial's audio was heard by the listener in another, training
trial. A model can profit from this only if it can recognise the audio and map it to a
label, as an audio-only or stimulus-identity classifier could. To measure what a content-sharing split does, we retrain every
model on KU Leuven and DTU with one group per \emph{trial}, the split used in most prior
work, so test stories also occur in training (Table~\ref{tab:leak}). No model gains
more than \leakMax{} points: \leakMin{} to \leakMax{} for the linear decoder, VLAAI, the match--mismatch
model and CoupleFlow alike. Sharing stories across trials is therefore \emph{not} what
inflates these decoders. Each of them reaches the decision through the neural response
to the candidate audio, with no path that can memorise a story's identity. The small KU Leuven gain for CoupleFlow (\leakMax) is consistent
with its trial split exposing all four story pairs in training instead of three.

\emph{Trial-identity leakage} (Table~\ref{tab:earlier}). In a within-trial split, every
trial contributes windows to both training and test: our earlier decoders used a random
5-fold split of overlapping windows, the side classifiers' protocol holds out the last 10\%
of each trial. Each test window then has training windows from the \emph{same trial}, with
the same label, overlapping it or seconds away. EEG has slow, trial-specific structure: electrode impedances, drifts,
posture and gaze, and an attended side that is constant within the trial. A decoder that
recognises which trial a window comes from therefore knows its label without decoding
attention. This is the much larger effect: \oldDropAllMin--\oldDropAllMax{} points for our earlier
decoders, and \spKULBmin--\spKULBmax\% at 1\,s on KU Leuven for the side classifiers
(Table~\ref{tab:app_spatial}; on DTU our runs under their split come out about 12 points
below the published numbers, plausibly because of our stricter preprocessing: 0.5-Hz
high-pass, per-trial z-scoring, no electrooculography (EOG) channels). NJU, whose EEG is not
time-locked to the audio at all,
reached \oldSegNJU\% this way. The two tables therefore
agree: what inflates AAD accuracy is evaluating on windows of trials seen in training, not
on stories heard in training. Content-disjoint folds rule out both.
\begin{table}[h]\centering\small
\caption{Content leakage: 5-s competing accuracy (\%) when folds share stories (trial
split: whole trials held out, but the stories of test trials also occur in training trials)
against content-disjoint folds. Every model is retrained on each split. CoupleFlow: seed 0,
125-ms lag window on both splits.}\label{tab:leak}
\begin{tabular}{llccc}\toprule
Dataset & model & trial split & content-disj. & inflation \\\midrule
KU Leuven & Linear & 66.2 & 64.9 & +1.2 \\
 & VLAAI & 63.5 & 63.8 & -0.2 \\
 & Dilated MM & 66.5 & 65.9 & +0.6 \\
 & CoupleFlow & 74.8 & 72.8 & +2.0 \\
\midrule
DTU & Linear & 72.2 & 72.4 & -0.3 \\
 & VLAAI & 64.6 & 64.8 & -0.2 \\
 & Dilated MM & 69.0 & 69.1 & -0.1 \\
 & CoupleFlow & 75.5 & 75.4 & +0.1 \\
\bottomrule\end{tabular}

\end{table}
\begin{table}[h]\centering\small
\caption{Recent side classifiers (authors' models in our training pipeline), accuracy (\%)
under their within-trial split (last 10\% of each trial held out for testing; earlier windows
split 9:1 into training and validation) and under our content-disjoint folds,
subject-dependent and subject-independent. On KU Leuven the content-disjoint side task is
ill-posed (one class per listener in each test fold).}\label{tab:app_spatial}
\begin{tabular}{llccc|ccc|ccc}\toprule
 & & \multicolumn{3}{c|}{their split} & \multicolumn{3}{c|}{content-disj., subj.-dep.} & \multicolumn{3}{c}{content-disj., subj.-indep.} \\
Dataset & model & 1\,s & 2\,s & 5\,s & 1\,s & 2\,s & 5\,s & 1\,s & 2\,s & 5\,s \\\midrule
KU Leuven & DARNet & 96.8 & 96.9 & 98.0 & 52.3 & 53.1 & 53.8 & 54.2 & 58.5 & 58.9 \\
 & DBPNet & 97.1 & 97.7 & 97.9 & 46.8 & 45.5 & 45.0 & 51.6 & 53.8 & 53.4 \\
 & ListenNet & 98.1 & 97.8 & 97.0 & 41.9 & 43.9 & 44.0 & 55.1 & 55.7 & 57.4 \\
\midrule
DTU & DARNet & 67.7 & 68.0 & 67.5 & 54.5 & 53.2 & 53.2 & 49.2 & 49.9 & 47.3 \\
 & DBPNet & 66.5 & 67.6 & 68.8 & 50.8 & 52.2 & 51.4 & 47.7 & 48.2 & 48.2 \\
 & ListenNet & 74.5 & 74.4 & 70.8 & 54.1 & 55.0 & 55.2 & 49.1 & 48.1 & 49.6 \\
\midrule
ESAA & DARNet & 85.4 & 85.7 & 86.0 & 56.0 & 58.6 & 59.4 & 55.8 & 54.9 & 54.5 \\
 & DBPNet & 87.8 & 89.6 & 89.1 & 59.8 & 61.6 & 61.6 & 55.8 & 56.8 & 58.1 \\
 & ListenNet & 91.6 & 91.2 & 88.5 & 56.5 & 58.8 & 59.5 & 56.0 & 57.9 & 59.3 \\
\midrule
Fuglsang'20 & DARNet & 82.7 & 82.8 & 82.8 & 67.1 & 68.7 & 68.9 & 65.7 & 65.3 & 67.8 \\
 & DBPNet & 83.6 & 83.6 & 81.8 & 68.3 & 68.8 & 68.6 & 66.8 & 68.4 & 69.7 \\
 & ListenNet & 88.6 & 87.6 & 85.4 & 70.5 & 71.2 & 71.3 & 68.9 & 71.0 & 71.2 \\
\midrule
KUL-255 & DARNet & 98.5 & 98.9 & 99.0 & 41.9 & 40.9 & 40.2 & 44.6 & 41.8 & 44.4 \\
 & DBPNet & 99.4 & 99.6 & 99.6 & 40.8 & 43.3 & 42.3 & 43.1 & 41.9 & 42.1 \\
 & ListenNet & 98.6 & 98.4 & 98.1 & 47.1 & 46.5 & 45.6 & 48.4 & 47.1 & 48.8 \\
\bottomrule\end{tabular}

\end{table}
\begin{table}[h]\centering\small
\caption{Our earlier decoders (a flow decoder and an envelope decoder) under two splits, 5-s windows, one model
per listener, mean over listeners (\%). Within-trial: 5-fold split of randomly assigned
windows with 50\% overlap; test windows neighbour training windows of the same trial. Trial-disjoint: whole trials
held out, no shared samples. Everything else (data, listeners, recipe, folds) is identical.
Trial-disjoint is still not content-disjoint on KU Leuven, whose stories repeat across
trials. $^a$NJU shows no EEG--audio time-locking (App.~\ref{app:data}), so its
within-trial accuracy can only come from the split.}\label{tab:earlier}
\begin{tabular}{lccccccc}\toprule
& & \multicolumn{3}{c}{earlier flow decoder} & \multicolumn{3}{c}{earlier envelope decoder} \\
\cmidrule(lr){3-5}\cmidrule(lr){6-8}
Dataset & listeners & within-trial & trial-disj. & drop & within-trial & trial-disj. & drop \\\midrule
KU Leuven & 16 & 86.4 & 55.9 & 30.5 & 76.1 & 57.5 & 18.6 \\
DTU & 18 & 83.6 & 56.2 & 27.4 & 74.4 & 57.7 & 16.7 \\
NJU$^a$ & 21 & 86.0 & 54.5 & 31.5 & 70.1 & 50.7 & 19.4 \\
\bottomrule\end{tabular}

\end{table}

\FloatBarrier
\subsection{Ablation on MAESTRO}\label{app:mabl}
Table~\ref{tab:mabl} repeats the ablation on
MAESTRO, where the audio itself marks the attended talker. Two results stand out. First,
there is no discrepancy between accuracy and contribution for \emph{any} variant. Unmatched
and matched candidates give the same accuracy to within \maUnmatchedGap{} points, and every
null stays at chance. Removing the adversary, the null hinge, contribution-based selection or
matched training (each changes per-trial accuracy by at most \maSafeMax{} points) does not
open the shortcut. It is the architecture that closes it: the time-centred coupling score
gives the acoustic pathway no route to the decision (Prop.~\ref{prop:null}). The safeguards
are therefore redundant with it rather than necessary. Second, accuracy is carried by the
components that shape the brain--speech coupling. Per-trial accuracy changes by
\maAblRank{} points without the per-listener correction, \maAblNce{} without InfoNCE,
\maAblBanks{} without sub-band features, \maAblGlimpse{} without glimpse channels and
\maAblAux{} without the auxiliary losses.
\begin{table*}[h]\centering\small
\caption{Ablation on MAESTRO, where the audio offers a shortcut (the attended talker is always
the loudest): within listeners, seed 0, 4-way. For each variant: 5-s accuracy and
contribution with unmatched candidates (shortcut available) and with matched candidates;
the EEG-shuffle and circular-shift nulls; per-trial accuracy and contribution; and embedding
collapse (mean cosine between windows; 1 = every window alike).}\label{tab:mabl}
\resizebox{\textwidth}{!}{\begin{tabular}{lc|cc|cccccc|c}\toprule
 & & \multicolumn{2}{c|}{unmatched (5\,s)} & \multicolumn{6}{c|}{matched} & \\
Variant & folds & acc. & contrib. & acc. & null & contrib. & circ.\ null & trial & trial contrib. & collapse \\\midrule
full model & 5 & 45.6 & +20.6 & 45.9 & 25.0 & +20.9 & 25.0 & 70.2 & +45.4 & 0.00 \\
no flow head & 5 & 46.1 & +21.0 & 45.9 & 25.1 & +20.8 & 25.0 & 70.1 & +46.3 & 0.00 \\
no listener correction & 5 & 41.6 & +16.5 & 41.7 & 25.1 & +16.5 & 25.1 & 61.4 & +36.4 & 0.00 \\
unmatched training & 5 & 45.5 & +20.5 & 45.8 & 25.0 & +20.8 & 25.1 & 69.9 & +44.7 & 0.00 \\
no glimpse features & 5 & 44.9 & +19.9 & 45.0 & 25.1 & +20.0 & 25.1 & 68.8 & +43.8 & 0.00 \\
no sub-band features & 5 & 44.3 & +19.3 & 44.5 & 25.0 & +19.5 & 25.1 & 68.3 & +42.8 & 0.00 \\
no audio adversary & 5 & 45.8 & +20.7 & 45.9 & 25.1 & +20.8 & 25.1 & 70.0 & +45.3 & 0.00 \\
no null hinge & 5 & 45.4 & +20.4 & 45.3 & 25.0 & +20.3 & 25.1 & 70.9 & +46.7 & 0.00 \\
no InfoNCE & 5 & 43.3 & +18.2 & 43.3 & 25.1 & +18.2 & 25.1 & 66.6 & +42.1 & 0.00 \\
select on accuracy & 5 & 45.7 & +20.8 & 45.5 & 24.9 & +20.6 & 25.0 & 70.3 & +46.2 & 0.00 \\
no side head & 5 & 45.2 & +20.3 & 45.3 & 25.0 & +20.3 & 25.2 & 70.3 & +46.1 & 0.00 \\
no env/VICReg aux. & 5 & 45.2 & +20.2 & 45.1 & 25.0 & +20.0 & 24.9 & 68.9 & +44.3 & 0.00 \\
lag control (EEG before sound) & 5 & 33.6 & +8.5 & 33.5 & 25.1 & +8.4 & 25.2 & 46.3 & +20.8 & 0.00 \\
\bottomrule\end{tabular}
}
\end{table*}

\FloatBarrier
\subsection{Variants and the accumulation law}\label{app:lawdetails}
\paragraph{Task and sensor.} For the linear decoder the task prediction
$\kappa_{\rm comp}/\kappa_{\rm mm}=1-\gamma$ has mean error \taskMaeLin{} ($r=\taskRLin$,
$p=\taskPLin$), and for CoupleFlow \taskMaeCF{} ($r=\taskRCF$); the task factor
$1/(1-\gamma)$ is \taskFacMin--\taskFacMax{} (mean \taskFacMean). The law
$\kappa=\sqrt{f/2}\,\Delta\zeta$ with $f=\fEff$ fits the 14 linear values with mean error
\thMape\%. At matched task and listening condition (single talker, match--mismatch), MEG's
per-sample tracking is \megRratio$\times$ that of SparrKULee's EEG, which predicts a $\kappa$
ratio of \megDzPred; we observe \megKlinObs{} for the linear decoder and \megKcfObs{} for
CoupleFlow, which is compressed by its ceiling near 99\%. Single-talker listening is not what
raises EEG tracking: SparrKULee's attended correlation (\rSparr) lies inside the two-talker
range (\rTwoMin--\rTwoMax). CoupleFlow's read-out reaches \rCFattMin--\rCFattMax{} on the
two-talker EEG sets and \rCFmeg{} on MEG.

Table~\ref{tab:app_variants} lists CoupleFlow variants at 5\,s, and Table~\ref{tab:kth} gives
the tracking values behind the test of Prop.~\ref{prop:law} in Sec.~\ref{sec:exp}.
\begin{table}[h]\centering\small
\caption{CoupleFlow variants, 5-s accuracy (\%, seed 0). 125\,ms / 0\,ms: start of the neural
lag window. env.\ only: broadband envelope without sub-band features (Fuglsang'20 and
Broderick are envelope-only in every column). rank 8: larger per-listener correction.
listener FT: per-listener fine-tuning after pooled training. native ch.: all native channels
instead of the 64 canonical ones (125-ms window). n/a: the variant equals the base
(envelope-only releases; caps with 64 EEG channels). ``--'': not run. Rank 8 and listener FT
were run only on KU Leuven, DTU and ESAA, to test whether listener heterogeneity explained
the ESAA gap; neither did. MEG-MASC uses its 208 sensors throughout (208-channel input
layer).}\label{tab:app_variants}
\begin{tabular}{lcccccc}\toprule
Dataset & 125\,ms (base) & 0\,ms & env.\ only & rank 8 & listener FT & native ch. \\\midrule
KU Leuven & 72.8 & 73.0 & 72.1 & 73.4 & 73.2 & n/a \\
DTU & 75.4 & 76.4 & 74.4 & 75.6 & 75.1 & n/a \\
Fuglsang'20 & 79.9 & 79.5 & n/a & -- & -- & n/a \\
ESAA & 71.2 & 77.9 & 72.1 & 72.0 & 71.4 & n/a \\
KUL-255 & 67.1 & 69.2 & 67.0 & -- & -- & 69.9 \\
Broderick & 62.3 & 63.5 & n/a & -- & -- & 66.4 \\
SparrKULee & 86.6 & 92.3 & 85.3 & -- & -- & n/a \\
MEG-MASC & 97.6 & 99.0 & 97.4 & -- & -- & n/a \\
\bottomrule\end{tabular}

\end{table}
\begin{table}[h]\centering\small
\caption{Test of Prop.~\ref{prop:law}: tracking of the attended ($r_{\rm att}$) and
competing ($r_{\rm comp}$) envelope by the per-listener linear decoder (all listeners),
ignored-talker ratio $\gamma$, the predicted and the observed ratio of competing to
match--mismatch $\kappa$.}\label{tab:kth}
\begin{tabular}{lcccc|cc}\toprule
 & & & & pred. & \multicolumn{2}{c}{obs.\ $\kappa_{\rm comp}/\kappa_{\rm mm}$} \\
Dataset & $r_{\rm att}$ & $r_{\rm comp}$ & $\gamma$ & $1-\gamma$ & linear & CoupleFlow \\\midrule
KU Leuven & 0.099 & 0.027 & 0.28 & 0.72 & 0.71 & 0.71 \\
DTU & 0.130 & 0.023 & 0.18 & 0.82 & 0.84 & 0.79 \\
Fuglsang'20 & 0.171 & 0.018 & 0.10 & 0.90 & 0.85 & 0.83 \\
ESAA & 0.139 & 0.019 & 0.14 & 0.86 & 0.86 & 0.98 \\
KUL-255 & 0.101 & 0.018 & 0.18 & 0.82 & 0.85 & 0.77 \\
Broderick & 0.073 & 0.003 & 0.05 & 0.95 & 0.97 & 1.11 \\
SparrKULee & 0.132 & -- & -- & -- & -- & -- \\
MEG-MASC & 0.278 & -- & -- & -- & -- & -- \\
\bottomrule\end{tabular}

\end{table}

\FloatBarrier
\subsection{Further analyses}\label{app:further}
\paragraph{Where the attention signal lives.} Decomposing the PMI attention effect
($\PMI_{\rm att}-\PMI_{\rm comp}$) by modulation band of the candidate envelope (seed 0), the delta
band (0.5--4\,Hz) carries \deltaMin--\deltaMax\% of it on every dataset, theta (4--8\,Hz)
\thetaMin--\thetaMax\%, and alpha and beta together at most \abMax\%. This matches the known spectral profile of cortical
speech tracking \citep{ding2012emergence}, and it matches for MEG and EEG alike.

\paragraph{Hearing-impaired listeners.} On Fuglsang'20, 5-s accuracy is \hiAcc\% for the 22
hearing-impaired and \nhAcc\% for the 22 normal-hearing listeners (mean over 3 seeds;
Mann--Whitney $p=0.06$--$0.46$ across seeds). Decoding is not worse with hearing loss,
consistent with the enhanced envelope tracking reported for these listeners
\citep{fuglsang2020effects}. We use the release's envelopes of the unamplified stimuli for every listener.

\FloatBarrier
\subsection{Leave-one-listener-out evaluation}\label{app:loso}
The content-disjoint folds hold out stories but test listeners whose other stories were
seen in training. Here each fold holds out one listener \emph{and} one content fold: the test
set is that listener's trials on the held-out content, and training uses only the other
listeners' trials on the other content, so neither the test listener nor the test audio is
seen in training (validation: whole training groups when at least 10 remain, else trial
tails, as in App.~\ref{app:data}). Every fold passed the same split audit as the content
folds (no shared trial, content, stimulus file or sample, and no test-listener data in
training). Each dataset's paper configuration and validation-chosen lag are used unchanged,
with a single seed.

Table~\ref{tab:loso}. An unseen listener costs \losoDropMin--\losoDropMax{} points at 5\,s
on every dataset except Broderick, and accuracy keeps rising with the window (20-s column).
On Broderick
the decoder does not transfer to new listeners at all; a pooled linear backward model under
the same folds fails in the same way (last column), so this is a property of that recording,
not of the model. Where both transfer, CoupleFlow stays ahead of the pooled linear model.
\begin{table}[h]\centering\small
\caption{Leave-one-listener-out (LOSO) accuracy (\%), listener and content held out together,
\emph{one seed}. Content folds: the main-table 5-s accuracy (3 seeds) for reference. Null:
EEG from another trial of the same listener (another window of the same trial when the
listener has a single test trial). Linear: pooled ridge backward model on the same folds
(two-talker sets; penalty chosen on the first 40 validation segments). $^a$208-sensor input (App.~\ref{app:net}). MAESTRO's LOSO results are
in Sec.~\ref{sec:exp}.}\label{tab:loso}
\begin{tabular}{lcccccc}\toprule
& content folds & \multicolumn{4}{c}{LOSO, CoupleFlow (1 seed)} & LOSO, linear \\
\cmidrule(lr){3-6}
Dataset & 5\,s & 1\,s & 5\,s & 20\,s & null (5\,s) & 5\,s \\\midrule
KU Leuven & 72.7 & 58.0 & 68.0 & 81.4 & 49.8 & 59.7 \\
DTU & 76.2 & 58.5 & 70.4 & 84.2 & 50.0 & 64.0 \\
Fuglsang'20 & 79.6 & 61.0 & 73.7 & 89.0 & 50.2 & 67.5 \\
ESAA & 77.9 & 58.8 & 68.6 & 83.0 & 50.4 & 66.1 \\
KUL-255ch & 69.4 & 57.5 & 66.0 & 79.7 & 50.1 & 64.0 \\
Broderick & 63.6 & 49.9 & 50.1 & 50.0 & 49.7 & 49.2 \\
SparrKULee & 92.2 & 68.2 & 85.9 & 97.1 & 50.1 & -- \\
MEG-MASC$^a$ & 99.0 & 76.1 & 95.0 & 99.7 & 50.0 & -- \\
\bottomrule\end{tabular}

\end{table}

\section{Confidence and Sequential Decisions}\label{app:conf}
Everything so far scores fixed windows. Here we ask how the decoder's per-window scores should
be combined over time, when a decision should be made, and whether overlapping windows or a
different training window help. Nothing in this appendix retrains the paper's models except
Sec.~\ref{app:conf:trainlen}; the decision rules only read their per-window scores.

\subsection{Theory}\label{app:conf:theory}
\textbf{Setting and notation.} A test trial is cut into consecutive, non-overlapping
windows $i=1,2,\dots$ of length $w$ seconds. Window $i$ has the recording $x_i$ and the two
candidate envelopes $a_{1,i},a_{2,i}$; we write $a_{\cdot,i}=(a_{1,i},a_{2,i})$,
$x_{1:n}=(x_1,\dots,x_n)$ and $a_{1:n}=(a_{\cdot,1},\dots,a_{\cdot,n})$. The attended index
$y\in\{1,2\}$ is fixed within the trial. $P_k$ and $\mathbb E_k$ denote probability and
expectation given $y=k$ (and the candidates), $\log$ is the natural logarithm,
$\Lambda$ is the logistic function, $\Lambda(\cdot)=1/(1+\exp(-\,\cdot\,))$, and $\Phi$ and $\varphi$ are the standard
normal distribution and density functions. The per-window log-likelihood ratio is
\[\ell_i=\log p(x_i\mid y{=}1,a_{\cdot,i})-\log p(x_i\mid y{=}2,a_{\cdot,i});\]
under the assumptions of Prop.~\ref{prop:pmi} it equals $\PMI(a_{1,i};x_i)-\PMI(a_{2,i};x_i)$.
After $n$ windows the accumulated evidence is $S_n=\sum_{i=1}^n\ell_i$ ($S_0=0$). A real
statistic $\hat\ell$ is \emph{calibrated} when $P(y{=}1\mid\hat\ell)=\Lambda(\hat\ell)$. A decoder gives the
candidates scores $s_{1,i},s_{2,i}$ (Sec.~\ref{sec:theory}); we calibrate their margin
$\Delta s_i=s_{1,i}-s_{2,i}$ by one scale, $\hat\ell_i=\beta\,\Delta s_i$, with $\beta>0$ fitted by the
logistic loss on the fold's validation windows, and use $\hat\ell_i$ in place of $\ell_i$. A
positive scale never changes a single-window decision; it matters only when windows are
combined.

\begin{proposition}[additive evidence]\label{prop:add}
Assume (a) given $y$ and all candidates, the $x_i$ are independent and $x_i$ depends on the
candidates only through $a_{y,i}$; (b) $P(y{=}1)=\tfrac12$ independently of the candidates.
Then $\log\frac{P(y=1\mid x_{1:n},a_{1:n})}{P(y=2\mid x_{1:n},a_{1:n})}=S_n$, i.e.\ $S_n$ is
calibrated, and the decision with the fewest errors given $n$ windows is $\mathrm{sign}\,S_n$. Majority
voting over $\mathrm{sign}\,\ell_i$ is optimal only when all $|\ell_i|$ are equal.
\end{proposition}

\begin{proposition}[accumulation law as an information rate]\label{prop:gauss}
Let the $\ell_i$ of Prop.~\ref{prop:add} be independent and identically distributed
(i.i.d.)\ given $y$, with
$\ell_i\mid y{=}1\sim\mathcal N(\mu,\nu^2)$ for some $\mu>0$, $\nu^2>0$. Then (i)
$\nu^2=2\mu$ and $\ell_i\mid y{=}2\sim\mathcal N(-\mu,2\mu)$; (ii) $\mu$ is the
Kullback--Leibler divergence $D_{\mathrm{KL}}\big(p(x_i\mid y{=}1)\,\|\,p(x_i\mid y{=}2)\big)$,
the information one window carries about $y$; (iii) with the information rate $\mathcal I=\mu/w$ (nats
per second) and a fixed decision length of $T=nw$ seconds ($n$ windows),
\[P(\mathrm{correct}\mid T)=\Phi\big(\sqrt{\mathcal I T/2}\big)=\Phi(\kappa\sqrt T),\qquad
\kappa^2=\mathcal I/2 .\]
\end{proposition}
Prop.~\ref{prop:gauss} gives Prop.~\ref{prop:law} a decision-theoretic reading that does not
rely on the decoder being a correlation: $2\kappa^2$ is the information the brain response
carries about the attended talker, per second. It also gives a check of calibration and of
the Gaussian model: by (i), exact Gaussian log-likelihood ratios have
$\mathrm{Var}_1(\ell_i)/(2\,\mathbb E_1\ell_i)=1$, where $\mathrm{Var}_1$ is the variance
under $P_1$; we evaluate it on test windows with the log-odds oriented so that the attended
candidate is candidate 1.

\begin{proposition}[voting wastes evidence]\label{prop:vote}
Under Prop.~\ref{prop:gauss} let $q=\Phi(\sqrt{\mu/2})$ be the one-window accuracy. Summing
$n$ windows gives $\Phi(\sqrt n\,\Phi^{-1}(q))$. Majority voting gives, for large $n$,
$\Phi\big(\sqrt n\,(2q-1)/(2\sqrt{q(1-q)})\big)$. As $\mu\to0$ voting needs
$\pi/2\approx1.57$ times as many windows as summing for the same accuracy.
\end{proposition}

\begin{proposition}[sequential decisions]\label{prop:sprt}
Under Prop.~\ref{prop:gauss}, fix a threshold $A>0$ and let
$N=\min\{n:|S_n|\ge A\}$ be the number of windows the rule reads before
$|S_n|$ first reaches $A$; the rule decides $\mathrm{sign}\,S_N$ after $T_A=Nw$ seconds
(Wald's sequential probability ratio test with symmetric thresholds). Neglecting the overshoot
of $S_N$ beyond $\pm A$, (i) its error rate is $\alpha=1/(1+e^{A})$; (ii) its mean decision
time is
\[\mathbb E[T_A]=\frac{(1-2\alpha)\,\log\frac{1-\alpha}{\alpha}}{2\kappa^2}\ \text{s};\]
(iii) no rule with both error rates $\le\alpha$ has a smaller mean decision time under
either hypothesis \citep{wald1948optimum}. A fixed-length decision reaches the same error rate
$\alpha$ after $T_{\rm fix}=(\Phi^{-1}(1-\alpha)/\kappa)^2$ seconds (Prop.~\ref{prop:gauss}),
so $\mathbb E[T_A]/T_{\rm fix}=0.59,\,0.54,\,0.49,\,0.42$ at $\alpha=0.2,\,0.1,\,0.05,\,0.01$,
decreasing to $1/4$ as $\alpha\to0$.
\end{proposition}
The sequential rule reaches the fixed-window accuracy in about half the time, with the
same $\kappa$. This is the classical result of \citet{wald1945sequential}; its continuous-time
limit is the drift--diffusion model of perceptual decisions \citep{bogacz2006physics}. It
rests on two assumptions that EEG violates to some degree, independence of consecutive
windows and a constant information rate; we measure both below. If the information rate varies
between stretches of a trial (attention lapses, noise bursts), the sequential rule spends its
time on the weak stretches, which a fixed window cannot do.

\begin{proposition}[stopping does not bias the posterior]\label{prop:stop}
Under the assumptions of Prop.~\ref{prop:add}, let $N$ be any stopping time of the window
sequence, i.e.\ whether to stop after window $n$ depends only on $x_{1:n}$ and $a_{1:n}$, and
decide $\mathrm{sign}\,S_N$. Then $P(\text{correct}\mid x_{1:N})=\Lambda(|S_N|)$. In particular
every decision taken when $|S_N|\ge A$ is correct with probability at least $\Lambda(A)$,
\emph{whatever its length}, and the accuracy of the sequential rule is set by the threshold
alone; the length only adapts to how fast the evidence arrives.
\end{proposition}

\begin{proposition}[overlapping windows]\label{prop:overlap}
Let the evidence be additive in time: the log-likelihood ratio of the data in
$[t_0,t_1)$ is $\ell(t_0,t_1)=\int_{t_0}^{t_1}\dot\ell(t)\,dt$ for a log-likelihood-ratio density
$\dot\ell(t)$ (nats per second). Let windows of length $w$ start at $t_j=jh$, $j=0,1,\dots$,
every $h=w/J$ seconds for an integer $J\ge1$ ($J=1$: no overlap). Over a stretch of length $T\gg w$, the sum of the
overlapping windows' log-odds equals $J$ times the log-odds of the stretch, up to edge terms
of length $<w$. Hence (i) the sign, and so a fixed-length decision, is unchanged: overlap adds
no information; (ii) the summed log-odds overstate the evidence by the factor $J$ and must be
multiplied by $h/w$ to stay calibrated; (iii) with that correction the sequential rule is the
one of Prop.~\ref{prop:sprt} checked every $h$ instead of every $w$ seconds, which only
reduces the overshoot.
\end{proposition}

\paragraph{Proofs.}
\emph{Prop.~\ref{prop:add}.} By Bayes' rule and (b), the posterior ratio equals the likelihood
ratio $p(x_{1:n}\mid y{=}1,a_{1:n})/p(x_{1:n}\mid y{=}2,a_{1:n})$. By (a) each likelihood
factorises over windows into $\prod_i p(x_i\mid y,a_{\cdot,i})$, so the logarithm of the ratio
is $\sum_i\ell_i=S_n$; hence $P(y{=}1\mid x_{1:n},a_{1:n})=\Lambda(S_n)$. The minimum-error
decision picks the larger posterior, i.e.\ $\mathrm{sign}\,S_n$. The majority vote depends on
the signs of the $\ell_i$ only; it agrees with $\mathrm{sign}\,S_n$ for every configuration of
signs only if all $|\ell_i|$ are equal. \qed

\emph{Prop.~\ref{prop:gauss}.} Let $p_1,p_2$ be the densities of $\ell_i$ under $P_1,P_2$.
Since $\ell_i$ is the log-likelihood ratio of $x_i$, the change of measure from $P_1$ to $P_2$
multiplies densities by $e^{-\ell_i}$, so $p_2(\ell)=e^{-\ell}p_1(\ell)$. With
$p_1=\mathcal N(\mu,\nu^2)$, $p_2$ integrates to one only if
$\mathbb E_1 e^{-\ell_i}=e^{-\mu+\nu^2/2}=1$, i.e.\ $\nu^2=2\mu$; completing the square then
gives $p_2=\mathcal N(\mu-\nu^2,\nu^2)=\mathcal N(-\mu,2\mu)$, which is (i). (ii) holds by
definition, $D_{\mathrm{KL}}=\mathbb E_1\log\frac{p(x_i\mid y=1)}{p(x_i\mid y=2)}
=\mathbb E_1\ell_i=\mu$. For (iii), independence gives $S_n\mid y{=}1\sim\mathcal N(n\mu,2n\mu)$,
so $P_1(S_n>0)=\Phi(n\mu/\sqrt{2n\mu})=\Phi(\sqrt{n\mu/2})$; by (i) the same holds for
$P_2(S_n<0)$. Substituting $n\mu=\mathcal I T$ and $\kappa^2=\mathcal I/2$ gives the claim. \qed

\emph{Prop.~\ref{prop:vote}.} Summing: by Prop.~\ref{prop:gauss}(iii) the accuracy is
$\Phi(\sqrt{n\mu/2})$, and $\sqrt{\mu/2}=\Phi^{-1}(q)$ by the definition of $q$. Voting: the
number of correctly signed windows is binomial with parameters $n$ and $q$; by the central limit
theorem the majority is right with probability
$\Phi\big((nq-n/2)/\sqrt{nq(1-q)}\big)=\Phi\big(\sqrt n\,(2q-1)/(2\sqrt{q(1-q)})\big)$. For small
$\mu$, $q-\tfrac12\approx\varphi(0)\sqrt{\mu/2}$ with $\varphi(0)=1/\sqrt{2\pi}$, and
$q(1-q)\approx\tfrac14$, so the argument of $\Phi$ is $\sqrt n\,\sqrt{2/\pi}\,\sqrt{\mu/2}$ for
voting against $\sqrt n\,\sqrt{\mu/2}$ for summing; equal accuracy needs
$n_{\rm vote}/n_{\rm sum}=\pi/2$ windows (the asymptotic relative efficiency of the sign test).
\qed

\emph{Prop.~\ref{prop:sprt}.} $N$ is finite with probability one because $S_n$ drifts to
$\pm\infty$. (i) Changing measure as in the proof of Prop.~\ref{prop:gauss}, the probability of
a wrong decision under $y{=}2$ is $P_2(S_N\ge A)=\mathbb E_1[e^{-S_N};\,S_N\ge A]$. Without
overshoot $S_N\in\{-A,A\}$, so this equals $e^{-A}P_1(S_N\ge A)=e^{-A}(1-\alpha)$, where
$\alpha$ is the error rate; by the symmetry of (i) in Prop.~\ref{prop:gauss} the error rate is
the same under both hypotheses, so $\alpha=e^{-A}(1-\alpha)$, i.e.\ $\alpha=1/(1+e^A)$ and
$A=\log\frac{1-\alpha}{\alpha}$. (ii) Wald's identity gives $\mathbb E_1S_N=\mu\,\mathbb E_1N$,
and without overshoot $\mathbb E_1S_N=A(1-\alpha)-A\alpha$, so
$\mathbb E_1N=A(1-2\alpha)/\mu$ windows (the same under $P_2$). Then
$\mathbb E[T_A]=w\,\mathbb E N=A(1-2\alpha)\,w/\mu$, and $\mu=\mathcal I w=2\kappa^2w$ by
Prop.~\ref{prop:gauss} gives the stated time. (iii) is the Wald--Wolfowitz theorem for i.i.d.\
increments. $T_{\rm fix}$ solves $\Phi(\kappa\sqrt{T})=1-\alpha$, and the ratio follows by
division. As $\alpha\to0$, $\log\frac{1-\alpha}{\alpha}\sim\log\frac1\alpha$ and
$\Phi^{-1}(1-\alpha)^2\sim2\log\frac1\alpha$, so the ratio tends to $1/4$. With overshoot,
$|S_N|>A$: the realised error rate is below $1/(1+e^A)$ and the time above (ii), both by at
most one window's increment. \qed

\emph{Prop.~\ref{prop:stop}.} Because $N$ is a stopping time, the event $\{N=n\}$ is
determined by $(x_{1:n},a_{1:n})$, so conditioning on it adds nothing to them:
$P(y{=}1\mid x_{1:N},a_{1:N},N{=}n)=P(y{=}1\mid x_{1:n},a_{1:n})=\Lambda(S_n)$ by
Prop.~\ref{prop:add}. The decision $\mathrm{sign}\,S_N$ is therefore correct with probability
$\Lambda(|S_N|)$, which is at least $\Lambda(A)$ whenever $|S_N|\ge A$. This is the likelihood
principle: the posterior does not depend on the stopping rule. \qed

\emph{Prop.~\ref{prop:overlap}.} Consider the stretch $[0,T)$ and let $\chi(t)$ be the number of
windows $[t_j,t_j+w)$ inside it that contain the time $t$. Every $t$ at distance at least $w$
from both ends of the stretch lies in exactly $J=w/h$ windows, so $\chi(t)=J$ there, and
$\chi(t)<J$ only within $w$ of the ends. Summing the window log-odds,
$\sum_j\ell(t_j,t_j+w)=\int_0^T\dot\ell(t)\,\chi(t)\,dt=J\,\ell(0,T)+E$, where the edge term $E$
involves only data within $w$ of the ends. Hence the sum has the sign of the stretch's
log-likelihood ratio $\ell(0,T)$ up to $E$ (i), and it is $J$ times the calibrated value (ii).
Multiplied by $1/J=h/w$, the running sum after each new window equals the log-likelihood ratio
of the data read so far, up to the edge term; this is the statistic $S$ of
Prop.~\ref{prop:sprt} evaluated every $h$ instead of every $w$ seconds (iii). \qed

\paragraph{Related rules.} \citet{geirnaert2020interpretable} turn fixed-window accuracies
into the minimal expected switch duration of a gain-control Markov chain; the sequential rule
can feed that chain the same way. Bayesian state-space filtering \citep{miran2018realtime}
additionally models attention switches, which our datasets do not contain. Calibration by one
scale is temperature scaling \citep{guo2017calibration}; averaging the calibrated log-odds of
independently trained models is a deep ensemble \citep{lakshminarayanan2017simple}.

\subsection{Calibration and combining windows}\label{app:conf:calib}
Table~\ref{tab:conf_calib}. Raw margins are over-confident (expected calibration error, ECE, \confEceRawMin--\confEceRawMax{}
on the two-talker sets); one scale fitted on validation brings every dataset to
ECE $\le$\,\confEceCalMax. Averaging the calibrated log-odds of the three seeds adds
\confEnsMin--\confEnsMax{} points at 5\,s. Summing a trial's log-odds instead of majority voting changes accuracy by
\confSumMin{} to \confSumMax{} points (Prop.~\ref{prop:vote}). The calibrated 1-s log-odds have
$\mathrm{Var}/(2\,\mathrm{mean})=$\,\confVarMin--\confVarMax{} (exactly 1 for Gaussian
log-likelihood ratios) and lag-1 autocorrelation \confLagMin--\confLagMax: consecutive windows
are close to, but not exactly, independent.
\begin{table*}[h]\centering\small
\caption{Calibration and combination of windows (test folds; mean of 3 seeds unless marked).
ECE: expected calibration error, 10 bins; NLL: log-loss in bits (chance 1). ens.: average of
the 3 seeds' calibrated log-odds. Trial: one decision per trial by majority vote of the 5-s
windows or by the sign of their summed log-odds. Last two columns: diagnostics of
Prop.~\ref{prop:gauss} on the 1-s log-odds.}\label{tab:conf_calib}
\begin{tabular}{lcccccccccc}\toprule
& \multicolumn{2}{c}{ECE (5\,s)} & \multicolumn{2}{c}{NLL, bits (5\,s)} & \multicolumn{2}{c}{acc.\ 5\,s} & \multicolumn{2}{c}{trial} & \multicolumn{2}{c}{1-s log-odds} \\
Dataset & raw & cal. & raw & cal. & single & ens. & vote & sum & $\frac{\mathrm{var}}{2\,\mathrm{mean}}$ & lag-1 $r$ \\\midrule
KU Leuven & 0.113 & 0.022 & 0.91 & 0.78 & 72.7 & 74.3 & 97.6 & 99.3 & 1.11 & 0.05 \\
DTU & 0.111 & 0.012 & 0.86 & 0.70 & 76.2 & 78.3 & 94.6 & 97.0 & 1.04 & 0.05 \\
Fuglsang'20 & 0.078 & 0.016 & 0.72 & 0.64 & 79.6 & 81.2 & 96.8 & 98.5 & 0.99 & 0.05 \\
ESAA & 0.098 & 0.016 & 0.82 & 0.68 & 77.9 & 80.0 & 95.6 & 97.1 & 1.03 & 0.05 \\
KUL-255ch & 0.143 & 0.028 & 1.04 & 0.84 & 69.4 & 71.5 & 99.6 & 99.4 & 1.13 & 0.05 \\
Broderick & 0.171 & 0.019 & 1.13 & 0.90 & 63.6 & 65.7 & 78.9 & 83.1 & 1.03 & 0.05 \\
SparrKULee & 0.014 & 0.002 & 0.28 & 0.27 & 92.3 & 93.0 & -- & -- & 0.98 & 0.05 \\
MEG-MASC & 0.014 & 0.001 & 0.05 & 0.04 & 99.0 & 99.3 & -- & -- & 1.05 & 0.01 \\
\bottomrule\end{tabular}

\end{table*}

\subsection{Sequential decisions}\label{app:conf:seq}
\begin{figure*}[h]\centering
\includegraphics[width=0.9\textwidth]{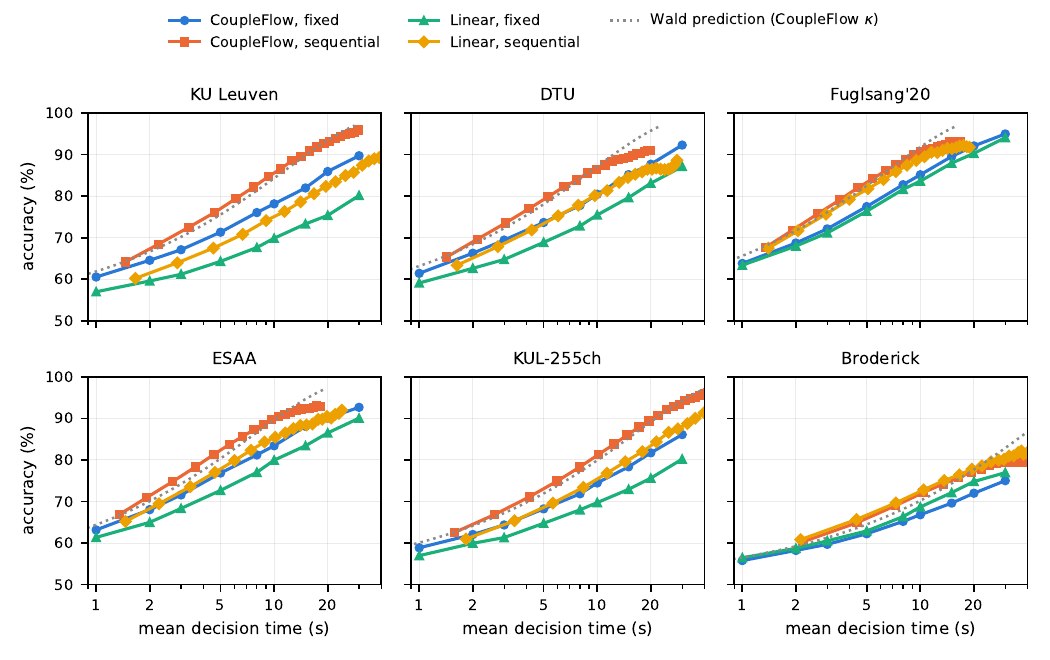}
\caption{Accuracy against mean decision time, two-talker datasets (test folds). Fixed:
sign of the summed calibrated log-odds of $T$ consecutive 1-s windows. Sequential: the rule
of Prop.~\ref{prop:sprt} at thresholds $A=0.25,\dots,5$. Dotted: Wald's prediction from
CoupleFlow's fitted $\kappa$ (Prop.~\ref{prop:sprt}). CoupleFlow: mean of 3 seeds.}
\label{fig:conf_seq}
\end{figure*}

Fig.~\ref{fig:conf_seq} and Table~\ref{tab:conf_seq} compare the sequential rule of
Prop.~\ref{prop:sprt} on calibrated 1-s log-odds with fixed-length decisions that sum the
same 1-s log-odds over $T$\,s, at equal mean decision time. Both rules see exactly the same
scores; only the stopping differs. Thresholds $A$ are fixed values, not tuned. For
CoupleFlow the sequential rule gains \confGainOneMin--\confGainOneMax{} points at $A{=}1$ and
\confGainTwoMin--\confGainTwoMax{} at $A{=}2$; for the linear backward model it gains
\confLinGainOneMin--\confLinGainOneMax{} and \confLinGainTwoMin--\confLinGainTwoMax. The gain
is therefore a property of the decision rule, available to any calibrated decoder; the
ranking of decoders is set by $\kappa$, which is larger for CoupleFlow on every dataset except
Broderick. The observed mean times are \confTheoryRatioMin--\confTheoryRatioMax{} times Wald's
prediction from the fitted $\kappa$ (dotted lines). Overshoot on the 1-s grid lengthens the
observed times; a varying information rate within trials shortens them, since the rule stops
early on strong stretches, which the i.i.d.\ prediction ignores.
``Time'' counts speech windows; every decision also needs the EEG of the neural lag
(up to 1.1\,s) after its last window, for fixed and sequential rules alike. Decision times are
spread: Table~\ref{tab:conf_seq} gives their 90th percentile next to the mean.
\begin{table*}[h]\centering\small
\caption{Sequential against fixed-length decisions on calibrated 1-s log-odds. $\kappa$: fit
of $\Phi(\kappa\sqrt T)$ to the fixed-length curve. Cells: sequential accuracy (\%) / mean
decision time (s) (90th percentile) / fixed-length accuracy at the same mean time (\%). Gain: sequential minus
fixed at $A{=}2$ (points).}\label{tab:conf_seq}
\resizebox{\textwidth}{!}{\begin{tabular}{lcccccccc}\toprule
& \multicolumn{4}{c}{CoupleFlow (mean of 3 seeds)} & \multicolumn{4}{c}{Linear backward model} \\
\cmidrule(lr){2-5}\cmidrule(lr){6-9}
Dataset & $\kappa$ & $A{=}1$ & $A{=}2$ & gain & $\kappa$ & $A{=}1$ & $A{=}2$ & gain \\\midrule
KU Leuven & 0.240 & 76.0 / 4.6 (9) / 70.5 & 86.6 / 10.9 (22) / 78.8 & +7.7 & 0.158 & 70.8 / 6.7 (13) / 66.1 & 80.6 / 16.7 (34) / 74.0 & +6.6 \\
DTU & 0.267 & 77.0 / 4.1 (8) / 71.8 & 85.5 / 8.8 (17) / 78.8 & +6.7 & 0.213 & 75.2 / 6.0 (12) / 70.2 & 83.3 / 13.3 (27) / 78.2 & +5.1 \\
Fuglsang'20 & 0.318 & 79.1 / 3.5 (7) / 73.5 & 87.5 / 7.3 (14) / 81.5 & +6.0 & 0.299 & 79.2 / 4.0 (8) / 73.7 & 87.4 / 8.4 (16) / 82.0 & +5.4 \\
ESAA & 0.292 & 78.3 / 3.6 (7) / 73.2 & 87.3 / 7.7 (15) / 80.8 & +6.5 & 0.250 & 77.0 / 4.6 (9) / 71.9 & 85.4 / 10.1 (20) / 80.0 & +5.4 \\
KUL-255ch & 0.203 & 74.9 / 5.9 (12) / 69.4 & 86.0 / 14.7 (30) / 78.1 & +7.9 & 0.159 & 73.4 / 8.4 (17) / 68.3 & 84.3 / 21.5 (43) / 76.3 & +8.0 \\
Broderick & 0.131 & 72.1 / 10.4 (22) / 66.8 & 77.7 / 22.0 (50) / 72.4 & +5.3 & 0.146 & 72.8 / 10.4 (21) / 69.0 & 78.6 / 22.2 (46) / 75.3 & +3.3 \\
\bottomrule\end{tabular}
}
\end{table*}

\subsection{Anatomy of the sequential decisions}\label{app:conf:anatomy}
\begin{figure*}[h]\centering
\includegraphics[width=\textwidth]{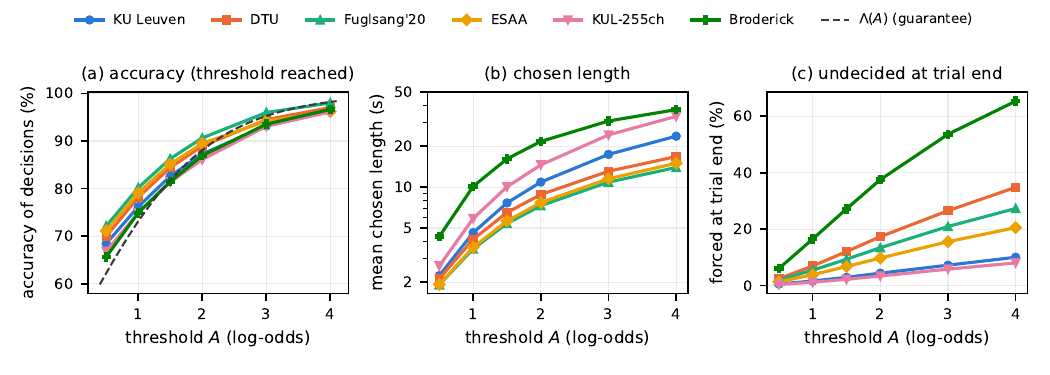}
\caption{Effect of the threshold $A$. (a) Accuracy of the decisions that reached the
threshold, against the guarantee $\Lambda(A)$ (dashed). (b) Mean chosen length. (c) Share of
decisions cut by the end of a trial.}\label{fig:conf_threshold}
\end{figure*}

\begin{figure*}[h]\centering
{\small (a) distribution of the chosen length}\\[2pt]
\includegraphics[width=0.85\textwidth]{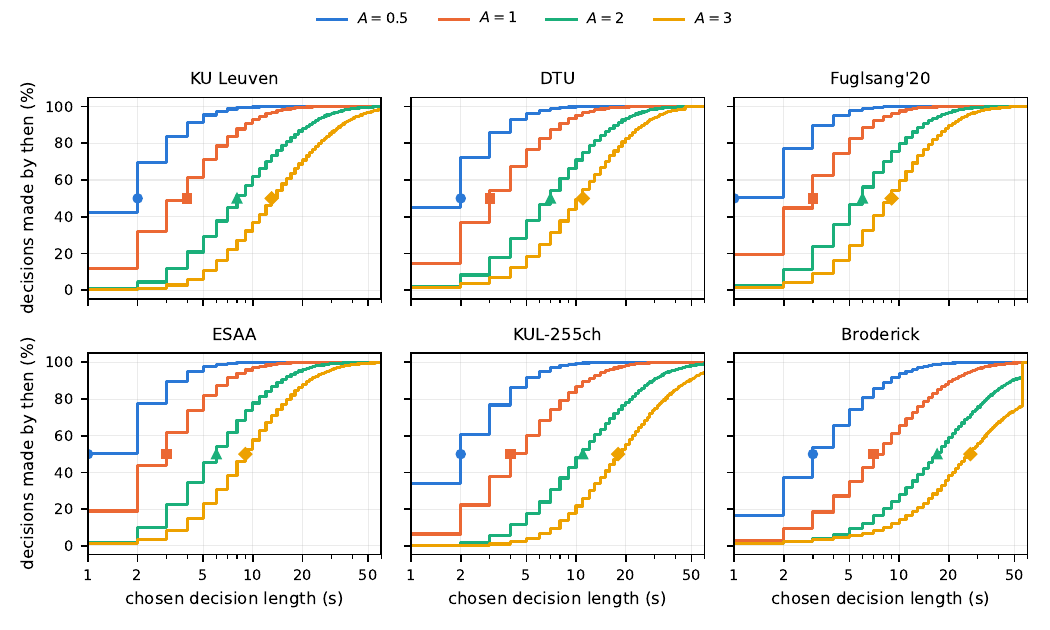}\\[6pt]
{\small (b) accuracy by chosen length}\\[2pt]
\includegraphics[width=0.85\textwidth]{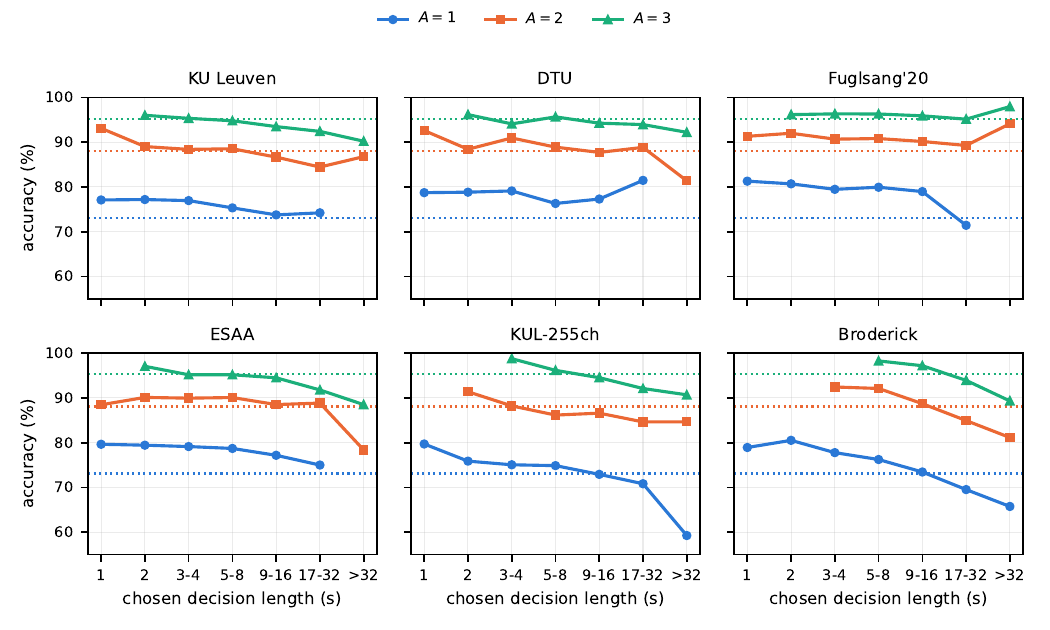}
\caption{The chosen decision length. (a) Share of decisions taken within a given number of
seconds, for four thresholds; markers: medians. (b) Accuracy of the decisions that reached the
threshold, grouped by the length they used, for $A=1,2,3$. Dotted: $\Lambda(A)$, the accuracy
Prop.~\ref{prop:stop} guarantees for every length. Bins with fewer than 20 decisions are
omitted.}\label{fig:conf_len}
\end{figure*}

Table~\ref{tab:conf_seq_detail} and Figs.~\ref{fig:conf_threshold} and~\ref{fig:conf_len}
open up the sequential rule on CoupleFlow's calibrated 1-s log-odds (decisions of the three
seeds pooled, test folds). Four things stand out.

\emph{The threshold sets the accuracy.} Decisions that reach the threshold are about as
accurate as the guarantee $\Lambda(A)$ of Prop.~\ref{prop:stop}
(Fig.~\ref{fig:conf_threshold}a; deviation \confDetAboveMin{} to \confDetAboveMax{} points over
all datasets and thresholds). At small $A$ they are above it, because a 1-s step overshoots
the threshold; at large $A$ slightly below it, the signature of the mild positive dependence
between consecutive windows, which makes long sums a little over-confident. At $A{=}1$
decisions are \confDetOneAccMin--\confDetOneAccMax\% correct ($\Lambda(1)=$\,\confDetOneNominal\%),
at $A{=}2$ \confDetTwoAccMin--\confDetTwoAccMax\% ($\Lambda(2)=$\,\confDetTwoNominal\%), on every
dataset: a target accuracy fixes $A$, to within a few points, without looking at the data.

\emph{The chosen length adapts widely.} The same threshold yields short and long decisions
(Fig.~\ref{fig:conf_len}a). At $A{=}1$ the median decision uses
\confDetOneMedMin--\confDetOneMedMax\,s and the 90th percentile \confDetOnePNineMin--\confDetOnePNineMax\,s;
\confDetOneOneSec--\confDetOneOneSecMax\% of decisions are taken after a single second. At
$A{=}2$ the median is \confDetTwoMedMin--\confDetTwoMedMax\,s and the 90th percentile
\confDetTwoPNineMin--\confDetTwoPNineMax\,s. The mean length grows with $A$ roughly as
$A(1-2\alpha)/(2\kappa^2)$ (Prop.~\ref{prop:sprt}), so datasets with small $\kappa$ (Broderick,
KUL-255ch) wait longest (Fig.~\ref{fig:conf_threshold}b).

\emph{Accuracy hardly depends on the chosen length.} As Prop.~\ref{prop:stop} predicts, fast
and slow decisions are about equally accurate (Fig.~\ref{fig:conf_len}b): decisions of
at most 8\,s are at most \confDetShortDevMax{} points below $\Lambda(2)$. The exception is the
longest decisions ($\ge$17\,s), up to \confDetLongDevMax{} points below $\Lambda(2)$. They come
from stretches in which the recording tracks the speech poorly, and one global calibration
scale over-states the evidence of such windows: calibration holds on average, not
conditionally on the local signal quality. A calibration that depends on the recent evidence
rate would close this gap.

\emph{Finite trials force some decisions.} When a trial ends before $|S|$ reaches $A$, the
leftover is decided anyway (Fig.~\ref{fig:conf_threshold}c). These forced decisions are less
accurate (\confDetForcedAccMin--\confDetForcedAccMax\% at $A\in\{1,2\}$) and frequent on the
short trials of \confDetForcedWorst{} (up to \confDetTwoForcedMax\% at $A{=}2$); the ``all''
column includes them. In continuous listening there is no trial end and they do not arise.

\begin{table*}[h]\centering\small
\caption{Anatomy of the sequential rule (CoupleFlow, calibrated 1-s log-odds, 3 seeds
pooled, test folds). $\Lambda(A)$: accuracy guaranteed by Prop.~\ref{prop:stop} for decisions
that reach the threshold. Accuracy of all decisions and of those that reached the threshold
(decided). Chosen length: mean, median and 10th--90th percentile of the number of seconds a
decision used. Forced: share and accuracy of decisions cut by the end of a trial. Fixed:
fixed-length accuracy at the same mean length; gain in points.}\label{tab:conf_seq_detail}
\resizebox{\textwidth}{!}{\begin{tabular}{lccccccccccc}\toprule
& & \multicolumn{3}{c}{accuracy (\%)} & \multicolumn{3}{c}{chosen length (s)} & \multicolumn{2}{c}{forced at trial end} & \multicolumn{2}{c}{vs.\ fixed length} \\
\cmidrule(lr){3-5}\cmidrule(lr){6-8}\cmidrule(lr){9-10}\cmidrule(lr){11-12}
Dataset & $A$ & $\Lambda(A)$ & all & decided & mean & median & 10--90\% & share (\%) & acc.\ (\%) & fixed (\%) & gain \\\midrule
KU Leuven & 0.5 & 62.2 & 68.4 & 68.5 & 2.2 & 2 & 1--4 & 0.5 & 60.1 & 65.1 & +3.3 \\
 & 1 & 73.1 & 76.0 & 76.2 & 4.6 & 4 & 1--9 & 1.6 & 65.3 & 70.5 & +5.5 \\
 & 1.5 & 81.8 & 82.1 & 82.6 & 7.6 & 6 & 2--15 & 2.9 & 66.6 & 75.5 & +6.7 \\
 & 2 & 88.1 & 86.5 & 87.3 & 10.9 & 8 & 3--22 & 4.4 & 70.9 & 78.8 & +7.7 \\
 & 3 & 95.3 & 91.8 & 93.1 & 17.4 & 13 & 5--35 & 7.2 & 74.1 & 83.9 & +7.9 \\
 & 4 & 98.2 & 94.3 & 96.4 & 23.7 & 19 & 8--47 & 10.0 & 75.9 & 87.3 & +7.0 \\
\midrule
DTU & 0.5 & 62.2 & 69.6 & 70.0 & 2.1 & 2 & 1--4 & 2.4 & 55.5 & 66.6 & +3.0 \\
 & 1 & 73.1 & 77.0 & 78.2 & 4.1 & 3 & 1--8 & 6.9 & 60.5 & 71.8 & +5.1 \\
 & 1.5 & 81.8 & 82.2 & 84.4 & 6.5 & 5 & 2--13 & 12.1 & 66.4 & 75.7 & +6.6 \\
 & 2 & 88.1 & 85.5 & 88.9 & 8.8 & 7 & 3--18 & 17.3 & 69.3 & 78.8 & +6.7 \\
 & 3 & 95.3 & 88.8 & 94.5 & 13.1 & 11 & 4--26 & 26.5 & 72.9 & 83.3 & +5.4 \\
 & 4 & 98.2 & 90.2 & 97.0 & 16.8 & 14 & 6--33 & 34.8 & 77.4 & 86.0 & +4.1 \\
\midrule
Fuglsang'20 & 0.5 & 62.2 & 71.8 & 72.1 & 1.9 & 1 & 1--4 & 2.0 & 54.6 & 68.3 & +3.5 \\
 & 1 & 73.1 & 79.1 & 80.2 & 3.5 & 3 & 1--7 & 5.4 & 59.7 & 73.5 & +5.6 \\
 & 1.5 & 81.8 & 84.2 & 86.3 & 5.4 & 4 & 2--11 & 9.3 & 63.8 & 78.2 & +6.0 \\
 & 2 & 88.1 & 87.5 & 90.6 & 7.3 & 6 & 2--14 & 13.4 & 67.4 & 81.5 & +6.0 \\
 & 3 & 95.3 & 91.3 & 96.0 & 10.9 & 9 & 4--21 & 20.9 & 73.4 & 85.9 & +5.4 \\
 & 4 & 98.2 & 92.4 & 98.0 & 13.9 & 12 & 5--26 & 27.3 & 77.7 & 88.6 & +3.9 \\
\midrule
ESAA & 0.5 & 62.2 & 70.9 & 71.1 & 1.9 & 1 & 1--4 & 1.3 & 55.8 & 67.6 & +3.3 \\
 & 1 & 73.1 & 78.3 & 79.1 & 3.6 & 3 & 1--7 & 3.8 & 58.7 & 73.2 & +5.1 \\
 & 1.5 & 81.8 & 83.8 & 85.1 & 5.6 & 4 & 2--11 & 6.8 & 65.0 & 77.8 & +6.0 \\
 & 2 & 88.1 & 87.3 & 89.5 & 7.7 & 6 & 2--15 & 9.8 & 66.6 & 80.8 & +6.5 \\
 & 3 & 95.3 & 91.0 & 94.2 & 11.5 & 9 & 4--22 & 15.5 & 73.6 & 84.8 & +6.2 \\
 & 4 & 98.2 & 92.3 & 96.2 & 15.0 & 12 & 5--28 & 20.5 & 77.4 & 88.1 & +4.3 \\
\midrule
KUL-255ch & 0.5 & 62.2 & 66.8 & 66.8 & 2.6 & 2 & 1--5 & 0.4 & 61.2 & 63.5 & +3.3 \\
 & 1 & 73.1 & 74.8 & 74.9 & 5.9 & 4 & 2--12 & 1.2 & 66.7 & 69.3 & +5.5 \\
 & 1.5 & 81.8 & 81.2 & 81.4 & 10.0 & 8 & 3--21 & 2.2 & 71.8 & 74.5 & +6.7 \\
 & 2 & 88.1 & 85.9 & 86.1 & 14.6 & 11 & 4--29 & 3.4 & 79.6 & 78.0 & +7.9 \\
 & 3 & 95.3 & 91.9 & 93.1 & 24.2 & 18 & 8--49 & 5.8 & 73.5 & 83.5 & +8.4 \\
 & 4 & 98.2 & 94.7 & 96.1 & 33.1 & 26 & 11--64 & 8.0 & 78.7 & 86.1 & +8.6 \\
\midrule
Broderick & 0.5 & 62.2 & 65.0 & 65.7 & 4.3 & 3 & 1--9 & 6.0 & 54.1 & 61.4 & +3.6 \\
 & 1 & 73.1 & 72.2 & 74.9 & 10.1 & 7 & 3--21 & 16.5 & 58.4 & 66.9 & +5.3 \\
 & 1.5 & 81.8 & 75.9 & 81.4 & 16.1 & 12 & 4--34 & 27.2 & 61.1 & 70.1 & +5.8 \\
 & 2 & 88.1 & 77.9 & 86.9 & 21.6 & 17 & 6--49 & 37.6 & 62.8 & 72.5 & +5.4 \\
 & 3 & 95.3 & 79.7 & 93.5 & 30.6 & 27 & 8--56 & 53.6 & 67.7 & 75.0 & +4.7 \\
 & 4 & 98.2 & 79.9 & 96.6 & 37.1 & 38 & 11--56 & 65.4 & 71.1 & 75.0 & +4.9 \\
\bottomrule\end{tabular}
}
\end{table*}

\subsection{Overlapping windows}\label{app:conf:overlap}
\begin{figure*}[h]\centering
\includegraphics[width=0.95\textwidth]{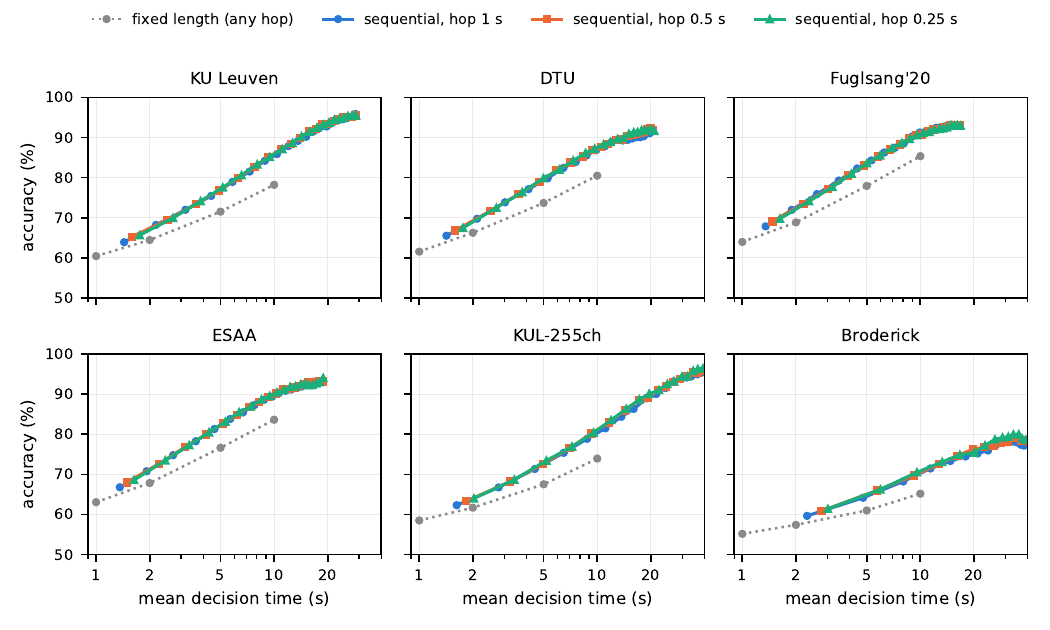}
\caption{Overlapping 1-s windows (CoupleFlow, seed 0, test folds): the sequential rule on
$h/w$-scaled log-odds with hop $h=1$, 0.5 and 0.25\,s, restarting after each decision at the
first window disjoint from it, and fixed-length decisions (nearly identical for every hop; the hop-1\,s curve is shown).}\label{fig:conf_overlap}
\end{figure*}

Table~\ref{tab:conf_overlap} and Fig.~\ref{fig:conf_overlap} repeat the analysis with 1-s
windows every 0.5\,s and 0.25\,s (CoupleFlow, seed 0). The data follow
Prop.~\ref{prop:overlap} closely. The log-odds of windows $jh$ apart correlate as
$1-jh/w$ (KU Leuven, hop 0.25\,s: \confOvCorOne, \confOvCorTwo, \confOvCorThree{} at
$j=1,2,3$; prediction 0.75, 0.5, 0.25), fixed-length accuracy changes by at most
\confOvFixedMax{} points, and the summed log-odds are over-confident (5-s ECE up to
\confOvEceRawMax) until scaled by $h/w$ (ECE $\le$\,\confOvEceCorrMax). The corrected
sequential rule gains \confOvSeqMin--\confOvSeqMax{} points over the non-overlapping one at
the same mean time, the small effect of checking the threshold more often. One detail
matters: after a decision the rule must restart at the first window that shares no data with
that decision. Restarting at the adjacent window re-uses the window that crossed the
threshold and inflates accuracy at $A{=}2$ by \confOvAdjMin--\confOvAdjMax{} points without any
new evidence.
\begin{table*}[h]\centering\small
\caption{Overlapping 1-s windows (CoupleFlow, seed 0). Fixed 5\,s: accuracy of the summed
log-odds of the windows inside each 5-s stretch. ECE of those sums without (raw) and with the
$h/w$ correction. Seq.: corrected sequential rule at $A{=}2$, accuracy (\%) / mean time
(s).}\label{tab:conf_overlap}
\resizebox{\textwidth}{!}{\begin{tabular}{lccccccccc}\toprule
& \multicolumn{3}{c}{hop 1\,s} & \multicolumn{3}{c}{hop 0.5\,s} & \multicolumn{3}{c}{hop 0.25\,s} \\
\cmidrule(lr){2-4}\cmidrule(lr){5-7}\cmidrule(lr){8-10}
Dataset & fixed 5\,s & ECE raw/corr. & seq.\ $A{=}2$ & fixed 5\,s & ECE raw/corr. & seq.\ $A{=}2$ & fixed 5\,s & ECE raw/corr. & seq.\ $A{=}2$ \\\midrule
KU Leuven & 71.5 & 0.04/0.04 & 85.9 / 10.4 & 71.8 & 0.12/0.02 & 87.1 / 11.0 (+0.5) & 71.6 & 0.19/0.02 & 87.1 / 11.1 (+0.4) \\
DTU & 73.7 & 0.02/0.02 & 85.5 / 8.7 & 74.2 & 0.10/0.01 & 86.9 / 9.4 (+0.6) & 74.3 & 0.17/0.01 & 87.2 / 9.7 (+0.6) \\
Fuglsang'20 & 77.9 & 0.01/0.01 & 87.5 / 7.2 & 77.3 & 0.10/0.01 & 88.2 / 7.7 (+0.2) & 77.2 & 0.15/0.01 & 88.6 / 7.9 (+0.3) \\
ESAA & 76.6 & 0.02/0.02 & 87.3 / 7.8 & 76.6 & 0.10/0.01 & 88.0 / 8.3 (+0.1) & 76.6 & 0.16/0.01 & 88.7 / 8.5 (+0.4) \\
KUL-255ch & 67.5 & 0.03/0.03 & 86.3 / 16.0 & 68.0 & 0.11/0.02 & 88.5 / 17.2 (+1.2) & 67.6 & 0.20/0.01 & 88.7 / 17.3 (+1.3) \\
Broderick & 61.0 & 0.03/0.03 & 75.9 / 24.0 & 61.1 & 0.10/0.02 & 77.2 / 25.7 (+0.3) & 60.8 & 0.19/0.02 & 78.7 / 26.2 (+1.6) \\
\bottomrule\end{tabular}
}
\end{table*}

\subsection{Training-window length}\label{app:conf:trainlen}
\begin{figure*}[h]\centering
\includegraphics[width=0.95\textwidth]{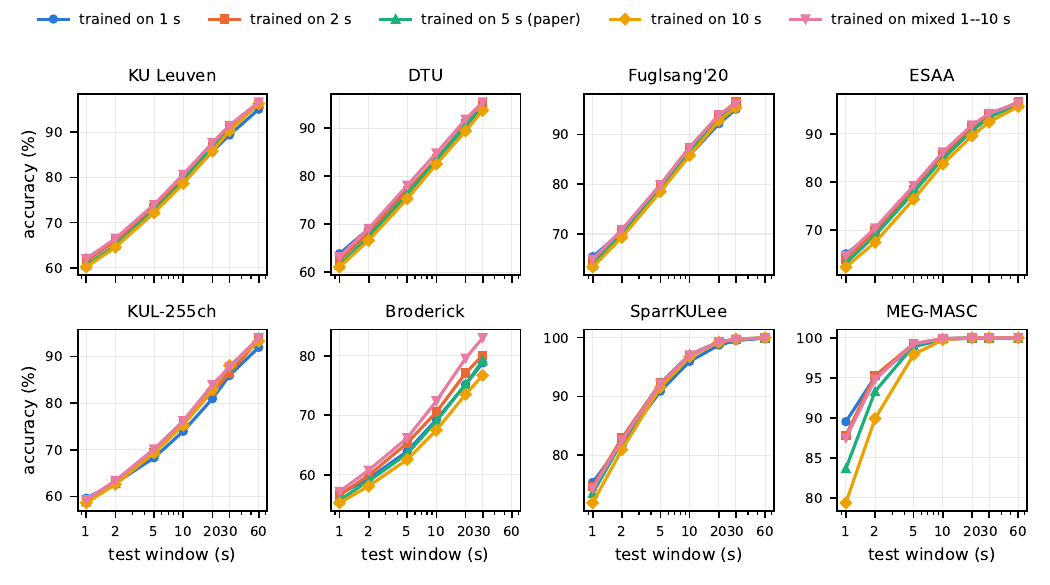}
\caption{Test accuracy against test window for CoupleFlow trained on 1-, 2-, 5- (paper),
10-s crops or a length redrawn from $\{1,2,5,10\}$\,s every epoch; everything else as in the
paper (selection on 5-s validation windows). Mean of 3 seeds, test folds; competing talker for
two-talker sets, match--mismatch for SparrKULee and MEG-MASC.}\label{fig:conf_trainlen}
\end{figure*}

The paper's models are trained on 5-s crops and applied at every test window. We retrained
CoupleFlow with 1-, 2- and 10-s crops, and with a crop length redrawn from $\{1,2,5,10\}$\,s
at every epoch (``mixed''), keeping everything else fixed: the same folds, seeds, losses,
epochs, and model selection on 5-s validation windows (Fig.~\ref{fig:conf_trainlen},
Table~\ref{tab:conf_trainlen}). Differences below are against the paper's 5-s training, mean of
3 seeds, over all 8 datasets.

\emph{Short crops help short windows.} Training on 1-s crops raises 1-s accuracy by
\confTlOneOneSMin{} to \confTlOneOneSMax{} points ($n=$\,\confTlOneOneSN{} datasets), but
changes 5-s accuracy by \confTlOneFiveSMin{} to \confTlOneFiveSMax{} and 30-s accuracy by
\confTlOneThirtySMin{} to \confTlOneThirtySMax. \emph{Long crops help nothing}: 10-s training
changes 1-s accuracy by \confTlTenOneSMin{} to \confTlTenOneSMax{} and 5-s accuracy by
\confTlTenFiveSMin{} to \confTlTenFiveSMax{} points.

\emph{Mixed crops are best overall.} Redrawing the crop length raises 1-s accuracy on every dataset
(\confTlMixOneSMin{} to \confTlMixOneSMax{} points, $n=$\,\confTlMixOneSN) and lowers neither
5-s accuracy (\confTlMixFiveSMin{} to \confTlMixFiveSMax) nor 30-s accuracy
(\confTlMixThirtySMin{} to \confTlMixThirtySMax) by more than 0.1 points. In the terms of Prop.~\ref{prop:gauss} it raises
the information rate of the 1-s evidence: $\kappa$ of the summed 1-s log-odds grows by a factor
\confTlMixKapMin--\confTlMixKapMax, and the sequential rule at $A{=}2$ needs
\confTlMixTimeMin--\confTlMixTimeMax{} of the time (\confTlOneTimeMin--\confTlOneTimeMax{}
for 1-s training). A plausible reading, which we have not tested further: a model that only sees long crops learns a coupling that is reliable when
averaged over many seconds but noisy within one; short crops force every second to carry
evidence, and mixing lengths keeps the long-window accuracy. The paper's main results use
5-s training throughout; mixed-length training is a free improvement for any window.

\begin{table*}[h]\centering\small
\caption{Training-window ablation: accuracy (\%, mean$\pm$s.d.\ over 3 seeds) at 1-, 5- and
30-s test windows, $\kappa$ of the summed 1-s log-odds, and the sequential rule at $A{=}2$
(two-talker sets).}\label{tab:conf_trainlen}
\resizebox{0.8\textwidth}{!}{\begin{tabular}{llccccc}\toprule
Dataset & training window & 1\,s & 5\,s & 30\,s & $\kappa$ (1-s sums) & seq.\ $A{=}2$: acc.\ / time \\\midrule
KU Leuven & 1 s & 62.1{\scriptsize$\pm$0.3} & 72.9{\scriptsize$\pm$0.6} & 89.4{\scriptsize$\pm$0.2} & 0.251 & 85.4 / 8.1 \\
 & 2 s & 61.5{\scriptsize$\pm$0.2} & 73.4{\scriptsize$\pm$0.6} & 91.2{\scriptsize$\pm$0.3} & 0.252 & 86.2 / 9.3 \\
 & 5 s (paper) & 60.5{\scriptsize$\pm$0.1} & 72.7{\scriptsize$\pm$0.2} & 90.5{\scriptsize$\pm$0.4} & 0.240 & 86.6 / 10.9 \\
 & 10 s & 60.2{\scriptsize$\pm$0.1} & 72.2{\scriptsize$\pm$0.6} & 90.3{\scriptsize$\pm$0.5} & 0.231 & 86.4 / 11.8 \\
 & mixed 1--10 s & 62.0{\scriptsize$\pm$0.3} & 74.0{\scriptsize$\pm$0.4} & 91.5{\scriptsize$\pm$0.6} & 0.259 & 86.5 / 8.7 \\
\midrule
DTU & 1 s & 63.9{\scriptsize$\pm$0.0} & 77.1{\scriptsize$\pm$0.3} & 94.3{\scriptsize$\pm$0.2} & 0.308 & 86.9 / 6.6 \\
 & 2 s & 62.5{\scriptsize$\pm$0.1} & 76.9{\scriptsize$\pm$0.5} & 94.8{\scriptsize$\pm$0.8} & 0.287 & 86.2 / 7.7 \\
 & 5 s (paper) & 61.4{\scriptsize$\pm$0.3} & 76.2{\scriptsize$\pm$0.6} & 94.5{\scriptsize$\pm$0.8} & 0.267 & 85.5 / 8.8 \\
 & 10 s & 61.1{\scriptsize$\pm$0.2} & 75.3{\scriptsize$\pm$0.3} & 93.7{\scriptsize$\pm$0.8} & 0.259 & 85.3 / 9.8 \\
 & mixed 1--10 s & 63.2{\scriptsize$\pm$0.1} & 78.1{\scriptsize$\pm$0.2} & 95.4{\scriptsize$\pm$0.2} & 0.304 & 86.5 / 6.9 \\
\midrule
Fuglsang'20 & 1 s & 65.4{\scriptsize$\pm$0.1} & 78.8{\scriptsize$\pm$0.2} & 95.1{\scriptsize$\pm$0.5} & 0.331 & 87.5 / 6.0 \\
 & 2 s & 64.5{\scriptsize$\pm$0.3} & 79.9{\scriptsize$\pm$0.4} & 96.5{\scriptsize$\pm$0.2} & 0.330 & 87.5 / 6.6 \\
 & 5 s (paper) & 63.8{\scriptsize$\pm$0.3} & 79.6{\scriptsize$\pm$0.8} & 95.9{\scriptsize$\pm$1.3} & 0.318 & 87.5 / 7.3 \\
 & 10 s & 63.4{\scriptsize$\pm$0.2} & 78.5{\scriptsize$\pm$0.1} & 95.4{\scriptsize$\pm$0.3} & 0.312 & 87.7 / 7.8 \\
 & mixed 1--10 s & 64.9{\scriptsize$\pm$0.2} & 79.9{\scriptsize$\pm$0.4} & 96.2{\scriptsize$\pm$0.2} & 0.339 & 87.9 / 6.4 \\
\midrule
ESAA & 1 s & 65.1{\scriptsize$\pm$0.1} & 78.3{\scriptsize$\pm$0.4} & 94.2{\scriptsize$\pm$0.6} & 0.316 & 87.2 / 6.0 \\
 & 2 s & 64.0{\scriptsize$\pm$0.3} & 78.4{\scriptsize$\pm$0.4} & 93.9{\scriptsize$\pm$0.3} & 0.309 & 87.6 / 6.9 \\
 & 5 s (paper) & 63.2{\scriptsize$\pm$0.3} & 77.9{\scriptsize$\pm$0.3} & 93.4{\scriptsize$\pm$0.5} & 0.292 & 87.3 / 7.7 \\
 & 10 s & 62.3{\scriptsize$\pm$0.4} & 76.4{\scriptsize$\pm$0.1} & 92.4{\scriptsize$\pm$0.3} & 0.275 & 87.0 / 9.0 \\
 & mixed 1--10 s & 64.7{\scriptsize$\pm$0.1} & 79.2{\scriptsize$\pm$0.1} & 94.2{\scriptsize$\pm$0.4} & 0.310 & 87.4 / 6.4 \\
\midrule
KUL-255ch & 1 s & 59.6{\scriptsize$\pm$0.3} & 68.3{\scriptsize$\pm$0.2} & 85.9{\scriptsize$\pm$0.4} & 0.200 & 84.2 / 12.3 \\
 & 2 s & 59.0{\scriptsize$\pm$0.3} & 69.3{\scriptsize$\pm$0.6} & 86.4{\scriptsize$\pm$0.6} & 0.208 & 84.7 / 13.2 \\
 & 5 s (paper) & 58.9{\scriptsize$\pm$0.3} & 69.4{\scriptsize$\pm$0.4} & 87.7{\scriptsize$\pm$0.9} & 0.203 & 86.0 / 14.7 \\
 & 10 s & 58.6{\scriptsize$\pm$0.3} & 69.4{\scriptsize$\pm$0.4} & 88.1{\scriptsize$\pm$1.2} & 0.201 & 85.1 / 14.4 \\
 & mixed 1--10 s & 59.3{\scriptsize$\pm$0.2} & 70.1{\scriptsize$\pm$0.3} & 87.7{\scriptsize$\pm$0.6} & 0.214 & 85.5 / 12.7 \\
\midrule
Broderick & 1 s & 56.8{\scriptsize$\pm$0.3} & 64.0{\scriptsize$\pm$0.5} & 78.8{\scriptsize$\pm$0.3} & 0.149 & 79.4 / 17.6 \\
 & 2 s & 56.6{\scriptsize$\pm$0.5} & 65.4{\scriptsize$\pm$1.0} & 80.1{\scriptsize$\pm$2.6} & 0.146 & 80.0 / 18.9 \\
 & 5 s (paper) & 55.8{\scriptsize$\pm$0.7} & 63.6{\scriptsize$\pm$1.8} & 79.2{\scriptsize$\pm$2.6} & 0.131 & 77.7 / 22.0 \\
 & 10 s & 55.3{\scriptsize$\pm$0.9} & 62.6{\scriptsize$\pm$2.1} & 76.7{\scriptsize$\pm$3.5} & 0.119 & 76.6 / 24.1 \\
 & mixed 1--10 s & 57.2{\scriptsize$\pm$0.4} & 66.2{\scriptsize$\pm$0.6} & 83.0{\scriptsize$\pm$1.0} & 0.160 & 81.0 / 16.7 \\
\midrule
SparrKULee & 1 s & 75.4{\scriptsize$\pm$0.1} & 90.9{\scriptsize$\pm$0.1} & 99.5{\scriptsize$\pm$0.0} & -- & -- \\
 & 2 s & 74.6{\scriptsize$\pm$0.1} & 92.3{\scriptsize$\pm$0.1} & 99.7{\scriptsize$\pm$0.0} & -- & -- \\
 & 5 s (paper) & 73.5{\scriptsize$\pm$0.0} & 92.2{\scriptsize$\pm$0.0} & 99.7{\scriptsize$\pm$0.1} & -- & -- \\
 & 10 s & 71.9{\scriptsize$\pm$0.0} & 91.5{\scriptsize$\pm$0.1} & 99.7{\scriptsize$\pm$0.0} & -- & -- \\
 & mixed 1--10 s & 74.4{\scriptsize$\pm$0.2} & 92.2{\scriptsize$\pm$0.0} & 99.7{\scriptsize$\pm$0.0} & -- & -- \\
\midrule
MEG-MASC & 1 s & 89.5{\scriptsize$\pm$0.2} & 98.9{\scriptsize$\pm$0.1} & 100.0{\scriptsize$\pm$0.0} & -- & -- \\
 & 2 s & 87.8{\scriptsize$\pm$0.3} & 99.3{\scriptsize$\pm$0.1} & 100.0{\scriptsize$\pm$0.0} & -- & -- \\
 & 5 s (paper) & 83.7{\scriptsize$\pm$0.4} & 99.0{\scriptsize$\pm$0.1} & 100.0{\scriptsize$\pm$0.0} & -- & -- \\
 & 10 s & 79.4{\scriptsize$\pm$0.5} & 97.9{\scriptsize$\pm$0.1} & 100.0{\scriptsize$\pm$0.0} & -- & -- \\
 & mixed 1--10 s & 87.5{\scriptsize$\pm$0.7} & 99.2{\scriptsize$\pm$0.1} & 100.0{\scriptsize$\pm$0.0} & -- & -- \\
\bottomrule\end{tabular}
}
\end{table*}

\end{document}